\documentclass[11pt,a4paper]{article}
\usepackage{jheppub}
\usepackage{amsmath,amsfonts,amssymb,amsthm,bbm,enumerate,times}
\usepackage{mathtools}
\usepackage{xcolor}
\usepackage{float}
\usepackage{etoolbox}
\usepackage{comment}
\usepackage{xfrac}

\usepackage{dsfont}
\newcommand{\id}{\mathds{1}}

\newcommand{\tr}{\mathrm{tr}}

 \renewcommand{\i}{\,\ensuremath\mathrm{i}}

\newcommand{\sgn}{\operatorname{sgn}}

\newcommand{\ket}[1]{\left.\left|{#1}\right.\right\rangle}

\newcommand{\bra}[1]{\left.\left\langle{#1}\right.\right|}

\newcommand\vacket{{\ket{\emptyset}}}

\newcommand{\vectwo}[2]{\begin{pmatrix} {#1} \\ {#2} \end{pmatrix}}

\DeclareMathAlphabet{\mathpzcc}{OT1}{pzc}{m}{it}
\DeclareMathAlphabet{\mathpzc}{T1}{pzc}{m}{it}{\huge}
\newcommand{\msp}{\phantom{-}}
\newcommand{\m}{\operatorname{\gamma}}

\newcommand{\fe}{\operatorname{f}^{\phantom{\dagger}}}

\newcommand{\fd}{\operatorname{f}^\dagger}

\def\and{\quad\text{and}\quad}

\title{Boundary theories of critical matchgate \\tensor networks}
\author[a,b]{A.\ Jahn,}
\author[b,c]{M.\ Gluza,}
\author[b]{C.\ Verhoeven,}
\author[d,e]{S.\ Singh}
\author[b,f]{and J.\ Eisert}
\affiliation[a]{Institute for Quantum Information and Matter, Caltech, Pasadena, CA 91125, USA}
\affiliation[b]{Dahlem Center for Complex Quantum Systems, 
Freie Universit{\"a}t Berlin, 14195 Berlin, Germany}
\affiliation[c]{School of Physical and Mathematical Sciences, Nanyang Technological University, 637371 Singapore, Republic of Singapore}
\affiliation[d]{Max-Planck Institute for Gravitational Physics (Albert  Einstein  Institute), 
14476 Potsdam, Germany}
\affiliation[e]{School of Physics and Astronomy, Monash University, Victoria, Australia}
\affiliation[f]{Helmholtz-Zentrum Berlin f{\"u}r Materialien und Energie, 14109 Berlin, Germany}

\emailAdd{a.jahn@fu-berlin.de}
\emailAdd{marekgluza@zedat.fu-berlin.de}
\emailAdd{charlotte.verhoeven@fu-berlin.de}
\emailAdd{sukhi.singh@monash.edu}
\emailAdd{jense@zedat.fu-berlin.de}

\abstract{
Key aspects of the AdS/CFT correspondence can be captured in terms of tensor network models on hyperbolic lattices. For tensors fulfilling the matchgate constraint, these have previously been shown to produce disordered boundary states whose site-averaged ground state properties match the translation-invariant critical Ising model. In this work, we substantially sharpen this relationship by deriving disordered local Hamiltonians generalizing the critical Ising model whose ground and low-energy excited states are accurately represented by the matchgate ansatz without any averaging. We show that these Hamiltonians exhibit multi-scale quasiperiodic symmetries captured by an analytical toy model based on layers of the hyperbolic lattice, breaking the conformal symmetries of the critical Ising model in a controlled manner.
We provide a direct identification of correlation functions of ground and low-energy excited states between the disordered and translation-invariant models and give numerical evidence that the former approaches the latter in the large bond dimension limit. This establishes tensor networks on regular hyperbolic tilings as an effective tool for the study of conformal field theories. Furthermore, our numerical probes of the bulk parameters corresponding to boundary excited states constitute a first step towards a tensor network bulk-boundary dictionary between regular hyperbolic geometries and critical boundary states.
}

\begin{document}
\maketitle

\section{Introduction}

Tensor networks are a widely applied tool in the study of many-body systems both in numerical approaches~\cite{Orus-AnnPhys-2014,schollwock2011density,PAECKEL2019167998} and analytical techniques~\cite{cirac2020matrix}, providing a variational ansatz for a large class of quantum states.
A tensor network depends on the choice of tensors as well as the \emph{bulk} geometry of how they are contracted together, both determining the properties of the quantum state on the network's \emph{boundary}, i.e., its uncontracted indices.
The choice of bulk geometry restricts many of the qualitative properties of possible boundary states, in particular their entanglement structure.
Ground states of systems in $1{+}1$ dimensions with an energy gap, exhibiting an entanglement entropy area law \cite{Hastings_2007}, can be efficiently parametrized by a one-dimensional chain of tensors \cite{verstraete2006matrix}. Such a \emph{matrix product state} (MPS) thus possesses the same geometry as the boundary system it describes; its bulk geometry is trivial.
For critical systems without a characteristic length scale and polynomial decay of correlations, whose continuum limit can often be described by a \emph{conformal field theory} (CFT), such an ansatz is insufficient for describing the entanglement structure at large scales. Instead, a bulk geometry with an additional spatial dimension beyond the ones corresponding to the modeled quantum system become necessary. On an intuitive level, this additional dimension captures the various length scales at which the system can be probed, all of which contribute equally to the scale-invariant theory describing such tensor network states.

The \emph{multi-scale entanglement renormalization ansatz} (MERA) provides such a tensor network model, efficiently representing low-energy states of critical quantum systems in $1{+}1$ dimensions \cite{Vidal:2007hda,PhysRevLett.100.130501,MERAAlgorithms}. With the MERA one can accurately extract conformal data (central charge, scaling dimensions and operator product expansion coefficients) of the corresponding continuum CFT \cite{Pfeifer:2009criticalMERA}.
Moreover, MERA allows for an explicit representation of lattice CFT transformations \cite{Milstead:2018MERAconformal}. 
The shortest path between two boundary points in the MERA network scales logarithmically in boundary distance, which resembles the geometry of hyperbolic space \cite{Evenbly:2011TNgeometry} and leads to the correct entanglement entropy scaling of a CFT ground state.
This has led to the proposal that the MERA -- or more generally, the process of \emph{entanglement renormalization} \cite{Vidal:2007hda} -- serves as a discrete implementation of a \emph{holographic duality} \cite{PhysRevD.86.065007,Swingle:2012wq,Evenbly:2011TNgeometry,PhysRevD.97.026012,Qi:2013caa,Singh:2017tet}. 
Such a duality appears most prominently in the \emph{anti-de Sitter/conformal field theory (AdS/CFT)} correspondence \cite{Maldacena98,Witten:1998qj} relating (quantum) gravity on a negatively curved AdS space-time to a dual CFT in one lower dimension.
As in the MERA, AdS/CFT thus features a hyperbolic bulk related to a flat boundary.
Entanglement entropies of a \emph{holographic CFT} with a dual bulk description can be computed on the gravitational side using the Ryu-Takayanagi approach \cite{PhysRevLett.96.181602,Ryu_2006,Hubeny_2007}, leading to the same scaling as found for boundary states of the MERA.

\begin{figure}[t]
\centering
\includegraphics[width=0.75\textwidth]{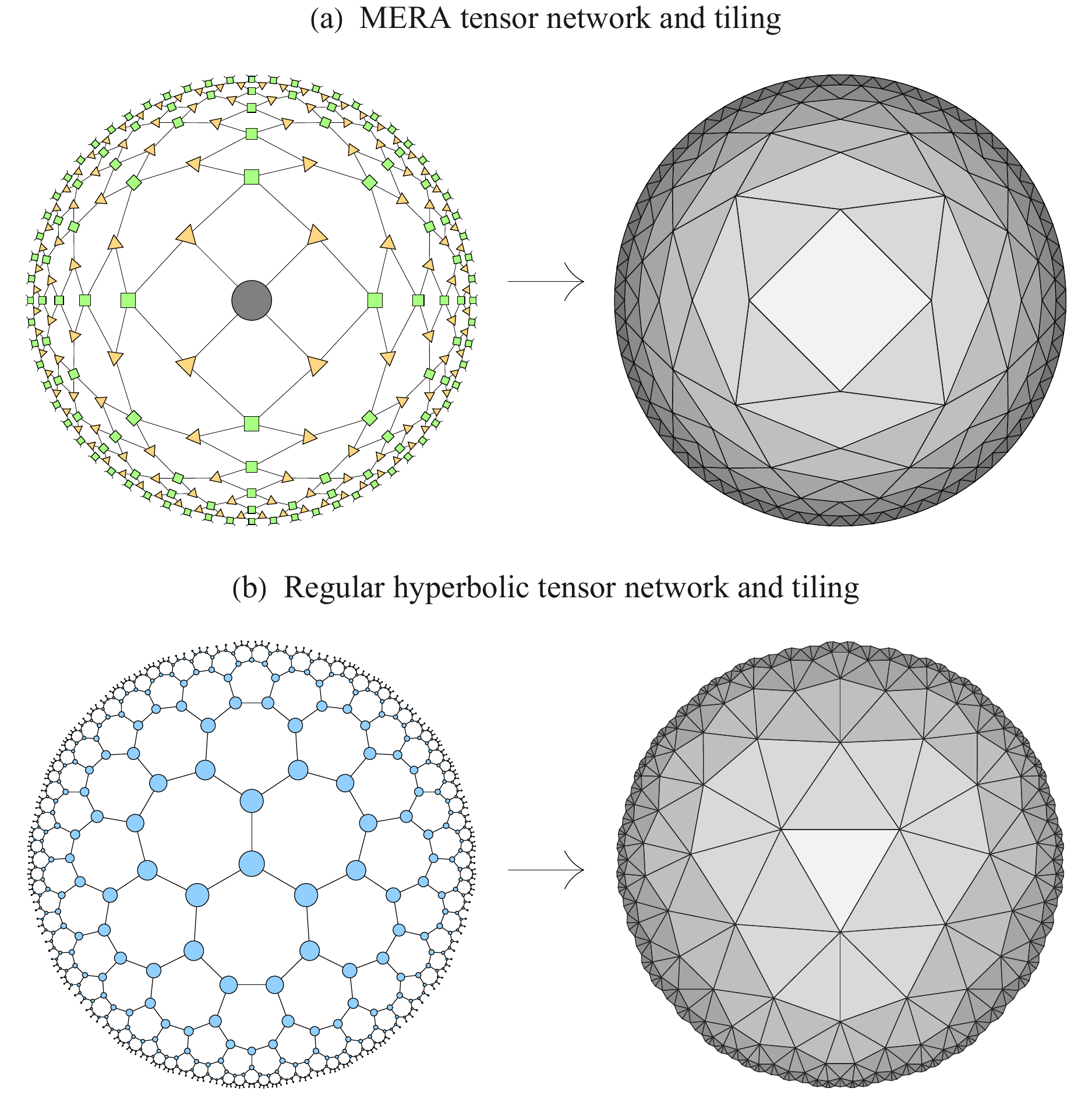}

\caption{The MERA and a regular hyperbolic tensor network with their corresponding tiling in the hyperbolic disk, organized into radial layers (shaded).
(a) The MERA consists of two types of tensors (isometries and disentanglers, drawn here as triangles and squares) arranged in layers around a central tensor.
It corresponds to an irregular, alternating triangle/quadrilateral tiling that breaks the symmetries of the hyperbolic disk but produces a smooth boundary after many layers $n$. 
(b) A regular triangular hyperbolic tensor network with one type of tensor (drawn as a circle) with the same geometrical structure around any lattice vertex, preserving a discrete subset of the symmetries of the hyperbolic disk.
Unlike the MERA, the tiling has no directionality or well-defined center; however, its boundary exhibits quasiperiodic irregularity at any $n$.
}
\label{FIG_MERA_VS_REG}
\end{figure}

While the connectivity of the MERA resembles hyperbolic AdS time-slices, it cannot be regularly embedded into the hyperbolic disk.
The resemblance to discretized time-like slices in positively curved \emph{de Sitter spacetime} (dS) has been pointed out \cite{Beny:2011vh}, though MERA has also been interpreted as a discretized path integral on light-like slices in AdS \cite{Milsted:2018san}.
Instead, tensor networks on \emph{regular} discretizations of AdS time-slices have been proposed. These first appeared in tensor network approaches to holographic dualities distinct from entanglement renormalization: Building on the notion of \emph{holographic quantum error correction} \cite{Almheiri15} describing AdS/CFT as an encoding map between bulk and boundary information, it was found that discrete holographic codes on such geometries can be implemented in a large number of toy models of holography~\cite{Pastawski2015,Latorre:2015xna,Harlow:2016vwg,Donnelly2017,Hayden2016,Osborne:2017woa,PhysRevA.98.052301,Jahn:2019nmz,Gesteau:2020hoz,Cao:2020ksw,Review}. 
As visualized in Fig.\ \ref{FIG_MERA_VS_REG}, regular hyperbolic lattices no longer possess the inherent directionality of the MERA but obey a discrete subset of the symmetries of the hyperbolic disk: Around each vertex, the lattice exhibits the same (discrete) geometry.
Specifically, regular hyperbolic tilings break the continuum $PSL(2,\mathbb{R})$ symmetries of the hyperbolic disk into a \emph{Fuchsian group} \cite{Osborne:2017woa,Jahn:2020ukq,Boettcher:2021njg}.

While regular tilings appear to be a more natural discrete geometry for studying AdS/CFT, they come with a trade-off: Rather than the smooth boundary of the MERA geometry, specifically designed to produce translation-invariant states through the use of local \emph{disentanglers} on every layer, the geometrical boundary of a regular hyperbolic tiling is inherently \emph{quasiperiodic}, i.e., breaking periodicity on every scale up to aperiodically repeating features \cite{Boyle:2018uiv}.
To elucidate the general behaviour of boundary states resulting from such bulk geometries, some of the authors previously considered a generic \emph{matchgate tensor network} (MTN), which is numerically efficient to contract and thus suitable for studying the whole parameter space of such a setup \cite{Jahn:2017tls}. This circumvents the need for isometric or causal constraints to reduce computational effort, as employed in the MERA as well as \emph{hyperinvariant tensor network} approaches \cite{PhysRevLett.119.141602,Steinberg:2020bef}, while also incorporating the \emph{hyperbolic pentagon code} \cite{Pastawski2015} into a more general framework.
Boundary states from MTNs on regular hyperbolic tilings were found to exhibit aperiodic disorder in their correlation functions, but surprisingly, the parameter space of these models contains boundary states whose correlation functions, when \emph{averaged} over all boundary sites, accurately reproduce ground states of the critical Ising model \cite{Jahn:2017tls}. This is surprising, as critical Ising ground states are translation-invariant and one should expect any disorder to alter their correlation functions in an uncontrolled manner.

In this work, we explain why and how ground states of translation-invariant, critical models can be well approximated by boundary states of regular hyperbolic tensor network, again relying on the MTN framework for concrete numerical results.
We begin in Sec.~\ref{S_BACKGROUND} with a review of regular tilings and their iterative construction via inflation rules as well as of matchgate tensors and the covariance matrix formalism for Gaussian states.
In addition, we recap the general properties of hyperbolic MTN boundary states and their resemblance to ground states of the Ising model.

Complementing these previous numerical observations with analytical techniques, we introduce in Sec.~\ref{SS_DISORDERED_ISING} a class of disordered Ising models with nearest-neighbor couplings whose ground state covariance matrix can be explicitly mapped to that of its translation-invariant relative.
We then show in Sec.~\ref{SS_MQI} that a sub-class of these models with coupling terms determined by a highly constrained \emph{multi-scale quasicrystal ansatz} (MQA), built from the bulk symmetries of a given regular tiling, produces ground states closely resembling those found in the hyperbolic MTN setup. The MQA effectively depends on only a single free parameter determining the disorder strength.
The relationship between the resulting \emph{multi-scale quasicrystal Ising} (MQI) model, the more general \emph{mode-disordered Ising} (MDI) model, and the original Ising model is summarized in Fig.\ \ref{FIG_MODEL_SPACE}.
In Sec.\ \ref{SS_BOUNDARY_SYMMETRIES} we discuss the boundary symmetries of these constructions and relate them to the recent proposal that regular hyperbolic TNs implement \emph{quasiperiodic CFTs} whose CFT-like state symmetries are determined by the bulk geometry \cite{Jahn:2020ukq}.
Specifically, we show that the qCFT symmetries realized by hyperbolic MTNs include a fractal self-similarity and an approximate translation invariance of correlation functions. In contrast to the qCFT construction in Ref.\ \cite{Jahn:2020ukq} using \emph{Majorana dimer states}, in the MTN setup we can define a nearest-neighbor Hamiltonian generalizing a well-known CFT lattice model -- the Ising model -- and are thus able to sharpen the relationship between qCFTs and CFTs. 

While critical models have been previously identified in systems with quasiperiodic couplings \cite{JuhaszZimboras2007,StrongDisorder,Crowley_2018,Crowley_2018_2,Jahn:2019mbb,2020NatCo.11.2225A}, we here provide the first tensor network ansatz leading to such an effective Hamiltonian.
We numerically demonstrate in Sec.\ \ref{SS_CONTINUUM_LIMIT} that the quasiperiodic disorder is weakened at large bond dimension, with MTN states converging towards the \emph{translation-invariant} ground states of the critical Ising model.
Furthermore, we show analytically in Sec.\ \ref{SS_DISORDERED_CORRELATION_FUNTIONS} that even at small bond dimension, the quasiperiodic nature of the disorder leads to site-averaged correlation functions that reproduce the continuum Ising model, explaining the results of Ref.\ \cite{Jahn:2017tls}. 
Finally, in Sec.\ \ref{SS_EXCITATIONS_AND_HOLOGRAPHY} we consider excitations beyond the ground state produced by deformations of the bulk tensors from their critical Ising value in the MTN ansatz, and find a ``holographic'' relationship between the location of the deformation in the hyperbolic lattice and the energy scale of the excitation. We further sharpen this relationship by numerically constructing the first low-lying excited states of the boundary Hamiltonian (visualized in Fig.\ \ref{FIG_EXC_PARAM}), demonstrating a concrete dictionary between bulk tensor configurations and boundary energy spectra, resembling the dictionary between bulk fields and boundary operators in AdS/CFT \cite{PhysRevD.88.026003,harlow2011operator}.

We close with a discussion on our results in Sec~\ref{S_DISCUSSION}. In particular, we note that our numerical results rely on the MTN ansatz  corresponding to a restriction to fermionic Gaussian states, but as many of its properties appear as a consequence of the tensor network geometry they should be expected in models with more general types of tensors, as well.

\section{Background}

\label{S_BACKGROUND}
\subsection{Regular hyperbolic tilings and their boundaries}
\label{SS_REG_TILINGS_QUASIP}

Throughout this work we consider tensor networks on regular hyperbolic tilings, with a strong focus on how the symmetries of the tiling boundary are reflected in the symmetries of the boundary states of the tensor network.
A useful notation for these highly symmetric tilings is given by the \emph{Schl\"afli symbol} $\{p,q\}$, denoting a tiling with regular $p$-gons with $q$ of them meeting at each corner. This implies an interior angle of $2\pi/q$ in each $p$-gon corner, which leads to a hyperbolic tiling with negative (Gaussian) curvature if $p q > 2(p+q)$.
While such a tiling is regular in the bulk, cutting it off after a finite number of suitably defined layers produces a boundary whose geometry is highly non-regular: Indeed one finds it to be\ \emph{quasiperiodic}, i.e., having no exact periodicity but containing aperiodically repeating features~\cite{Boyle:2018uiv}.
\begin{figure}[t]
\centering
\includegraphics[width=1.0\textwidth]{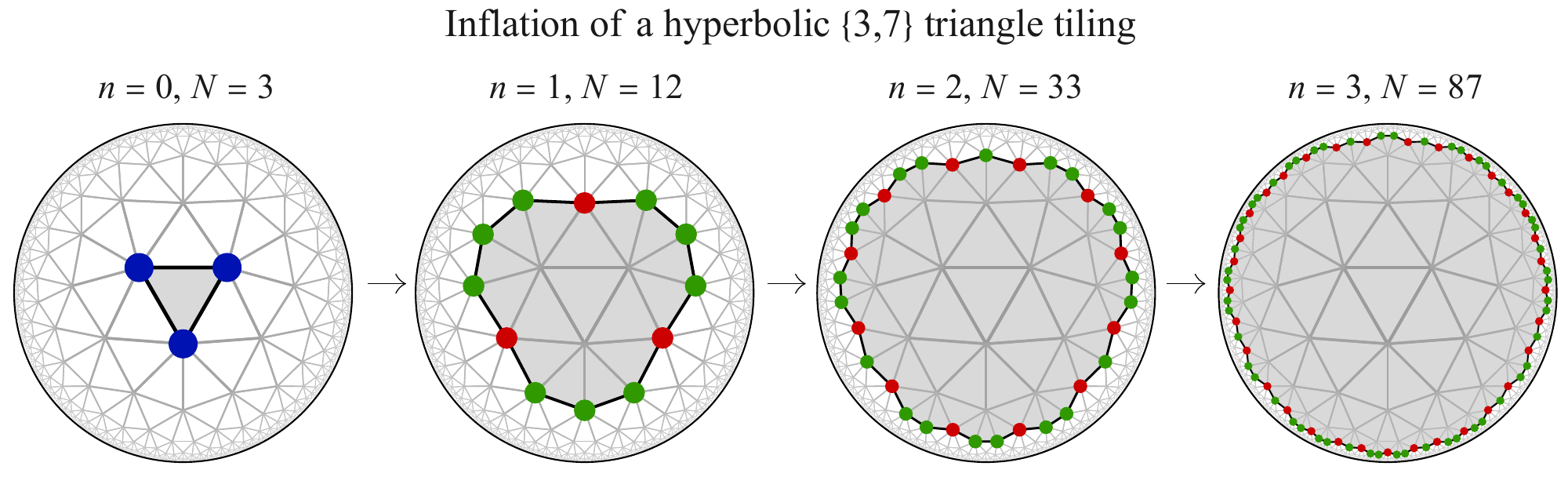}\\
\includegraphics[width=1.0\textwidth]{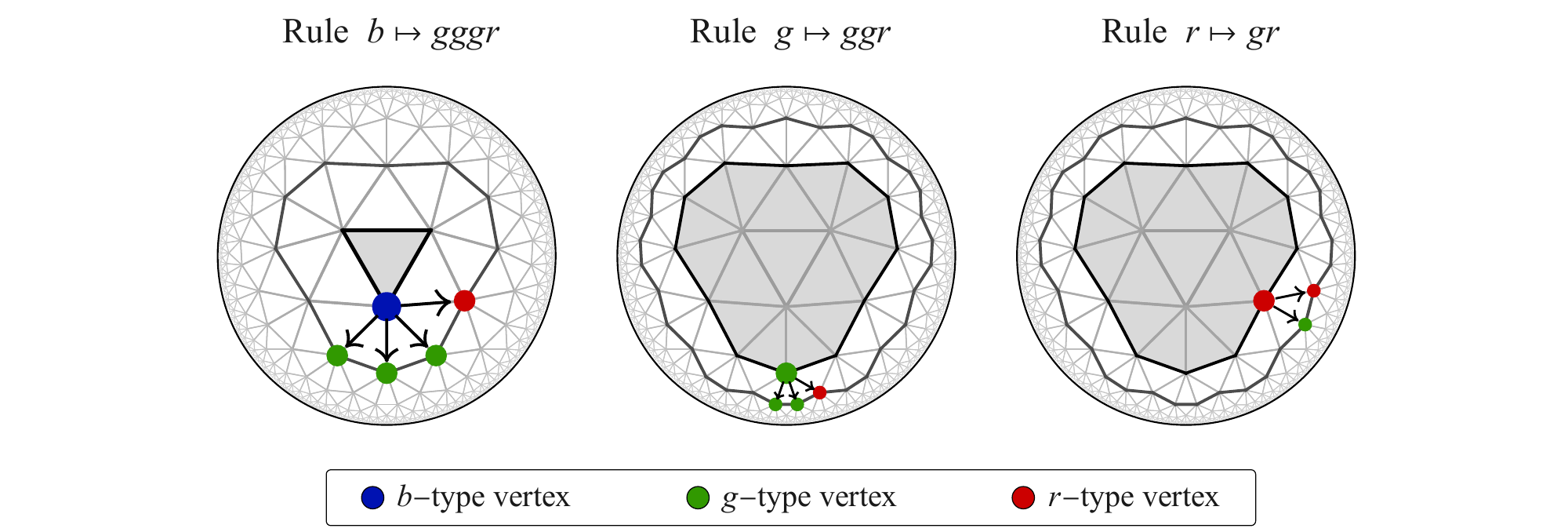}
\caption{Iterative construction of a regular hyperbolic tiling through inflation rules \eqref{EQ_V_INFLATION_37} of a $\{3,7\}$ triangle tiling. With each additional iteration $n$, a connected layer of tiles is added, leading to an exponentially increasing number $N$ of boundary vertices.
The types $b,g,r$ denote boundary vertices with $0,1,2$ adjacent edges connected to the previous inflation layer, respectively.
The inflation procedure for any tiling can be associated with an iterative tensor network contraction, with each $p$-gon tile representing an $p$-index tensor, and adjacent edges between tiles within the region covered by the inflation step (grey-shaded area) corresponding to contracted pairs of tensor indices. 
Note that this visualization is dual to the common convention of identifying vertices, rather than tiles, with individual tensors.
}
\label{FIG_TILING_INFLATION}
\end{figure}
Such quasiperiodicity, sometimes also referred to as \emph{quasicrystal symmetries}, can be characterized by \emph{inflation rules} parametrizing the construction of tiling layers.
These rules allow us to obtain the sequence of the types of corners exposed at the boundary of the tiling from the sequences describing the boundary of a smaller tiling embedded in its interior.
Here we will always consider this construction in terms of a starting tile onto which concentric layers of new tiles are ``glued'' upon, thus producing a progression of tilings of (exponentially) increasing size, motivating the name \emph{inflation}. 
Fig.~\ref{FIG_TILING_INFLATION} demonstrates two examples of this procedure for $\{p,q\}=\{3,7\}$ using \emph{vertex inflation}, where each iteration adds a closed layer of tiles that are adjacent to the vertices of the previous tiling boundary. These boundary vertices (small circles in Fig.\ \ref{FIG_TILING_INFLATION}) can be classified by their adjacency with respect to the tiles included up to the given inflation step: Blue vertices (denoted $b$) are connected only to other boundary vertices, while green and red vertices ($g$ and $r$, respectively) are each connected to one and two interior vertices on previous inflation layers. 

We briefly recapitulate how to define these inflation rules in terms of letter sequences:
Defining an alphabet of unique letters $\{l_i\}$, an inflation rule is a map $f$ from each letter $l_k$ to a string $\{l_i\}^{\times n_k}$ of $n_k$ letters.
Applying the rule $f$ to a string of letter is defined as applying $f$ to each string individually and concatenating the resulting strings.
We call this a \emph{local inflation} if $f$ is only applied to a subset of a total string, and a \emph{global inflation} if it is applied to the complete string. The string usually grows exponentially under repeated global inflation.
This general prescription is best exemplified by the \emph{Fibonacci quasicrystal}.
Its sequences are defined by considering two letters $a$ and $b$ and the inflation rule (letter replacement rules)
\begin{align}
 a \mapsto ab \quad \text{and} \quad b \mapsto a  \ ,
\end{align}
where we omitted the set notation for a string of letters.
For instance, starting from a seed string $aaa$, applying the inflation rule globally to all three letters produces $ababab$. Globally inflating the preceding string produces $abaabaaba$, and so on. Iterative applications of these rules produce strings of $a$'s and $b$'s with a quasiperiodic structure, in which the ratio of the number of $a$'s to the number of $b$'s converges to an asymptotic value as the string size grows. This asymptotic ratio is characteristic of the quasicrystal. For the Fibonacci case, this ratio asymptotes to the so-called ``golden ratio,'' $\frac{3 + \sqrt{5}}{2}$.

We now define the inflation rules for the $\{3,7\}$ tiling in Fig.\ \ref{FIG_TILING_INFLATION} in terms of the boundary vertex types $b,g,r$. This leads to the vertex inflation rule
\begin{align}
\label{EQ_V_INFLATION_37}
b &\mapsto gggr \ ,& 
g &\mapsto ggr \ , & 
r &\mapsto gr \ .
\end{align}
Note that the $b$ letter vanishes after the first inflation step and any larger sequences are fully characterized by $g$s and $r$s.
In fact, the remaining inflation rule for $g$ and $r$ is equivalent to a twofold application of the Fibonacci rule if we relabel letters $(g,r) \mapsto (a,b)$ and allow for an overall shift in the sequence. 
The boundary of the $\{3,7\}$ tiling under vertex inflation thus has the same geometric structure as a Fibonacci quasicrystal.

Inflation rules can be constructed for any regular hyperbolic tiling: In Appendix \ref{APP_45TILING} we consider the $\{4,5\}$ tiling that can be constructed in a similar fashion, shown in Fig.\ \ref{FIG_TILING_INFLATION2}.
In fact, the inflation rules for every $\{p,q\}$ tiling produce quasiperiodic sequences \cite{Boyle:2018uiv,Jahn:2019mbb}. 
While the inflated sequences share the symmetries of the initial sequence ($bbb$ with $\mathbb{Z}_3$ symmetry for the $\{3,7\}$ tiling), the scaling of the length of a sequence under a single inflation step quickly becomes independent of the starting seed sequence after a few inflation steps. Each inflation rule is thus characterized by a unique number $\lambda$ that defines the asymptotic scaling factor between steps. This factor can be computed analytically and is given by $\lambda_{\{3,7\}} = \frac{3 + \sqrt{5}}{2}$ for the $\{3,7\}$ tiling.

These purely geometrical considerations have physical implications: When constructing tensor networks with the geometry of regular hyperbolic tilings, correlations between boundary degrees of freedom are influenced by the tiling geometry.
Unsurprisingly, the quasiperiodic boundary geometry of these tilings implies that tensor networks on them will generally lead to boundary states that do not obey translation invariance.
However, it is not immediately obvious how exactly this quasiperiodicity is reflected in boundary correlation functions for generic choices of tensors.
As we will show, the critical states naturally encapsulated by such tensor networks indeed possess correlations appearing quasiperiodically disordered \emph{on all length scales} and are thus being influenced by the boundary symmetries at all inflation steps, rather than merely the final one.

\subsection{Matchgate tensor networks}
\label{S_MATCHGATE_TENSOR_NETWORKS}

We consider tensor networks composed of matchgate tensors \cite{Valiant2002}, whose contraction can be performed numerically very efficiently. The \emph{matchgate constraint} is an algebraic constraint on the entries of a given tensor which physically results in considering a variational ansatz restricted to \emph{fermionic Gaussian states}, i.e., ground states of non-interacting fermionic Hamiltonions \cite{Bravyi2008,Jahn:2017tls}.
In tensor network approaches contraction of individual tensors in the network is a key sub-routine and crucially the tensor obtained by contracting two matchgate tensor again retains the matchgate condition which facilitates efficient numerical contraction schemes~\cite{Bravyi2008}.
As an example, consider a tensor $T_{i,j,k}$ with three indices and bond dimension $2$, i.e., each index can take the value $0$ or $1$. If $T=T(A)$ is an (even-parity) matchgate tensor, it can be fully expressed by a $3 \times 3$ matrix $A$, which we call the \emph{generating matrix} (or \emph{squeezing matrix}) of the matchgate tensor. The relationship between $A$ and $T$ is most conveniently expressed in a fermionic Fock basis where the quantum state vector $\ket{\psi}_3$ on $3$ modes is expressed in terms of $T$ as 
\begin{align}
\ket{\psi}_3 = \sum_{i,j,k=0}^1 T_{i,j,k}\, \ket{i,j,k} \quad\text{with}\quad
\ket{i,j,k} := (\fd_1)^i (\fd_2)^j (\fd_3)^k \vacket_3 \ ,
\end{align}
where $\vacket_3$ is the fermionic vacuum on three modes with creation operators denoted by $\fd_k$. The extension to $N$ modes is straightforward. We can now express a matchgate tensor $T$ through $A$ via
\begin{align}
T_{i,j,k}(A) = \bra{i,j,k} c \exp\left( \frac{1}{2} \sum_{a,b=1}^3 A_{a,b} \fd_a \fd_b \right) \vacket_3 \ .
\end{align}
Here $c = T_{0,0,0}$ is a normalization constant. Due to anti-commutation relations $\{\fd_j,\fd_k\} = 0$, we find that the matrix $A$ can always be anti-symmetrized, so that only $3$ of its elements are relevant. This construction can also be easily extended to $N$ modes, and at large $N$, we immediately see the advantage of the matchgate representation: Instead of working with $2^N$ components of a tensor $T$, we instead use an antisymmetric $N \times N$ matrix $A$ with only $\frac{N(N-1)}{2}$ independent entries. Of course, matchgate tensors form a heavily restricted subclass of all possible tensors; specifically, they can only describe \emph{Gaussian} states that appear as states of non-interacting theories \cite{Bravyi2008,Jahn:2017tls}.

To give an example of the efficiency of matchgate tensor network (MTN) contractions, consider two 3-leg matchgate tensors $T_{i,j,k}(A)$ and $T_{l,m,n}(B)$ with $3 \times 3$ generating matrices $A$ and $B$. We specify an overall index ordering of $(i,j,k,l,m,n)$ that becomes necessary in the fermionic language of anti-commuting modes. We now perform a contraction of the index pair $(k,l)$, producing the tensor
\begin{equation}
T_{i,j,m,n}(C) \equiv T_{i,j,0}(A) T_{0,m,n}(B) + T_{i,j,1}(A) T_{1,m,n}(B) \ .
\end{equation}
Writing out the resulting expression side as a matchgate state, we find that its $4 \times 4$ generating matrix $C$ is given by
\begin{align}
C =
\begin{pmatrix}
0                & A_{1,2}         & A_{1,3} B_{4,5} & A_{1,3} B_{4,6} \\
-A_{1,2}         & 0               & A_{2,3} B_{4,5} & A_{2,3} B_{4,6} \\
-A_{1,3} B_{4,5} &-A_{2,3} B_{4,5} & 0               & B_{5,6}         \\
-A_{1,3} B_{4,6} &-A_{2,3} B_{4,6} &-B_{5,6}         & 0
\end{pmatrix} .
\end{align}
We find that this matrix $C$ is a direct sum of the ``uncontracted blocks'' of $A$ and $B$, along with diagonal elements that form products of matrix entries over the contracted indices. These contraction rules can be explicitly constructed for any $N$-mode matchgate tensor, including for self-contractions \cite{Jahn:2017tls}. As the computational cost for any single contraction is merely $O(N^2)$, this makes possible the contraction of matchgate tensor networks with thousands of bonds in a few seconds on standard desktop hardware.

MTN states, being Gaussian, are fully described by their \emph{covariance matrix} $\Gamma$ encoding two-point correlations, from which higher-order moments follow via the Wick's theorem. For $N$ physical fermions, this matrix can be conveniently expressed in terms of $2N$ self-adjoint Majorana operators $\m_k$ with $\{\m_j, \m_k\}=2\delta_{j,k}$ for Majorana labels $j,k$ which are related to the $N$ fermionic anihilation operators as $\fe_k = (\m_{2k-1} + \i\,\m_{2k})/2$.
The entries of the covariance matrix can then be written as
\begin{equation}
\label{EQ_COV_MATRIX_DEF}
\Gamma_{j,k} = \frac{\i\,}{2}\langle \psi | \left[ \m_j, \m_k \right] | \psi \rangle \ .
\end{equation}
For a matchgate tensor $T(A)$ with a general $N \times N$ generating matrix $A$, its covariance matrix $\Gamma=\Gamma(T(A))$ can be computed explicitly from $A$. Decomposing $\Gamma$ into submatrices $\Gamma^1$ to $\Gamma^4$ describing the correlations between even and odd Majorana modes, these submatrices are given by \cite{Windt:2020tra}
\begin{subequations}
\begin{align}
\Gamma^1_{j,k} &\equiv \Gamma_{2j-1,2k-1} 
= \Im\left( 2(\id + A)(\id + A^\dagger A)^{-1} \right) \ , \\
\Gamma^2_{j,k} &\equiv \Gamma_{2j-1,2k}
= \Re\left( (-\id - 2 A + A^\dagger A)(\id + A^\dagger A)^{-1} \right) \ , \\
\Gamma^3_{j,k} &\equiv \Gamma_{2j,2k-1} 
= \Re\left( (\id - A - 2 A^\dagger A)(\id + A^\dagger A)^{-1} \right)\ , \\
\Gamma^4_{j,k} &\equiv \Gamma_{2j,2k}
= \Im\left( 2(\id - A)(\id + A^\dagger A)^{-1} \right) \ ,
\end{align}
\end{subequations}
If $A$ is real, we thus find that correlations between two even or two odd Majorana modes vanish and only the even-odd correlations are non-trivial. The resulting checkerboard pattern of the covariance matrix is typical for eigenstates of quadratic Hamiltonians with coupling terms over an odd number of sites, such as the Ising model \cite{Jahn:2017tls}.

\subsection{Matchgate boundary states}
\label{SS_MATCHGATE_BOUNDARY_STATES}
We now briefly review previous results on matchgate tensor networks (MTNs) on regular tilings.
Ref.\ \cite{Jahn:2017tls} considered the simplest flat and hyperbolic triangular tilings, i.e., the $\{3,q\}$ case with $q=6$ (flat) and $q \geq 7$ (hyperbolic). For the minimal bond dimension case $\chi=2$ (with one spin or fermion per tiling edge), this leads to three degrees of freedom per matchgate tensor, given by the independent entries of its anti-symmetric $3\times 3$ generating matrix.
This setting is simplified further when assuming that the tensors respect the symmetries of the tiling itself: Regularity requires the geometry around any vertex to be indistinguishable and isotropic, which leads us to the construction of a tensor network of identical tensors $T$, each of which is invariant under cyclic permutations of its indices, i.e, $T_{j,k,l} = T_{k,l,j}$. 
On the level of the generating matrix $A$ defining each local matchgate tensor $T=T(A)$, this leads to the constraint $A_{1,2}=A_{1,3}=A_{2,3} \equiv a$. 
The state vector of an (even-parity) fermionic state constructed from $T(A)$ has the form
\begin{equation}
\label{EQ_matchgate_TRI}
\ket{\psi(a)}_3 = c \exp\left( a\, (\fd_1 \fd_2 + \fd_2 \fd_3 + \fd_1 \fd_3) \right) \vacket_3 \ .
\end{equation}
Here $c$ is a normalization constant.
While in general $a \in \mathbb{C}$ , it suffices to assume $a \in \mathbb{R}$ to produce ground states of local Hamiltonians (both for the three-site state and larger contractions thereof).
Clearly, by changing $a$ we can tune the coupling between fermionic modes, with $a=0$ corresponding to the vacuum state. 
Similar to how the fermionic vacuum can be represented in terms of maximally entangled pairs between the two Majorana modes composing each physical fermion -- a \emph{trival phase} in the language of topological order -- the $a=1$ case produces a \emph{topological phase} where pairs of Majorana modes between \emph{neighboring} sites are maximally entangled. For physical models with some decay of correlations over (tensor network) distance, the $0 < a < 1$ range is thus the relevant one for constructing physical MTN ground states.
We can visualize the full contraction of the MTN as a (fermionically ordered) projection of tensor products of \eqref{EQ_matchgate_TRI}, each contraction between a pair of indices being equivalent to a projection onto a Bell state of two fermions. 
In the tiling picture of Figs.\ \ref{FIG_MERA_VS_REG} and \ref{FIG_TILING_INFLATION}, each edge between two tiles corresponds to such a contraction, with the uncontracted boundary degrees of freedom being represented by the ``open'' edges on the boundary of the tiling.

What are the properties of the boundary states produced by this one-parameter MTN ansatz? As these states do not exhibit translation invariance, Ref.\ \cite{Jahn:2017tls} considered boundary properties averaged over all boundary sites. 
In particular, this includes the site-averaged correlation decay function
\begin{equation}
c(d) := \frac{1}{2N} \sum_{j=1}^{2N} |\Gamma_{j,j+d}| \ ,
\label{EQ_AVG_CORR_FALLOFF}
\end{equation}
computed from the $2N \times 2N$ covariance matrix $\Gamma = \Gamma(a)$ for a boundary state of a $\{p,q\}$ for a specific choice of $a$ on $N$ boundary sites. 
Here we defined covariance matrix indices up to modulo $2N$. The absolute value $|\Gamma_{j,j+d}|$ was taken to account for sign flips resulting from anti-periodic boundary conditions for fermionic systems, e.g.\ $\sgn\Gamma_{j,2N} = -\sgn\Gamma_{j+1,1}$.
Now, for the flat case $q=6$ one finds $c(d)$ to decay exponentially with $d$ for generic $a$ and polynomially only around a critical value $a \approx 0.58$.
In the hyperbolic case $q \geq 7$ the decay is always polynomial; still, there exists a special value of $a \approx 0.61$ where $c(d)$ decays with the smallest power $p=1$. 
This is also the decay power to be expected -- without averaging -- for the \emph{critical Ising model} corresponding to a CFT with central charge $c=\sfrac{1}{2}$ in the continuum limit.
Furthermore, the correlation decay between more complicated operators on the $\{3,7\}$ MTN boundary reproduces scaling dimensions and operator product expansion (OPE) coefficients consistent with the Ising CFT prediction, again under averaging over boundary sites.

Let us briefly review relevant properties of the critical Ising model, one of the simplest critical lattice models.
Written in terms of $2N$ Majorana operators $\m_k$, its Hamiltonian is given by
\begin{equation}
\label{EQ_H_ISING}
H_\text{I} = \frac{\i\,}{2} \sum_{k=1}^{2N-1} \m_k \m_{k+1} + \;\text{(boundary terms)} \ ,
\end{equation}
where each neighboring pair of Majorana modes is coupled with equal strength.
Restricting to states with positive fermionic parity and choosing antisymmetric boundary conditions $\m_{2N+1} \equiv -\m_1$ further allows us to neglect any boundary terms. Under a Jordan-Wigner transformation, this choice produces the same fermionic ground state as the Ising model as it is commonly defined in the spin picture.
In the fermionic picture the model is exactly solvable and leads to a translation-invariant ground state whose covariance matrix \eqref{EQ_COV_MATRIX_DEF} is
\begin{align}
\label{EQ_COV_MAT_ISING}
\Gamma^\text{I}_{j,k} =
\begin{cases}
0 & \text{for even } j-k \\
\frac{-1}{N \sin\left(\frac{\pi}{2N}(i-j) \right)} & \text{for odd } j-k
\end{cases} \ .
\end{align}
A full derivation of this result is included in App.\ \ref{APP_DISORDERED_ISING}.

We shall now establish a more precise connection between the ground states of the Ising model which respects translation invariance, and the hyperbolic MTN boundary states that explicitly break it. Following a characterization of the boundary state symmetries in an analytical model, we will show the connection between both models, explain the surprising effectiveness of site-averaging, and study the emergence of translation invariance for MTNs in the limit of large bond dimension.

\section{Results}
\label{S_RESULTS}
\subsection{Mode-disordered Ising models}
\label{SS_DISORDERED_ISING}

\begin{figure}[t]
\centering
\includegraphics[height=0.22\textheight]{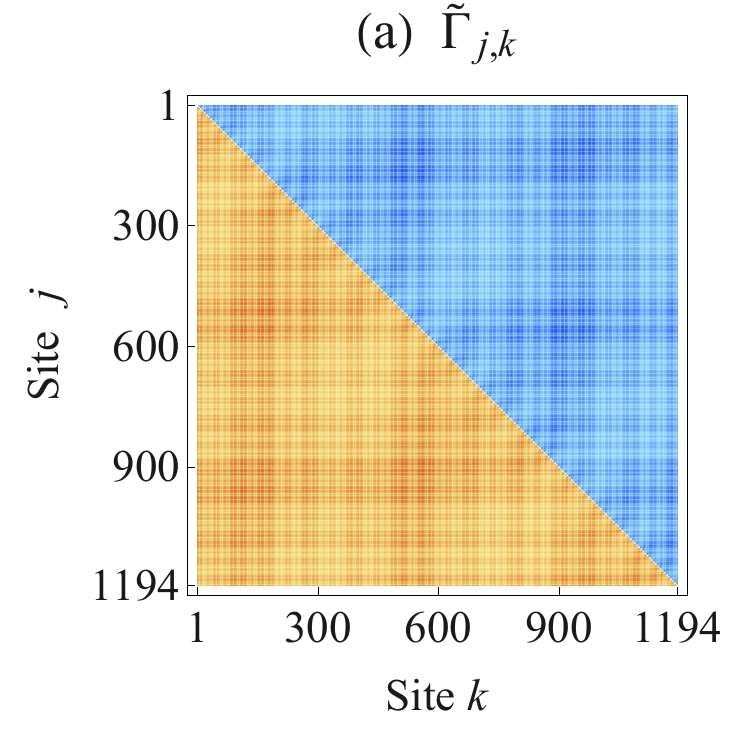}
\hspace{0.2cm}
\includegraphics[height=0.22\textheight]{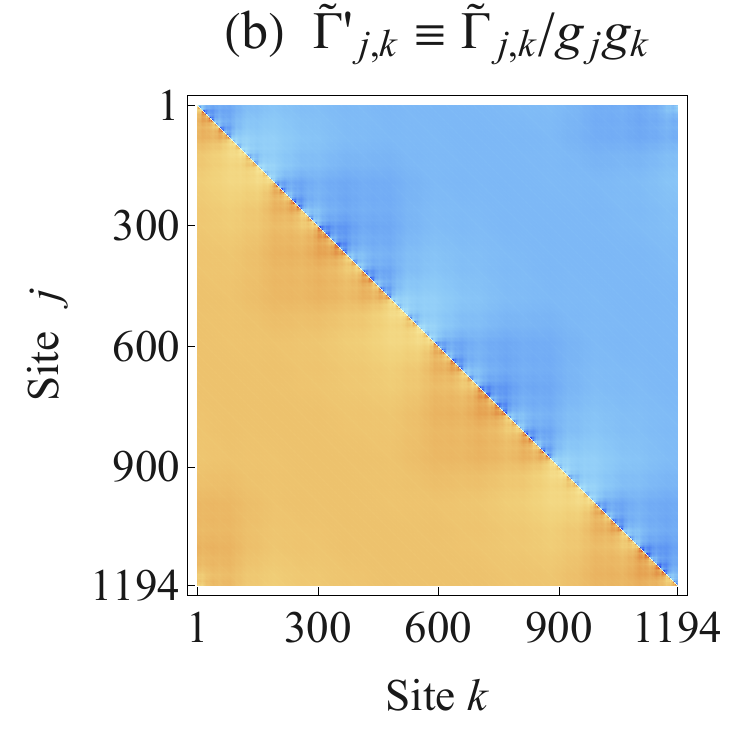}
\hspace{0.2cm}
\includegraphics[height=0.22\textheight]{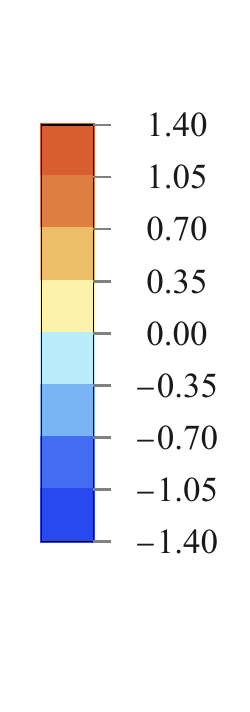} \\[10pt]
\includegraphics[height=0.22\textheight]{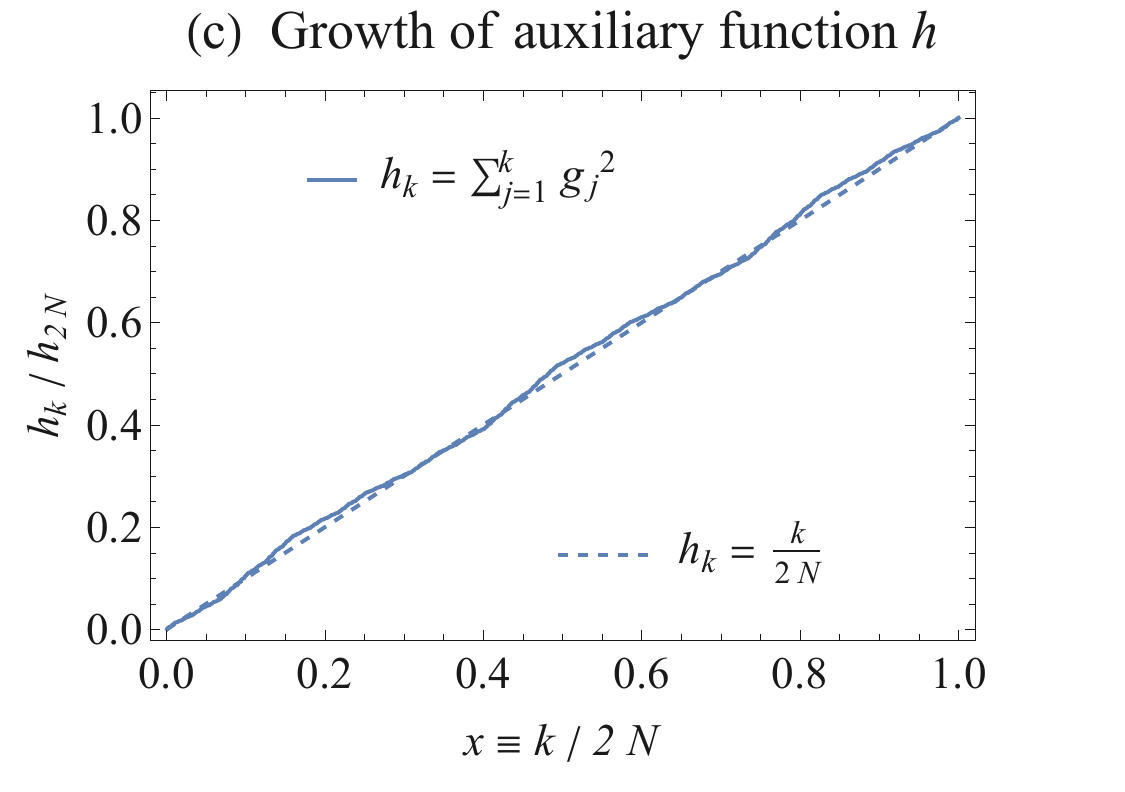}
\caption{(a)~Decay-adjusted covariance matrix $\tilde{\Gamma}$ (as defined in Eq.~\eqref{EQ_COV_MAT_ADJ}) for the boundary state of a $\{3,7\}$ matchgate tensor network (MTN) of $N=597$ sites ($n=5$ inflation steps) at the critical Ising point.
(b)~Nearly translation-invariant matrix $\tilde{\Gamma}^\prime$ computed by applying a site-dependent disorder vector, computed via Eq.~\eqref{EQ_DISORDER_37_EST}, on $\tilde{\Gamma}$.
(c)~The integrated squared disorder vector grows close to linearly on large scales and thus fulfills the condition \eqref{EQ_QR_H_CONDITION} for the validity of the mode-disordered Ising (MDI) parent Hamiltonian \eqref{EQ_ISING_MOD}.
}
\label{FIG_QR_SYMMETRIES}
\end{figure}

We now seek to construct the physical theory describing \emph{matchgate tensor network} (MTN) boundary states on the hyperbolic $\{3,7\}$ tiling at the critical Ising point.
As mentioned above, the site-averaged correlation decay \eqref{EQ_AVG_CORR_FALLOFF} follows the Ising model prediction $c(d) \propto 1/d$, where $d$ is the distance between boundary Majorana modes.\footnote{The physical distance between boundary sites (tiling edges) is therefore given by $\lfloor \frac{d}{2}\rfloor$}
To more clearly study the local deviations of correlations from this average, i.e., the state's \emph{disorder}, we modify the (boundary) state's covariance matrix \eqref{EQ_COV_MATRIX_DEF} into the \emph{decay-adjusted covariance matrix} $\tilde{\Gamma}$ defined as
\begin{equation}
\label{EQ_COV_MAT_ADJ}
\tilde{\Gamma}_{j,k} = \frac{\Gamma_{j,k}}{c(j-k)} \ .
\end{equation} 
For the covariance matrix of any translation-invariant state, such as the ground state of the Ising model, we would find values $|\tilde{\Gamma}_{j,k}| = 1$ for all sites $j,k$ where $\Gamma_{j,k}$ is nonzero. 
In the case of the $\{3,7\}$ MTN boundary states, however, $\tilde{\Gamma}$ encodes the disorder on the level of two-point correlations, shown in Fig.\ \ref{FIG_QR_SYMMETRIES}(a).
This disorder has a peculiar structure, which appears as a ``tartan pattern'' of site-dependent disorder.
We find that this disorder can be almost completely captured by a \emph{disorder vector} $g$, i.e., that we can find $2N$ real numbers $g_k$ so that
\begin{align}
\label{EQ_ISING_COV_MOD2}
\tilde{\Gamma}^\prime_{j,k} \equiv \frac{\tilde{\Gamma}_{j,k}}{g_j g_k}
\approx
\begin{cases}
0 & \text{for even } j-k \\
\sgn(j-k) &  \text{for odd } j-k
\end{cases}
 \ ,
\end{align}
where $\tilde{\Gamma}^\prime$ is approximately translation-invariant, i.e., $\tilde{\Gamma}^\prime_{j,k} \approx \tilde{\Gamma}^\prime_{j+d,k+d}$.
Indeed, as we see in Fig.\ \ref{FIG_QR_SYMMETRIES}(b), $\tilde{\Gamma}^\prime_{j,k}$ only deviates from this approximation in a small band around the diagonal $j=k$.
From definition \eqref{EQ_COV_MAT_ADJ} it immediately follows that we can rewrite \eqref{EQ_ISING_COV_MOD2} in terms of the original covariance matrix:
\begin{align}
\label{EQ_ISING_COV_MOD}
\Gamma^\prime_{j,k} \equiv \frac{\Gamma_{j,k}}{g_j g_k}
\approx
\begin{cases}
0 & \text{for even } j-k \\
\sgn(j-k)\, c(|j-k|) &  \text{for odd } j-k
\end{cases}
 \ .
\end{align}
As $c(d)$ matches the correlation decay of the Ising model, it follows that $\Gamma^\prime_{j,k}$ closely approximates its covariance matrix $\Gamma^\text{I}$ as given by \eqref{EQ_COV_MAT_ISING}.
Because both the Ising ground state and MTN boundary states are Gaussian and thus completely determined by their covariance matrices, \eqref{EQ_ISING_COV_MOD} allows us to directly map between both states and all of their correlation functions.
Making use of the $\{3,7\}$ tiling's $\mathbb{Z}_3$ symmetry, we can construct a suitable disorder vector $g$ while avoiding lattice effects by averaging over a third of the system at large distances. Explicitly,
\begin{equation}
\label{EQ_DISORDER_37_EST}
g_j \approx \frac{\sum_{k=1}^{2N/3} \tilde{\Gamma}_{j,\frac{2N}{3}+j+k}}{N_g} \ , \quad
N_g = \frac{3}{2N}\sum_{j,k=1}^{2N/3} \tilde{\Gamma}_{j,\frac{2N}{3}+j+k} \ ,
\end{equation}
where the normalization factor $N_g$ ensures that $\frac{1}{2N}\sum_j g_j = 1$.
This is the method used for determining $g$ as used in Fig.\ \ref{FIG_QR_SYMMETRIES}(b), leading to an almost translation-invariant $\tilde{\Gamma}^\prime$.

\begin{figure}[t]
\centering
\includegraphics[height=0.25\textheight]{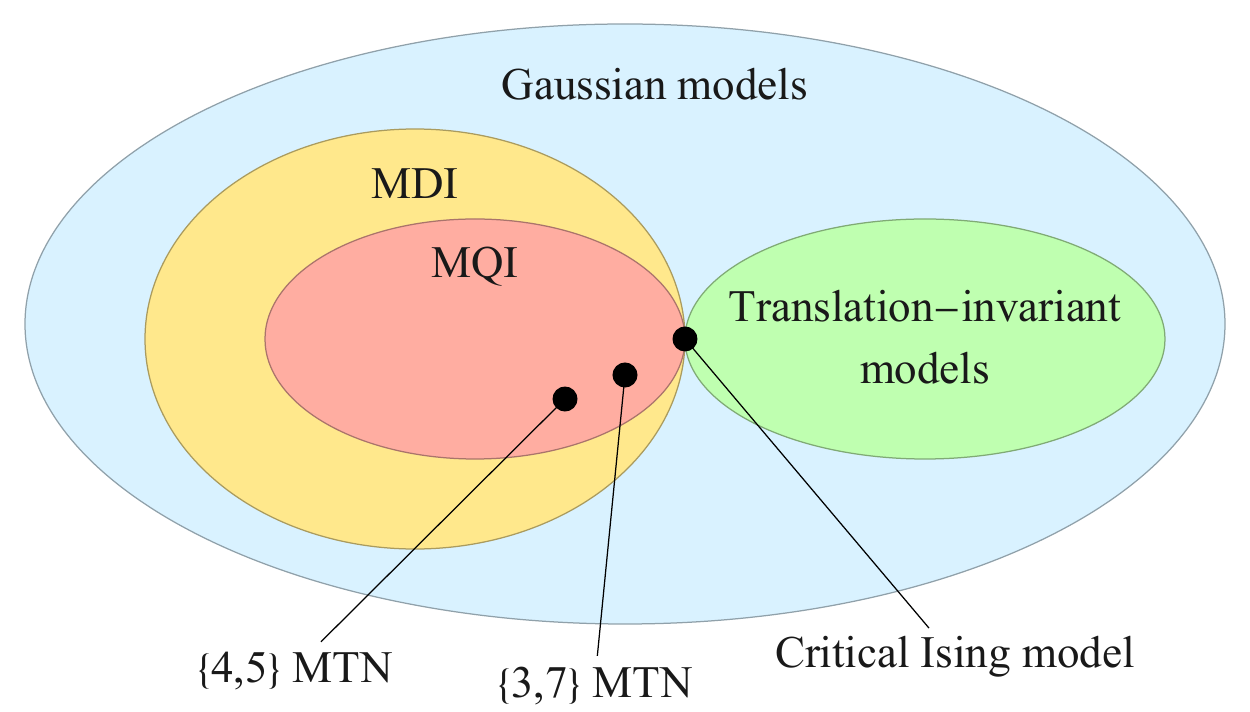}
\caption{The space of theoretical models considered in this paper. All are based on exactly solvable Gaussian fermionic models (describable by a Hamiltonian \eqref{EQ_H_FREE_FERMIONS}) with correlation functions that are fully encoded by a covariance matrix.
Only a subset of Gaussian models are translation-invariant, in particular including the (critical) Ising model.
This model is generalized into the \emph{mode-disordered Ising} (MDI) model described by a Hamiltonian \eqref{EQ_ISING_MOD}, with the critical Ising model resulting from a trivial (constant) disorder vector.
The MDI model includes the \emph{multi-scale quasicrystal Ising} (MQI) model where the disorder vector is specified by a \emph{multi-scale quasicrystal ansatz} (MQA) as introduced in Sec.\ \ref{SS_MQI}.
Instances of the MQI model with symmetries of regular tiling layers closely approximate the $\{3,7\}$ and $\{4,5\}$ matchgate tensor network (MTN) states at their critical Ising point, the latter being discussed in App.\ \ref{APP_45TILING}.
}
\label{FIG_MODEL_SPACE}
\end{figure}

This relationship between the ground state of the translation-invariant Ising model and the disordered MTN boundary states can be extended to a relationship between models with specific Hamiltonians.
In the following, we will introduce two models, the second a constrained version of the first:
\begin{enumerate}
\item The \emph{mode-disordered Ising} (MDI) model, where a disorder vector $g$ is applied to each Majorana mode.
\item The \emph{multi-scale quasicrystal Ising} (MQI) model, where $g$ follows a distribution determined by an analytical \emph{multi-scale quasicrystal ansatz} (MQA).
\end{enumerate}
The MDI model will give us a \emph{parent Hamiltonian} whose ground states are related to the critical Ising model via \eqref{EQ_ISING_COV_MOD}, which clearly includes the hyperbolic MTN boundary states, while the MQI model will closely reproduce the specific coupling terms appearing in the latter. We summarize the relationship between the models relevant for our work in Fig.\ \ref{FIG_MODEL_SPACE}.

We begin with the MDI model, which is defined as the class of Hamiltonians
\begin{align}
\label{EQ_ISING_MOD}
H^\text{MDI}[g] = \frac{\i\,}{2} \left( \sum_{k=1}^{2N-1} \frac{1}{g_k g_{k+1}} \m_k \m_{k+1} + \frac{1}{g_1 g_{2N}} \m_1 \m_{2N} \right) \ ,
\end{align}
dependent on a disorder vector $g$ with $2N$ components $g_k>0$.
As we prove in App.\ \ref{APP_DISORDERED_ISING}, the ground state covariance matrix $\Gamma^\text{MDI}[g]$ of this Hamiltonian is related to the Ising model's $\Gamma^\text{I}$ via
\begin{equation}
\label{EQ_COV_MAT_ADJ2}
\Gamma^\text{MDI}_{j,k}[g] = g_j g_k\, \Gamma^\text{I}_{j,k}[g] \ .
\end{equation}
This relationship holds only if $g$ is approximately constant over large length scales, i.e., varies only on smaller ones. Specifically, the derivation of \eqref{EQ_COV_MAT_ADJ2} assumes that the auxiliary function
\begin{equation}
h_k = \sum_{j=1}^k g_j^2 \ , \quad 1 \leq k \leq 2 N \ , 
\end{equation}
grows approximately linearly with $k$, and thus \eqref{EQ_COV_MAT_ADJ2} breaks down at scales below which $h$ grows nonlinearly. 

We can immediately test this conjecture with the MTN covariance matrix at the critical Ising point with $g$ provided by \eqref{EQ_DISORDER_37_EST}: As we see in Fig.\ \ref{FIG_QR_SYMMETRIES}(c), the corresponding auxiliary function $h$ indeed grows approximately linearly at distances of more than a few dozen sites (of $N=597$ total ones), and indeed $\tilde{\Gamma}_{j,k}/(g_j g_k)$ is very close to translation-invariant at these scales.
We therefore conclude that \eqref{EQ_ISING_MOD} with the disorder vector $g$ extracted from the MTN covariance matrix via \eqref{EQ_DISORDER_37_EST} describes a theory whose ground state is given by our MTN ansatz.
In App.\ \ref{APP_COUPLING_OPT}, we give further support to this result by showing that a numerical optimization over all nearest-neighbor Hamiltonians indeed yields the couplings specified in \eqref{EQ_ISING_MOD}. We also argue that this Hamiltonian is indeed the unique parent Hamiltonian for these boundary states.

\subsection{Multi-scale quasiperiodic Ising models}
\label{SS_MQI}

\begin{figure}[t]
\centering
\includegraphics[height=0.25\textheight]{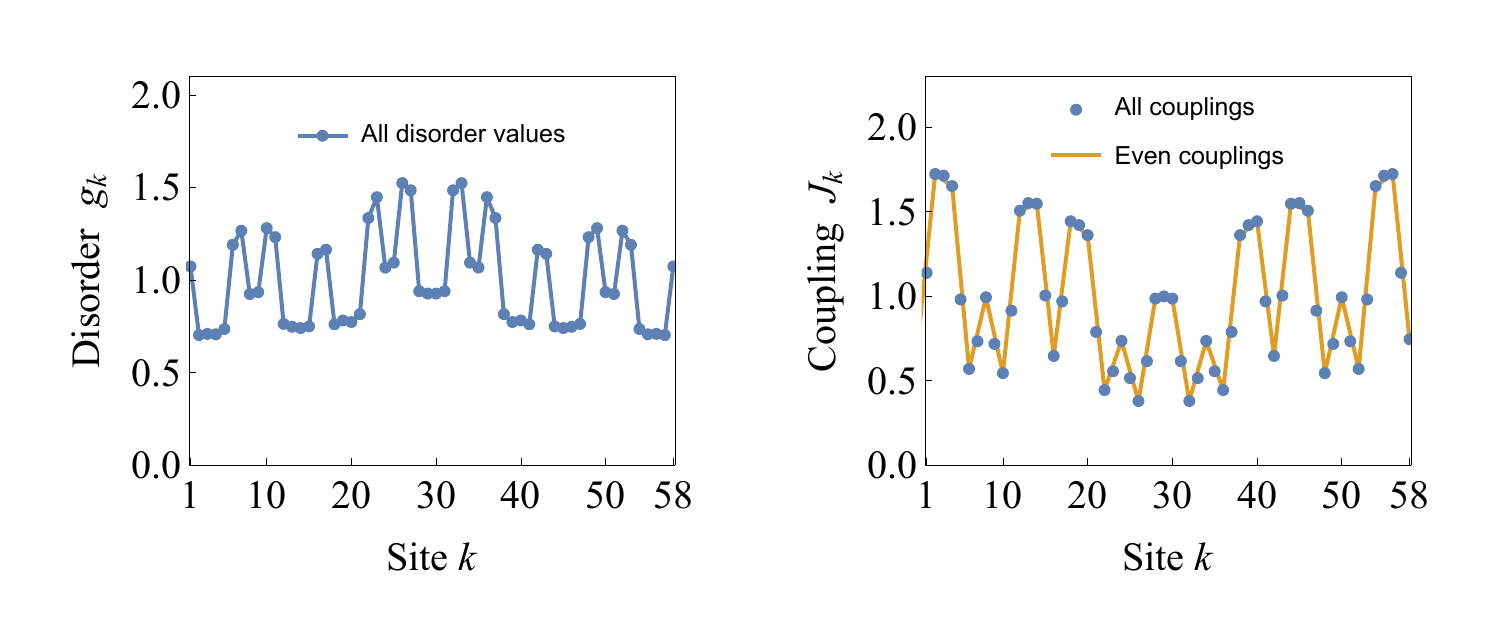}
\caption{Disorder values $g_k$ and coupling terms $J_k=1/(g_k g_{k+1})$ for the $\{3,7\}$ MTN at the critical Ising point for $N=174$ boundary sites, $N/3$ of which are shown here.
The disorder values are extracted from the MTN boundary covariance matrix following \eqref{EQ_DISORDER_37_EST}.
For the coupling terms, a linear interpolation is shown between the even terms $J_{2i}$, showing that $J_{2i+1} \approx (J_{2i} + J_{2i+2})/2$.
}
\label{FIG_DISORDER_AND_COUPLING}
\end{figure}

Possessing a technique for constructing effective parent Hamiltonians for the boundary states of our hyperbolic matchgate tensor networks at their critical Ising point, let us now study the relationship between the resulting coupling terms and the quasiperiodic symmetries inherent to the underlying tensor network geometry.
We first rewrite the Hamiltonian $H^\text{MDI}[g]$ of the disordered Ising model into the form
\begin{equation}
\label{EQ_H_QR}
H^\text{MQI}[J] = \frac{\i\,}{2} \sum_k J_k \m_k \m_{k+1} \ ,
\end{equation}
with couplings $J_k=1/(g_k g_{k+1})$ and a subscript that will be explained shortly. Note that there are effectively two types of couplings $J_k$: Those for odd $k$ describe coupling within the $\frac{k+1}{2}$th fermionic mode, while those for even $k$ describe coupling between the $\frac{k}{2}$th and $\frac{k+2}{2}$th fermionic modes.
For the disorder vectors resulting from our matchgate tensor network ansatz, we find that the odd couplings can be well approximated by
\begin{equation}
\label{EQ_COUPLING_MIDPOINT}
J_{2i-1} \approx \frac{J_{2i-2}+J_{2i}}{2} \ ,
\end{equation}
as is shown in Fig.\ \ref{FIG_DISORDER_AND_COUPLING}. We can see that this corresponds to a condition $g_{2i} \approx g_{2i+1}$ on the level of the disorder values, which is equivalent to \eqref{EQ_COUPLING_MIDPOINT} up to a subleading $( g_{2i+1}^{-1} - g_{2i-1}^{-1})^2 / 2$ correction.
As a result, it is sufficient to focus our attention on the $N$ even coupling terms $J_{2i} \equiv \bar{J}_i$, or geometrically, on effects from the vertices of the tiling boundary.

\begin{figure}[ht]
\centering
\includegraphics[width=1.0\textwidth]{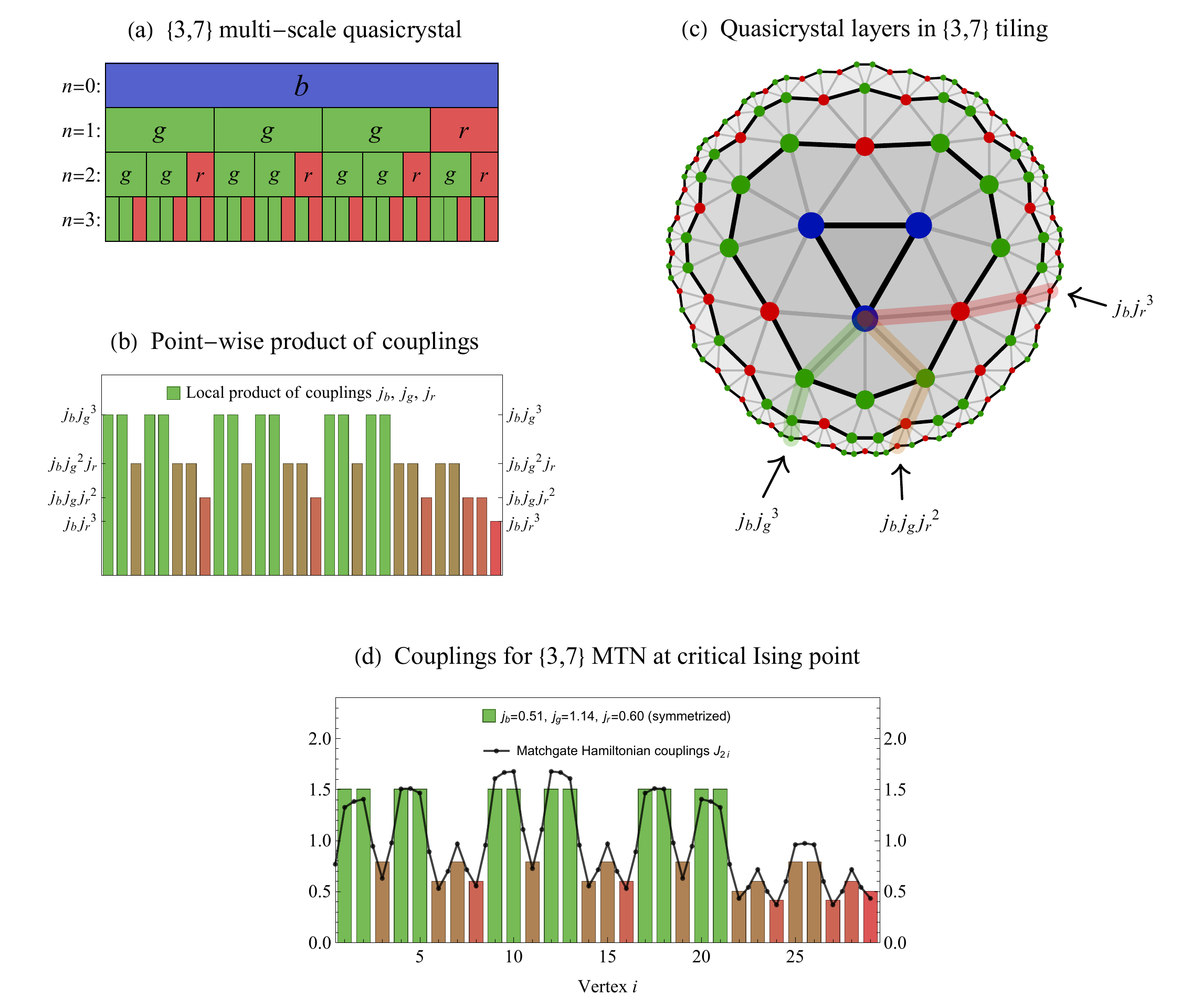}
\vspace{-0.5cm}
\caption{Construction of a multi-scale quasicrystal ansatz (MQA) for coupling sequences with the symmetries of $\{3,7\}$ tiling layers.
(a) The MQA of Ref.\ \cite{Jahn:2020ukq} composed of inflation layers following the rule \eqref{EQ_V_INFLATION_37}. The sequence directly below each letter is generated by applying \eqref{EQ_V_INFLATION_37} to that letter.
(b) A point-wise (vertical) product of couplings $j_l$ corresponding to each letter $l \in (b,g,r)$ of the inflation sequence. Each bar corresponds to the product of blocks directly above it in (a).
(c) The inflation layers embedded into the $\{3,7\}$ tiling, with the path between layers highlighted for three point-wise products.
(d) For a suitable choice of the $j_l$ the couplings closely reproduce the effective Hamiltonian couplings for the $\{3,7\}$ MTN states at the critical Ising point.
As the inflation rules are not reflection-symmetric, the couplings are taken as a symmetrized sum.
The $j_b$ coupling is chosen to provide an overall normalization $\langle v \rangle = 1$.
Note that there are $2N$ Hamiltonian coupling terms for a boundary with $N$ vertices.
}
\label{FIG_MQI_37}
\end{figure}

As explained in Sec.\ \ref{SS_REG_TILINGS_QUASIP}, the geometry of the tiling boundary is characterized by a quasiperiodic sequence of vertex types (denoted $b,g,r$ for the $\{3,7\}$ tiling), so is it possible to relate these vertex types to the even coupling terms $\bar{J}_i$?
It is immediately clear from Fig.\ \ref{FIG_DISORDER_AND_COUPLING} that there exists no one-to-one correspondence between the two, as the couplings take a range of values that can be only poorly approximated by assigning a fixed coupling strength to each boundary vertex type.
This discrepancy worsens when increasing the number $n$ of inflation steps, where the range of values assumed by the couplings appears to approach a continuum.
As we are considering the couplings of a model with (averaged) critical behavior, it should perhaps not be surprising that we cannot completely characterize our model by the symmetries on a fixed length scale alone.
With the inflation layers interpreted as a discretized renormalization group (RG) flow, and the symmetries of each RG layer being characterized by a quasiperiodic sequence, we should instead expect the boundary state to contain symmetry contributions from all length scales in the RG process.
Following this intuition, we define a simple model to construct such a Hamiltonian analytically, and show that it reproduces the qualitative features of the couplings obtained from our tensor network ansatz.
This model, which is an extension of the \emph{multi-scale quasicrystal ansatz} (MQA) proposed in Ref.\ \cite{Jahn:2020ukq} for describing quasiperiodic CFT (qCFT) symmetries, relies on the same inflation rules used above, but stacks the letter sequences produced at each iterations in sequence. 
Representing each letter as a rectangular block, we arrange the blocks so that the letters produced by an application of the inflation rule are directly under the letter from which originated, as visualized in Fig.\ \ref{FIG_MQI_37}(a) for the inflation rule \eqref{EQ_V_INFLATION_37} corresponding to the $\{3,7\}$ tiling.
We can equivalently denote the MQA as a sequence of sets of letters, encoded as a matrix. For example, the iterated inflation
\begin{align}
\label{EQ_37_INFL_EX}
g \mapsto ggr \mapsto ggrggrgr \ ,
\end{align}
which is essentially a mapping between vectors of different lengths, can be turned into an MQA simply by including ``parent'' letters with each element of the sequence,
\begin{align}
\label{EQ_37_INFL_EX2}
g \mapsto 
\begin{pmatrix}
g & g & g \\
g & g & r
\end{pmatrix}
\mapsto 
\begin{pmatrix}
g & g & g & g & g & g & g & g \\
g & g & g & g & g & g & r & r \\
g & g & r & g & g & r & g & r 
\end{pmatrix}  \ .
\end{align}
The last row in each iteration step in \eqref{EQ_37_INFL_EX2} contains the sequence \eqref{EQ_37_INFL_EX} at that step.
We now map this list of sequences onto physical couplings $\bar{J}_i$ in \eqref{EQ_H_QR} by associating each letter $l$ with a positive real number $j_l$ and taking the product across  columns of the MQA.
From \eqref{EQ_37_INFL_EX2}, for example, we produce the list of couplings
\begin{equation}
\bar{J} = \begin{pmatrix}
j_g^3, & j_g^3, & j_g^2 j_r, & j_g^3, & j_g^3, & j_g^2 j_r, & j_g^2 j_r, & j_g j_r^2 
\end{pmatrix}  .
\end{equation}
We thus produce coupling terms with a multi-scale quasiperiodic symmetry.
For this reason, we refer to the class of Hamiltonians $H^\text{MQI}(J)$ in \eqref{EQ_H_QR} with couplings produced by an MQA as the \emph{multi-scale quasicrystal Ising} (MQI) model.
The only free parameters of the MQA are the coupling terms $j_l \geq 0$ for each letter $l$, making the MQI model highly restrained.

Let us now apply the MQI model to the $\{3,7\}$ tiling: As shown in Fig.\ \ref{FIG_MQI_37}(b), the couplings $\bar{J}$ resulting from the MQA take a range of values between $j_b j_r^{n}$ and $j_b j_g^{n}$, $n$ being the number of iterations.
Each local coupling value contains the factor $j_b$ once, as this letter only appears at the $0$th inflation step. In the geometrical tiling picture (Fig.\ \ref{FIG_MQI_37}(c)), this corresponds to every geodesic path from a boundary vertex to the center ending at a $b$-type vertex.
As the overall normalization of the couplings $\bar{J}$ is irrelevant for the physical (ground) states, we are thus free to choose $j_b$ so that the expectation value $\langle \bar{J} \rangle$ over all site equals one.
With this normalization, only two effective parameters $j_g,j_r$ remain; as we find in Fig.\ \ref{FIG_MQI_37}(c), these are sufficient to produce a coupling pattern that closely matches the actual MTN couplings computed using \eqref{EQ_DISORDER_37_EST}. In other words, hyperbolic MTNs can produce Ising-like boundary states with multi-scale quasiperiodic disorder quantified by the MQA, sharpening the relationship suggested in Ref.\ \cite{Jahn:2020ukq}.
Note that $j_g>j_r$, which we can explain as the result of the two boundary edges on a $g$-type vertex being geometrically closer than those on an $r$-type one, with an angle of $\frac{4\pi}{7}$ and $\frac{6\pi}{7}$ between them, respectively.

Before studying the specific symmetries of the $\{3,7\}$ MQI model, we address the question of continuum limit consistency of this class of Hamiltonians. 
We will find that this reduces the number of free parameters by a further constraints, leaving only a single effective parameter for a consistent model.
One may worry that the couplings $\bar{J}$ defined as above diverge with $n \to \infty$ as we inflate the tiling, as they tend towards infinite products. Indeed, for generic values of the weights $j_l$ the continum limit of the Hamiltonian couplings cannot be described by a smooth function with finite values at every point: 
In order for the couplings $\bar{J}$ not to converge to zero in this limit, we have to demand $j_l>1$ for at least one letter $l$, leading to diverging couplings $(j_l)^n \to \infty$ at a set of boundary points.
Fortunately, the ratio between the number of sites with divergent couplings to the total number of sites generally converges to zero exponentially in $n$. For example, in the $\{3,7\}$ MQA the ratio of sites with coupling $j_b (j_g)^n$ converges as $\propto (2/\lambda_{\{3,7\}})^n$.
As a result, despite local divergences it is still possible to define an MQI model with finite \emph{average} couplings.
To compute this average analytically, we first recapitulate the description of inflation steps via a substitution matrix $M$ with rows and columns corresponding to vertex types, following the notation of Ref.\ \cite{Jahn:2019nmz}. For the $\{3,7\}$ tiling, the corresponding rule \eqref{EQ_V_INFLATION_37} corresponds to
\begin{equation}
M_{\{3,7\}} =
\begin{pmatrix}
0 & 3 & 1 \\
0 & 2 & 1 \\
0 & 1 & 1 \\
\end{pmatrix} \ .
\end{equation}
For example, the first row indicates that each inflation step turns one $b$ letter into a sequence containing zero $b$s, three $g$s and one $r$. Representing the number of letters in a given sequence at inflation step $n$ by a row vector $\vec{v}^{(n)}$ (with entries $(N_b,N_g,N_r)$ in the ${\{3,7\}}$ example), an inflation step acts as a matrix multiplication $\vec{v}^{(n)} \mapsto \vec{v}^{(n+1)} = \vec{v}^{(n)} M$. After sufficiently many inflation steps $n$, $\vec{v}^{(n)}$ will eventually be proportional to the left eigenvector $\vec{l}$ fulfilling $\vec{l} M = \lambda M$ for the largest eigenvalue $\lambda$, and each successive inflation step increases the size of the sequence by $\lambda$.

This logic can be extended to describe the inflation steps of a product of sequences: Assume we start with a sequence consisting of a single letter $g$ as in \eqref{EQ_37_INFL_EX2}. 
We thus start with an initial vector $\vec{v}^{\prime (0)}_{\{3,7\}} = (0, j_g, 0)$, denoting a single coupling term $j_g$ on one $g$-type vertex. We then apply the modified inflation step
\begin{align}
\vec{v}^{\prime (0)}_{\{3,7\}} &\mapsto \vec{v}^{\prime (1)}_{\{3,7\}} = \vec{v}^{\prime (0)}_{\{3,7\}} M^\prime_{\{3,7\}} = \left( 0,\, 2j_g^2,\, j_b j_g \right) \ ,
&
M^\prime_{\{3,7\}} = 
\begin{pmatrix}
0 & 3j_g & j_r \\
0 & 2j_g & j_r \\
0 & j_g & j_r \\
\end{pmatrix} \ ,
\end{align}
where $M^\prime_{\{3,7\}}$ is just $M_{\{3,7\}}$ with each column multiplied by the corresponding coupling term.
This inflation step produces a new vector $\vec{v}^{\prime (1)}_{\{3,7\}}$ that encodes the  product of couplings after the second step of \eqref{EQ_37_INFL_EX2}. If we wish to compute the average $\langle j \rangle^{(n)}$ of all final couplings at step $n$, we simply compute
\begin{equation}
\langle j \rangle^{(n)} = \frac{\sum_i v_i^{\prime (n)}}{\sum_i v_i^{(n)}} \ ,
\end{equation}
which here corresponds to $\langle j \rangle^{(1)}_{\{3,7\}} = \frac{2j_g^2 + j_g j_r}{3}$. Does $\langle j \rangle^{(n)}$ remain finite at arbitrarily large $n$? Just as $M$, the modified substitution matrix $M^\prime$ has a left eigenvector $\vec{l}^\prime$ for its largest eigenvalue $\lambda^\prime$ describing the asymptotic scaling after computing $\vec{v}^{\prime (n)}\, = \vec{v}^{\prime (0)}\, (M^\prime)^{n}$ for large $n$. This leads to the following asymptotic scaling for the average coupling:
\begin{equation}
\langle j \rangle^{(n)} \propto \left(\frac{\lambda^\prime}{\lambda}\right)^n \ .
\end{equation}
We immediately find that $\lambda_{\{3,7\}}=\lambda_{\{3,7\}}^\prime$ for translation-invariant couplings $j_g=j_r$ (where $M^\prime_{\{3,7\}} \propto M_{\{3,7\}}$), and hence the coupling average remains constant.
But for $j_b \neq j_c$, the constraint $\lambda_{\{3,7\}}=\lambda^\prime_{\{3,7\}}$ ensuring a finite coupling average restricts the values of the couplings. In the $\{3,7\}$ case, this constraint can be written as
\begin{equation}
\label{EQ_J_CONSTR_37}
j_r = \frac{7 + 3\sqrt{5} - 2\left(\sqrt{5}+3\right) j_g}{3 + \sqrt{5} - 2j_g} \ .
\end{equation}
Note that if we restrict ourselves to coupling terms with $0 < j_r < j_g$, following our observation that $g$-type vertices are geometrically closer than $r$-type ones and are thus coupled more strongly, this implies 
\begin{equation}
1 < j_g < \frac{3\sqrt{5}+5}{2\sqrt{5}+6} \approx 1.31 \ .
\end{equation}
Indeed, as seen in Fig.\ \ref{FIG_MQI_37}(d), a choice $j_g \approx 1.14, j_r \approx 0.60$ fulfilling the constraint \eqref{EQ_J_CONSTR_37} produces MQA couplings that closely match with the even couplings $\bar{J}$ resulting from the $\{3,7\}$ matchgate tensor network ansatz at the critical Ising point. Because the inflation rule \eqref{EQ_V_INFLATION_37} is not reflection-symmetric (i.e., $g \mapsto ggr \neq rgg$), we actually take the average of the sequence and its reflection around the symmetry axis of the geometry (vertex $i=11$ in Fig.\ \ref{FIG_MQI_37}(d)).
The optimal values for $j_g$ and $j_r$ appear convergent as the number of inflation steps $n$ is increased further: For $n=3,4,5$ we find the optimal coupling $j_g=1.139,1.124,1.115$ (the $n=3$ result was quoted above and used in Fig.\ \ref{FIG_MQI_37} as well).

Similar consistency conditions can be calculated for any $\{p,q\}$ tiling and the MQI constructed from its inflation rule; in App.\ \ref{APP_45TILING}, we show that the $\{4,5\}$ works analogously to the $\{3,7\}$ one, with an MQI for suitable weights $j_l$ again well approximating the couplings obtained from the Ising-critical MTN ansatz.
Vertex inflation for a $\{p,q\}$ tiling can always be written in terms of two letters up to initial conditions, where a third letter type may become necessary (such as $b$ for the $\{3,7\}$ tiling) \cite{Jahn:2019mbb}. As a result, the MQI produced from $\{p,q\}$ inflation rules only has a single free parameter when constraint equations such as \eqref{EQ_J_CONSTR_37} for a well-defined scaling limit are imposed.
This free parameter determines the disorder strength, with a choice of unity corresponding to a translation-invariant model.

The small deviations between the coupling terms of $H^\text{MQI}$ and those obtained from the MTN boundary states via \eqref{EQ_DISORDER_37_EST} are due to a smearing of contributions on each inflation layer under tensor contraction, for which the sharp block structure shown of the MQA in Fig.\ \ref{FIG_MQI_37}(a) is only an approximation. Yet, as this smearing happens equally on all layers of the tensor network, the symmetries of the MQA are preserved in the MTN boundary states, as we will explore in the next section.

We wish to stress that the MQI model is somewhat different from previously studied ``aperiodic'' Hamiltonians (as in Ref.\ \cite{JuhaszZimboras2007}) where the couplings are fixed to a finite number of distinct values by a \emph{single} letter sequence.
In contrast, in our model the scaling limit produces a fractal sequence of unbounded couplings which are disordered on arbitrarily small distances. 

\subsection{Boundary symmetries}
\label{SS_BOUNDARY_SYMMETRIES}
\begin{figure}[t]
\centering
\includegraphics[height=0.25\textheight]{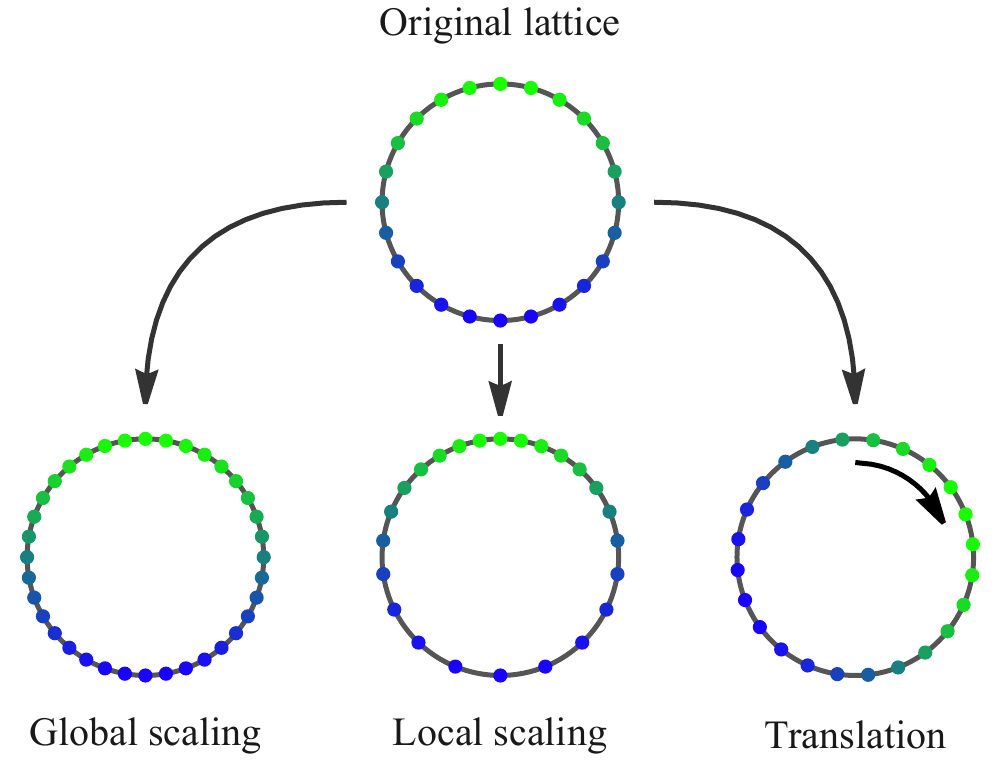}
\caption{Visualization of symmetry transformation of a CFT ground state with periodic boundary conditions on a regular discretization: While translations and local (non-uniform) scale transformations merely distort the lattice under constant number of discretization points, global (uniform) scale transformations change its resolution and are thus equivalent to a renormalization group (RG) transformation. Figure adapted from Ref.\ \cite{Jahn:2020ukq}.
}
\label{FIG_CFT_SYMMETRIES}
\end{figure}

As we have seen in the last section, boundary states of the Ising-critical MTN exhibit multi-scale quasiperiodic symmetries that are well approximated by the MQI model. 
We will now discuss the relationship of these symmetries to those of continuum CFT states, connecting to the recent proposal that tensor networks on regular hyperbolic tilings produce (ground) states of a \emph{quasiperiodic CFTs} (qCFTs) \cite{Jahn:2020ukq} only invariant under a discrete subgroup of the full CFT symmetries.
The symmetries of discretized CFT ground states are shown in Fig.\ \ref{FIG_CFT_SYMMETRIES}: These include global (i.e., \emph{uniform}) scale invariance, local (non-uniform) scale invariance, and translation invariance. Local scale invariance, under a metric deformation with a local, position-dependent scale factor, is specific to CFTs. 

As was shown in Ref.\ \cite{Jahn:2020ukq}, boundary states of tensor networks with regular hyperbolic geometry -- as the MTN setup considered here -- fulfill symmetries corresponding to discretized versions of the above:
\begin{enumerate}
\item Global scale invariance is represented by the inflation rules, which add short-range degrees of freedom to the boundary state and thus define an RG step. The scale factor is restricted to powers of $\lambda_{\{p,q\}}$, the asymptotic ratio between two boundary vertex type sequences under application of the inflation rule.  
\item Local scale invariance turns into a fractal self-similarity between boundary subsystems of different size, with the scale factor between them again given by $\lambda_{\{p,q\}}$. The self-similarity only becomes exact in the large inflation limit $n \to \infty$.
\item Exact translation invariance is broken into an approximate subsystem translation invariance, resulting from cumulative self-similarity at sizes smaller than the subsystem size.
\end{enumerate}
We now discuss how these qCFT symmetries appear in $\{3,7\}$ MTN boundary states. As we have seen in Sec.\ \ref{SS_DISORDERED_ISING}, the deviations of the covariance matrix of these boundary states from the Ising ground state covariance matrix are fully captured by the disorder vector $g$ (or equivalently, the effective couplings $J$). It is thus sufficient to study only the symmetries of $g$. 

\begin{figure}
\centering
\small
(a) \hspace{10pt} Global self-similarity of disorder vector

\includegraphics[width=1.0\textwidth]{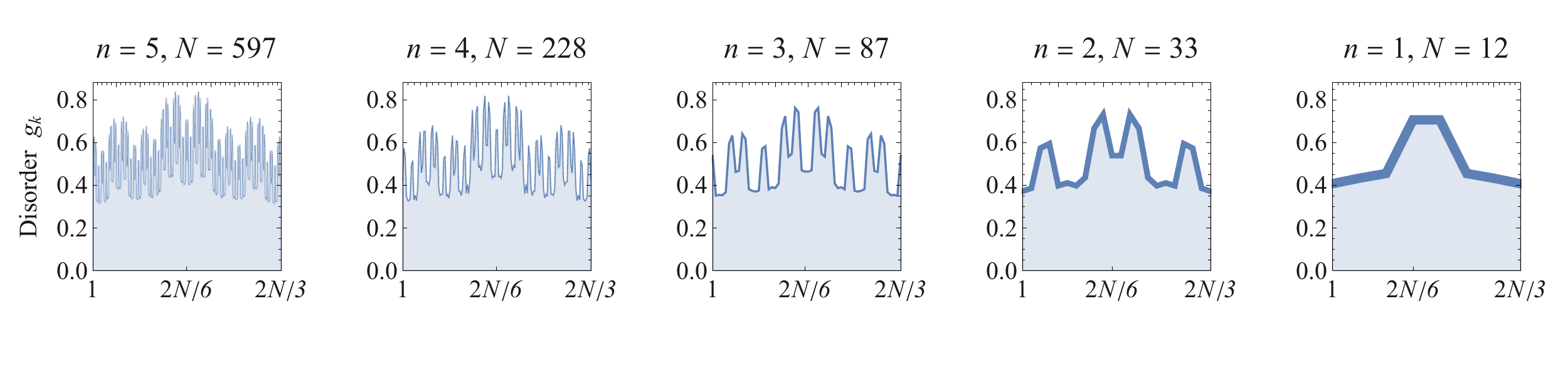}
\vspace{-0.5cm}

(b) \hspace{10pt} Local self-similarity of disorder vector

\includegraphics[width=1.0\textwidth]{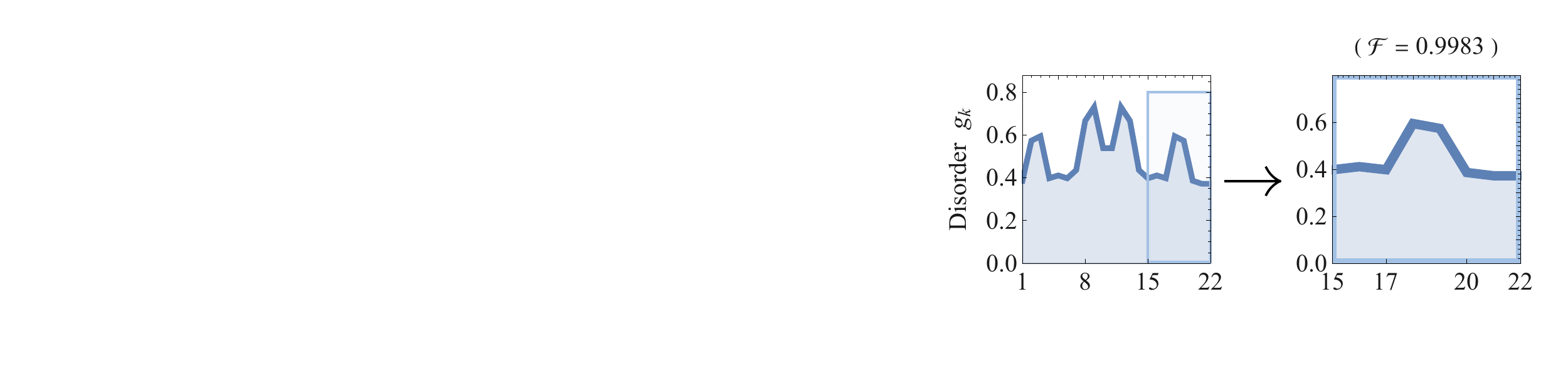}
\vspace{-0.5cm}

\includegraphics[width=1.0\textwidth]{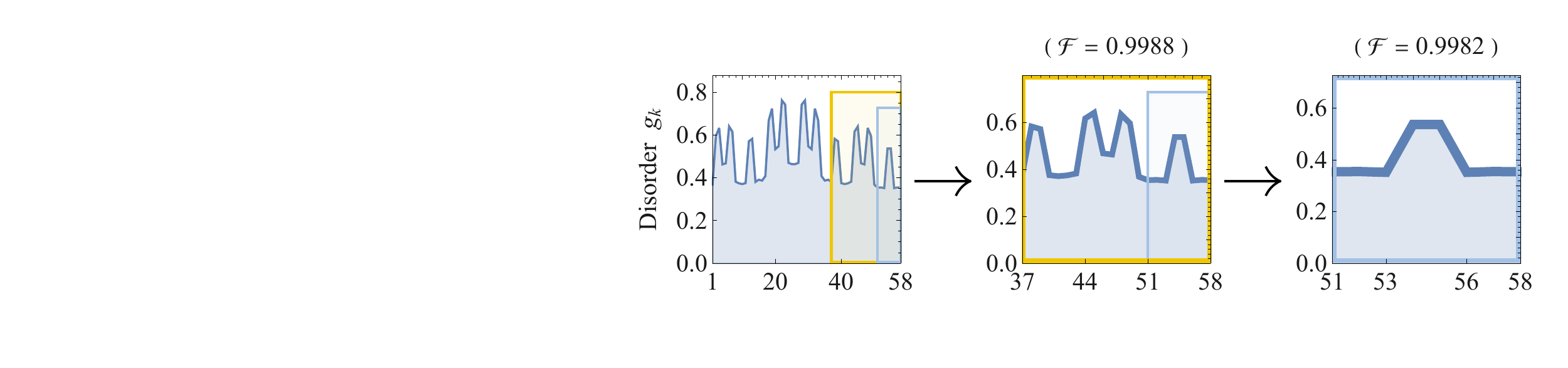}
\vspace{-0.5cm}

\includegraphics[width=1.0\textwidth]{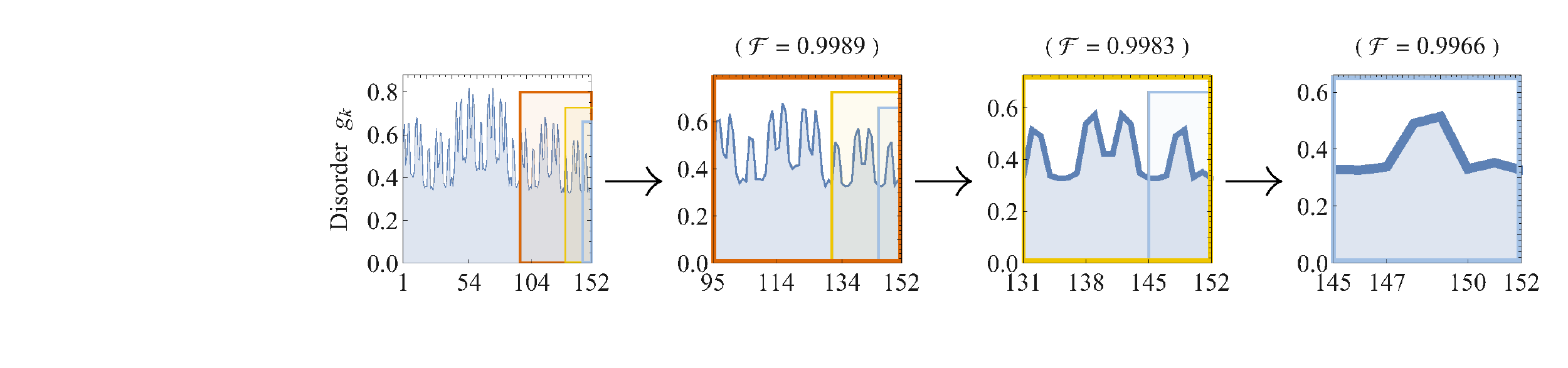}
\vspace{-0.5cm}

\includegraphics[width=1.0\textwidth]{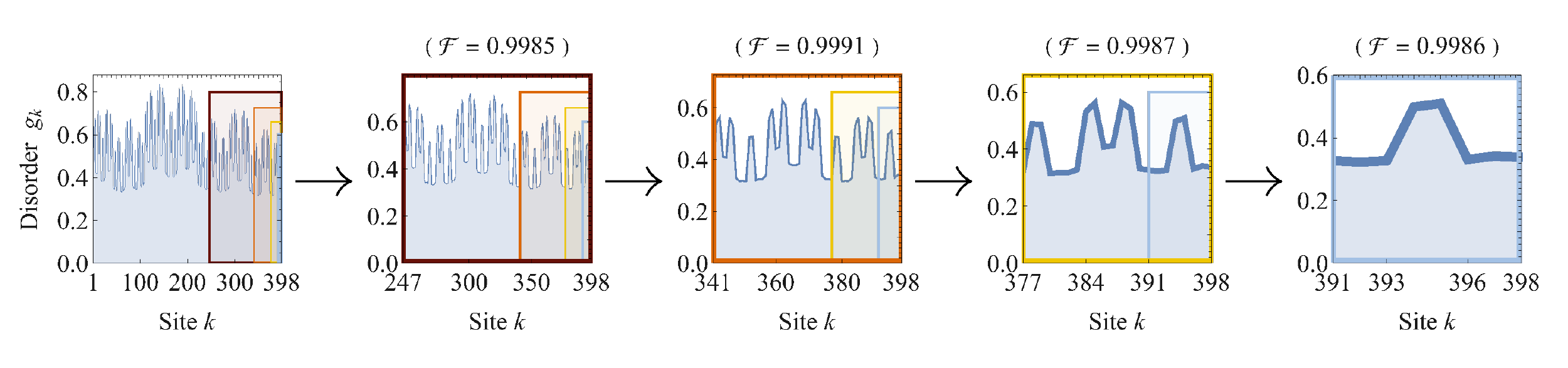}
\vspace{-0.5cm}
\caption{(a) The disorder vector $g$ after $n=2,3,4,$ and $5$ inflation vertex inflation steps (right to left) of the $\{3,7\}$ tiling at the critical Ising point. Only a third of the total system is shown, as the $\mathbb{Z}_3$ symmetry leads to a repeating pattern on the remaining sites. 
We a find a gradual refinement of the disorder vector at a larger number of inflation steps.
(b) Subsystems (coloured frames) of the disorder vector for $N=597,228,87,$ and $33$, slightly site-shifted for easier visualization.
We find a self-similarity between total system and subsystems resembling the scaling between inflation steps, up to a global rescaling factor.
The fidelity $\mathcal{F}$ between subsystems and rescaled total system (top row) is quoted above each plot (with the site shift taken into account).
}
\label{FIG_GLOB_LOC_SCALING}
\end{figure}

We begin with global scale transformations. Cosider the disorder values $g_k$ produced by different numbers of vertex inflation steps: As Fig.\ \ref{FIG_GLOB_LOC_SCALING}(a) shows, a higher number of inflation steps effects a \emph{fine-graining transformation} on $g$, adding additional detail on small scales while preserving averages on larger ones.
This implies that adding or removing inflation steps implements a global scale transformation similar to the effect of a layer of the MERA tensor network in previous approaches to entanglement renormalization \cite{Vidal:2007hda}. We show in the next section that this transformation also acts as a renormalization step on the spectrum of the boundary Hamiltonian.
The scale factor of the number of sites per inflation step can be computed analytically for vertex inflation on any $\{p,q\}$ tiling, and for the $\{3,7\}$ case is given by
\begin{equation}
\lambda_{\{3,7\}} = \frac{3 + \sqrt{5}}{2} \approx 2.62 \ ,
\end{equation}
in the limit of large system sizes (independent of the original subregion after sufficiently many inflation rules are applied).

In addition to an invariance under global rescaling, we find that the disorder values $g_k$ also possess a self-similarity that represents the qCFT equivalent of a \emph{local scale transformations}: For example, as shown in Fig.\ \ref{FIG_GLOB_LOC_SCALING}(b), various subsystems of the $g_k$ at $N=597$ sites (five vertex inflation steps) correspond to a coarse-grained version of the whole, with the scaling factor equivalent to that of a global scale transformation.
We  quantify this self-similarity and equivalence between local and global rescaling by introducing a fidelity measure between two disorder sequences $g$ and $g^\prime$ of equal length $L$, defined as
\begin{equation}
\mathcal{F}(g|g^\prime) = \frac{\sum_{k=1}^{L} g_k g^\prime_k}{\Vert g \Vert\, \Vert g^\prime \Vert} \ ,
\label{EQ_MOD_FIDELITY}
\end{equation}
where $\Vert g \Vert$ and $\Vert g^\prime \Vert$ are the $2$-norm of each sequence.
This natural definition gives $\mathcal{F}=1$ for two functions that are completely identical up to a total scale and fulfills $\mathcal{F}>0$ for any well-defined disorder vector with positive values.
By this definition, the disorder values $g_k$ at $N=12$ (Fig.\ \ref{FIG_GLOB_LOC_SCALING}(a), last column) and the ones obtained by choosing a specific $12$-site subsystem of the disorder values after four further vertex inflation steps (Fig.\ \ref{FIG_GLOB_LOC_SCALING}(b), last row, last column) have a fidelity of $\mathcal{F}=0.9986$. The fidelity values given in Fig.\ \ref{FIG_GLOB_LOC_SCALING}(b) for other inflation numbers and subsystem sizes are very close to 1 as well, confirming that local and global inflation steps have almost the same effect up to numerical variations and small-size effects. As expected, the scale factor of local self-similarity is also $\lambda_{\{3,7\}}$, i.e., a global deflation step from $N$ to $N^\prime < N$ sites is approximately equivalent to taking a suitable subsystem of $N^\prime$ sites.

\begin{figure}
\centering
\small
(a) \hspace{10pt} Local self-similarity of multi-scale quasicrystal

\includegraphics[width=0.9\textwidth]{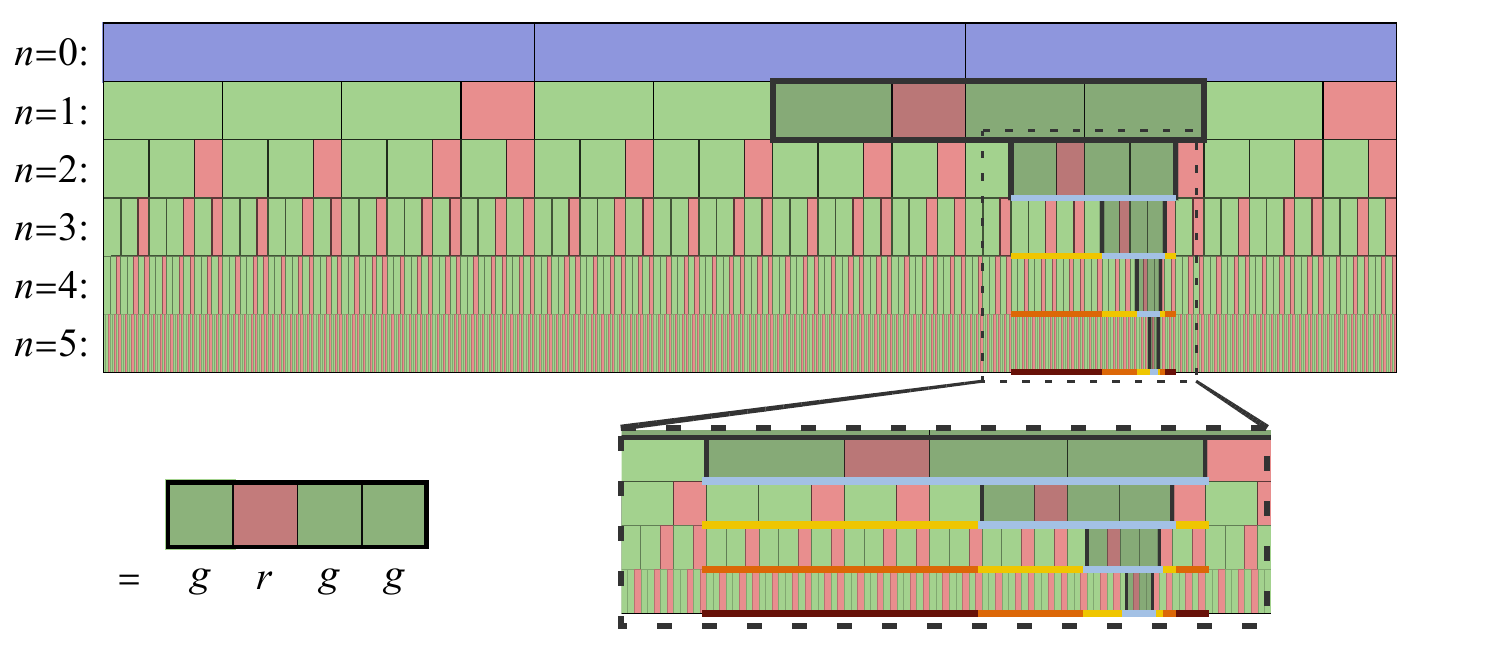}\\

(b) \hspace{10pt} Approximate translation invariance of multi-scale quasicrystal

\includegraphics[width=0.9\textwidth]{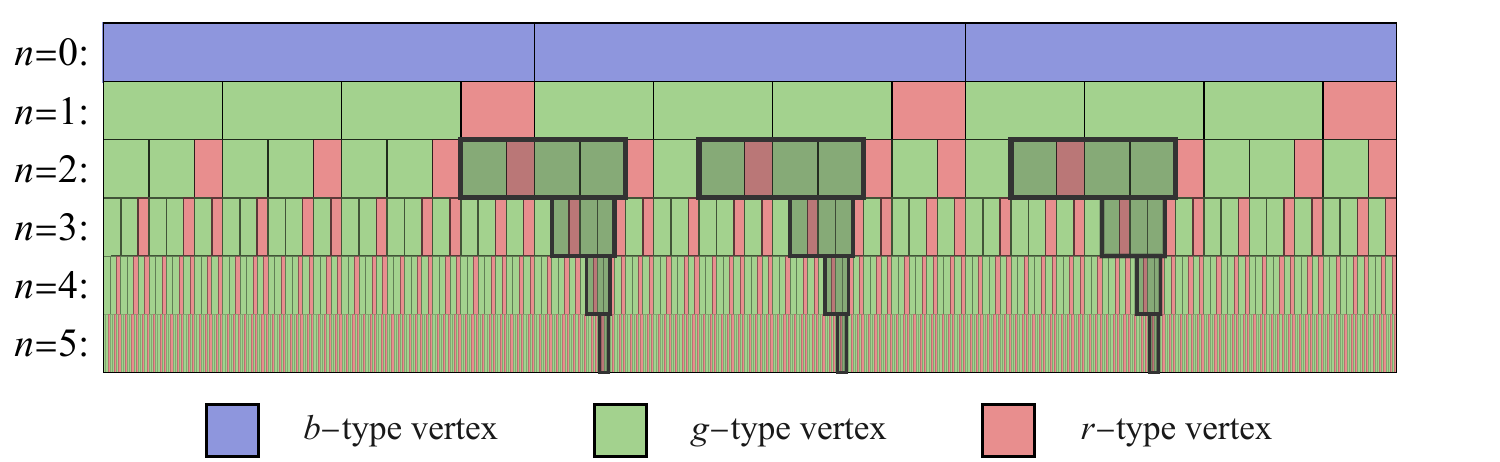}
\caption{Self-similarity of the $\{3,7\}$ multi-scale quasicrystal ansatz (MQA). The letters $b,g,r$ for each block are omitted, only using the colors blue, green, and red.
(a)~The sequence $grgg$ (shaded blocks) characterizing one third of the total system at $n=1$ inflation steps appears again as subsequences at $n>1$.
This leads to a fractal self-similarity of the resulting disorder vector. The colored lines at the bottom of each inflation layer indicate the boundary subsystems of the same color in each row of Fig.\ \ref{FIG_GLOB_LOC_SCALING}(b).
(b)~Appearance of the sequence $grgg$ in other subsystems at $n=2$ inflation steps, with equal self-similarity on further inflation layers. In addition to the three subsystems shown, $\mathbb{Z}_3$ symmetry leads to six additional repetitions. 
With exponentially many repeating sequences at large $n$, any sufficiently large boundary region is approximately invariant under translations.
}
\label{FIG_QCFT_SCALING}
\end{figure}

\begin{figure}
\centering
\includegraphics[width=0.75\textwidth]{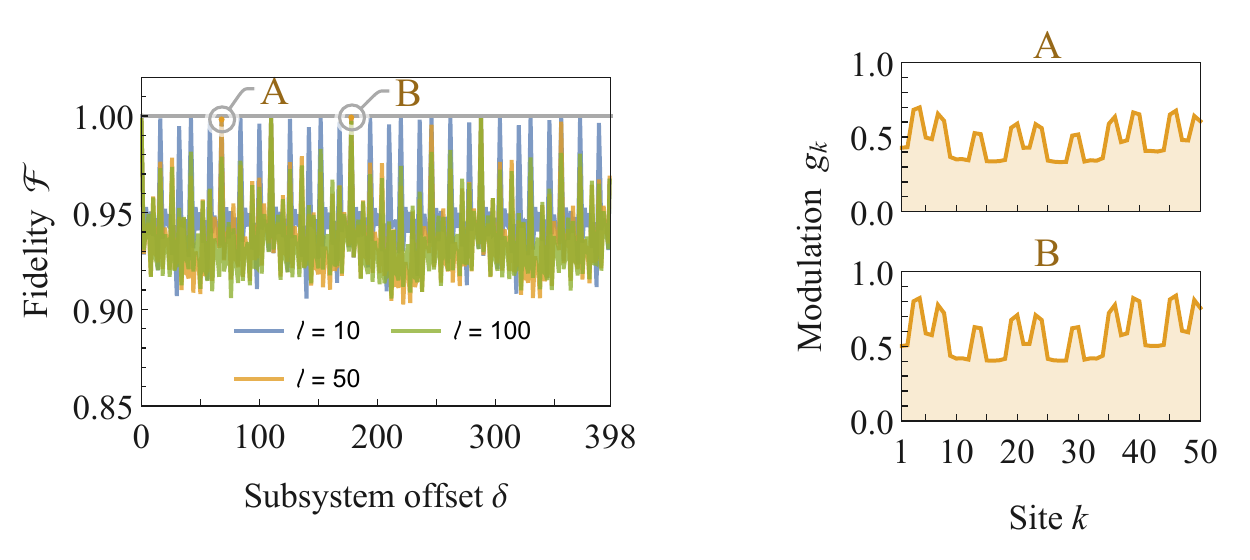}
\caption{Approximate translation invariance of a subsystem of $l$ values of the disorder vector $g$ on $N=597$ total sites (Fig.\ \ref{FIG_GLOB_LOC_SCALING}, top left) with another subsystem at offset $\delta$. The fidelity $\mathcal{F}$ between both subsystems of $g$ is shown on the left for three different subsystem sizes, with two examples subsystems $A$ and $B$ with fidelity close to $1$ shown on the right.
}
\label{FIG_QFT_TRL}
\end{figure}


This self-similarity also directly follows from the structure of the multi-scale quasicrystal ansatz (MQA) introduced in Fig.\ \ref{FIG_MQI_37}(a), which as we saw in Sec.\ \ref{SS_MQI} closely describes the boundary disorder (equivalently expressed by either $g$ or the couplings $J$).
The appearance of an invariance under discrete local scale transformations in the $\{3,7\}$ MQA is visualized in Fig.\ \ref{FIG_QCFT_SCALING}(a).
We first consider the quasiperiodic letter sequences appearing on each boundary layer. After a single inflation step under rule \eqref{EQ_V_INFLATION_37}, the sequence characterizing the full boundary geometry is given by
\begin{equation}
bbb \mapsto gggrgggrgggr \ ,
\end{equation}
which is composed of three repetitions of the subsequence $grgg$ (allowing for cyclic permutations), which thus characterizes the non-repeating part of the boundary geometry. Applying \eqref{EQ_V_INFLATION_37} to this subsequence leads to a new sequence that contains the old one twice,
\begin{equation}
\label{EQ_BCBB}
grgg \mapsto ggr\underline{grg}\overline{\underline{g}rgg}r \ ,
\end{equation}
where both appearances are over- and underlined, respectively. This means that successively applying the inflation rules leads to a sequence that recursively contains the non-repeating part of the starting sequence, as can be seen in Fig.\ \ref{FIG_QCFT_SCALING}(a), where we only show the second appearance (overlined in the above equation) on five inflation layers.
The effective scale factor between the original and the rescaled sequence is necessarily the same as the global one, $\lambda_{\{3,7\}} \approx 2.62$, after many inflation steps. For example, in the above sequence the ratio between the lengths of both sequences is $\frac{11}{4} = 2.75$. After two further inflation steps, we find
\begin{align}
\label{EQ_BCBB2}
\underbrace{grgg}_{\text{4 letters}}  \mapsto \underbrace{ggr\underline{grg\overline{g}}\overline{rgg}r}_{\text{11 letters}} 
\mapsto \underbrace{g\underline{grgg}r\underline{grgg}r\underline{grg\overline{g}}\overline{rgg}r\underline{grg\overline{g}}\overline{rgg}rgr}_{\text{29 letters}} 
\mapsto \underbrace{g\underline{grgg}rg \dots r\underline{grgg}rgr}_{\text{76 letters}}
\ ,
\end{align}
leading to a ratio $(\frac{76}{4})^{1/3} \approx 2.67$ per scale, gradually approaching $\lambda_{\{3,7\}}$. This is because inflating any sequence of length $\ell$ for $N$ times leads to a sequence of length $\sim \ell\, \lambda_{\{3,7\}}^N$ at large $N$. If we consider a subsequence equal to the original sequence (and thus of equal length $\ell$), the ratio between the length of the total sequence after $N$ inflation steps and that of the subsequence thus scales as $\sim \lambda_{\{3,7\}}^N$ in the same limit. Global and local self-similarity thus behave equivalently in quasiperiodic sequences.

We now extend this insight from letter sequences describing a single inflation layer to multi-scale blocks thereof, that is, to the full MQA: 
First we note that as the inflation rules are deterministic, all the further sequences ``below'' each appearance of $grgg$ (i.e., after more inflation steps) are the same, no matter on which inflation layer it appears, up to different cutoffs after a finite number of inflation steps. As a result, the boundary subsystems characterized by each $grgg$ sequence exhibit the same symmetries up to differing contributions from inflation layers ``above'' each sequence, which again become negligible as long as we allow for sufficiently many inflation steps below each sequence.\footnote{Recall that the local product of couplings in the MQA follows the structure of a Markov chain, eventually becoming independent of the starting point.} 
This leads to a discrete self-similarity: The disorder vector of the entire system (or rather, its non-repeating part) appears as a coarse-grained subsequence of itself on every inflation layer, again associated with the scaling factor $\lambda_{\{3,7\}}$. The number of such local rescalings that leave the system invariant grows exponentially with the number of inflation steps, as the subsequences appear more than once in inflated versions of themselves. Fig.\ \ref{FIG_QCFT_SCALING} (b) demonstrates this by showing three of the nine appearances of $grgg$ on the second inflation layer, each of which again contain itself recursively. As a result, a constant fraction of the entire system consists of self-similar sequences.
The specific subregion self-similarity of the $\{3,7\}$ disorder vector in Fig.\ \ref{FIG_GLOB_LOC_SCALING}(b) exactly matches the self-similarity of the $grgg$ sequence: In the $N=12$ case (fifth column of Fig.\ \ref{FIG_GLOB_LOC_SCALING}(a)), the central peak of $g_k$ corresponds to a weakened coupling term $J_k=1/(g_k g_{k+1})$ characterized by the $r$ letter (recall that $j_g \approx 1.14, j_r \approx 0.60$ in the $\{3,7\}$ MQA). Cyclically permuting the disorder values to move the $g_k$ peak to the left (matching the $grgg$ sequence), we find exactly the self-similarity of subsequences of Eq.\ \eqref{EQ_BCBB2}. This one-to-one correspondence is shown graphically through colored lines denoting equivalent subsystems between Figs.\ \ref{FIG_GLOB_LOC_SCALING}(b) and \ref{FIG_QCFT_SCALING}(a). Using our argument from Eq.\ \eqref{EQ_BCBB2}, we can thus quantitatively describe the self-similarity of $g_k$ and explain why its scaling factor matches $\lambda_{\{3,7\}}$.

Finally, this self-similarity also implies an approximate translation invariance: Each boundary translation that maps a subsequence on one inflation layer onto a repetition of itself leaves the disorder vector in that subsystem invariant. As with the number of repeating subsystems, the number of such boundary translations grows exponentially with the number of inflation steps (compare Fig.\ \ref{FIG_QCFT_SCALING}(b)) and remains a constant fraction of all possible translations.
In Fig.\ \ref{FIG_QFT_TRL}, we show how this approximate translation invariance manifests itself in the disorder vector: By taking a subset of $l$ values and computing the fidelity $\mathcal{F}$ (as defined in \eqref{EQ_MOD_FIDELITY}) with another subset obtained by shifting its position by an offset $0 \leq \delta \leq 2N/3$, we find a close but not exact matching between subsets with $\mathcal{F} \approx 0.94$ for most $\delta$, independent of $l$. 

However, for certain offsets the fidelity jumps very close to $\mathcal{F}=1$; these correspond exactly to two subregions related to matching subsequences in the MQA. Indeed, one finds disorder vector values at these offsets to be almost exactly equal up to a small difference in overall scale; this scale corresponds to contributions ``above'' the matching sequences in the MQA, acting uniformly on each subsystem. 
To summarize, the self-similarity properties of the disorder vector -- and by extension, the multi-scale quasicrystal Ising (MQI) model -- can be completely captured by the multiscale quasiperiodic ansatz (MQA), which also provides a concrete setting of discretized conformal transformation of the qCFT proposal.

\subsection{Continuum limit}
\label{SS_CONTINUUM_LIMIT}

The MTN ansatz is inherently discrete and finite-dimensional, with boundary states of the $\{3,7\}$ setup exhibiting disorder at all length scales.
This prohibits a clean identification with a continuum CFT, even though boundary states appear to approximate properties of the Ising CFT.
In the following, we will discuss two limits that appear to produce a physical continuum limit, i.e., in which the $\{3,7\}$ MTN boundary states converge towards the Ising ground state. These are:
\begin{enumerate}
\item The limit of an infinite number of inflation steps $n$, or \emph{scaling limit}, producing an infinite number of tensor network boundary sites $N$.
\item The limit of an infinite (bulk) bond dimension $\chi$, which restricts the space of parametrizable boundary states in our ansatz.
\end{enumerate}
We begin with the first limit. As seen in the previous section, the disorder vector $g$ encoding the deviation of the boundary states from the translation-invariant critical Ising ground state is successively fine-grained under successive inflation rules, exhibit fractal disorder on all length scales.
Instead of considering the $n\to\infty$ behavior of the disorder or coupling terms of the effective Hamiltonian whose ground state is produced by the MTN ansatz, we focus our attention on the scaling behavior of the \emph{spectrum} of these Hamiltonians. Here we use the mode-disordered Hamiltonian $H^\text{MDI}[g]$ with $g$ extracted from the boundary states via \eqref{EQ_DISORDER_37_EST}, though one may also use the couplings from the MQI model Hamiltonian $H^\text{MQI}[J]$ (with suitably chosen free MQA parameter), leading to very similar results.

The translation-invariant ground state of the critical Ising model \eqref{EQ_H_ISING} on $N$ lattice sites is exactly solvable.
Antisymmetry of the Gaussian Hamiltonian in its coupling terms $\m_j \m_k = -\m_k \m_j$ in terms of the Majorana operators $\m_k$ yields a spectrum of $N$ eigenvalue pairs $\lambda^\pm_k \equiv \pm \lambda_k$ with opposite sign. As we compute in App.\ \ref{APP_DISORDERED_ISING},
\begin{equation}
\lambda^\pm_k = \pm \sin \frac{(2k-1)\pi}{2N} \ .
\end{equation}
In other words, the low-energy spectrum follows a linear dispersion relation $\lambda^\pm_k \propto k$ but saturates at large momenta, i.e., exhibits a \emph{finite bandwidth}.

\begin{figure}[tb]
\centering\small
(a) \hspace{10pt} Energy spectra for normalized Hamiltonians \\
\includegraphics[width=0.8\textwidth]{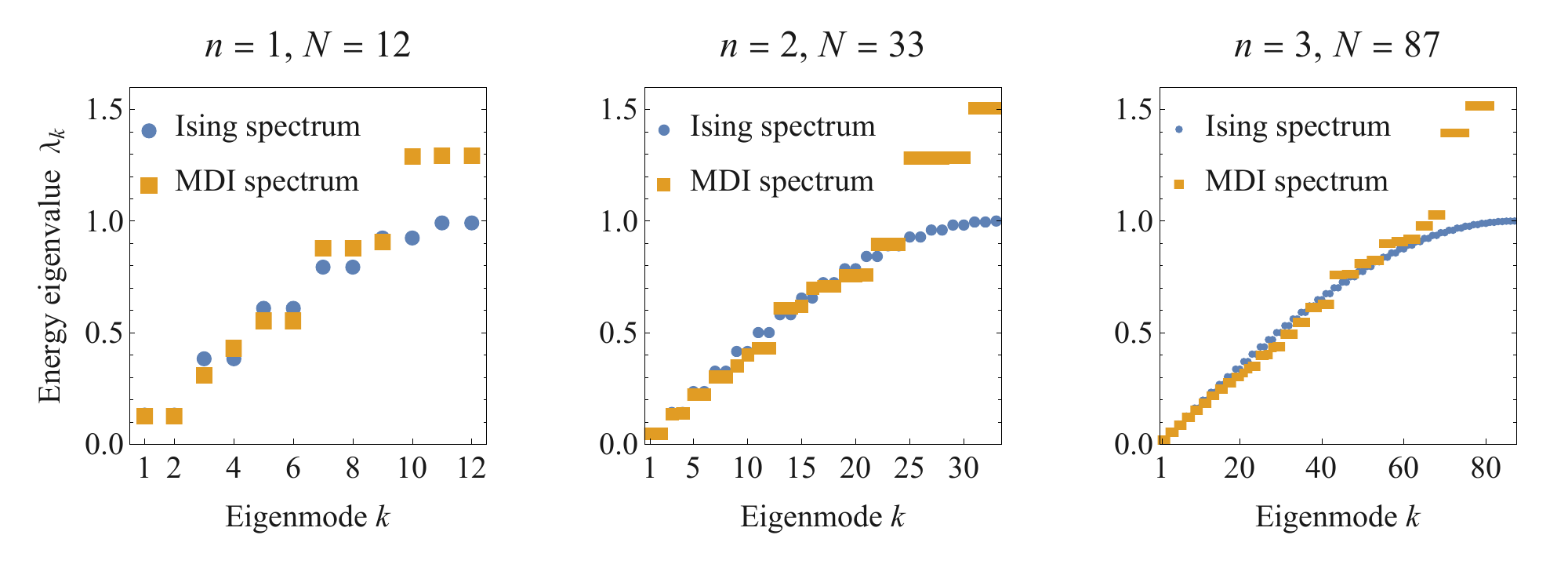} \\
(b) \hspace{10pt} Energy spectra for rescaled Hamiltonians \\
\includegraphics[width=0.48\textwidth]{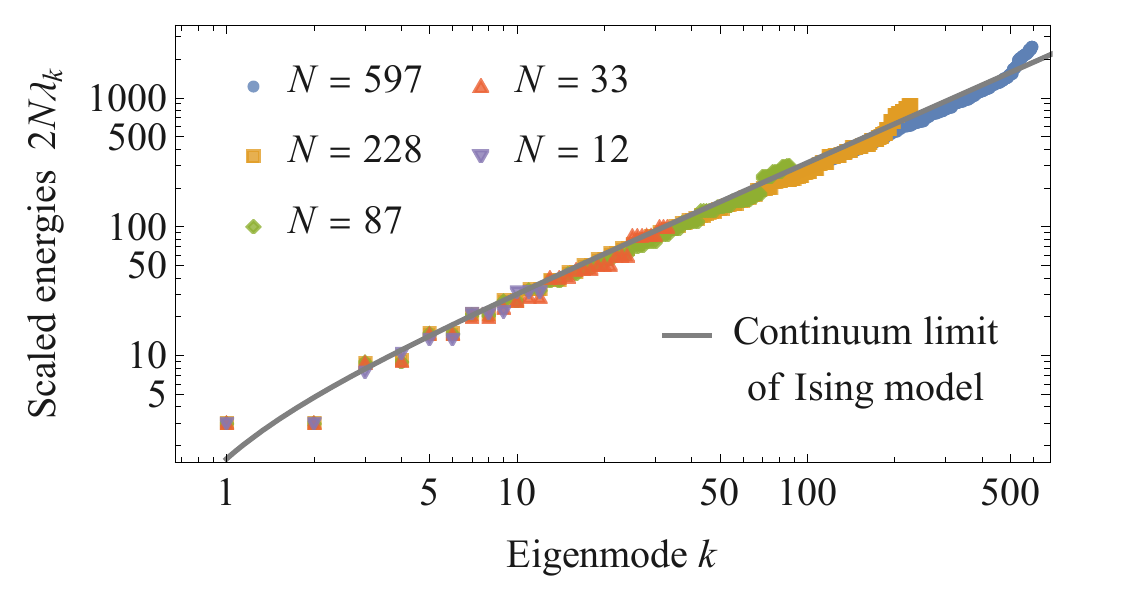}

\caption{(a)~Spectrum of the critical Ising Hamiltonian \eqref{EQ_H_ISING} vs.\ the mode-disordered Ising (MDI) Hamiltonian \eqref{EQ_ISING_MOD} for the $\{3,7\}$ MTN at the critical Ising point, the latter with normalized disorder $\frac{1}{2N}\sum_k g_k = 1$, at three successive inflation steps on $N=12,33,$ and $87$ sites ($n=1,2,3$ inflation steps, respectively).
(b)~Overlaid spectrum of MDI Hamiltonian for various $N$ with rescaled energies $2N\lambda_k$, along with the dispersion relation of the continuum Ising model.
The MDI spectra overlap up to finite $N$ effects, indicating an RG flow between inflation steps.
}
\label{FIG_QC_H_SPEC}
\end{figure}

In Fig.\ \ref{FIG_QC_H_SPEC}(a), we show how this property changes for the spectrum of the MDI model Hamiltonian for the $\{3,7\}$ MTN boundary states: At low energy and momentum, the MDI and Ising spectrum overlap, but unlike the latter, the former does not saturate at large momentum but continues an approximately linear growth.
At large system size $N$, we find that both spectra exhibit a vanishing energy gap, i.e., become critical.
We therefore arrive at the surprising result that both models describe the same low-energy physics, even though the disorder vector $g$ contains contributions on all length scales. 
This result was already foreshadowed by our observation that the disordered covariance matrices of the $\{3,7\}$ MTN boundary states can be mapped to the translation-invariant ones only up to deviations at small scales or equivalently, large energies (compare Fig.\ \ref{FIG_QR_SYMMETRIES}). In other words, the two models are dynamically equivalent in the IR.

Overlaying the MDI energy spectra for different system sizes $N$ and rescaling each by $N$, as done in Fig.\ \ref{FIG_QC_H_SPEC}(b), we find that spectra closely match up to small deviations at the high-momentum end of each set.
This supports our earlier claim that the vertex inflation steps define a renormalization group (RG) flow towards the high-energy (UV) limit: Each step only modifies the high-energy part of the model while leaving the low-energy part unchanged.
Thus, similarly to the MERA \cite{Vidal:2007hda,PhysRevLett.101.110501,Evenbly2015TNR1,Evenbly2015TNR2}, the emergent dimension (radial direction) of the critical Ising MTN can be interpreted as an energy/length scale.
At large $N$, the spectra approach the continuum limit of the critical Ising model; intriguingly, the spectrum of the disordered Hamiltonian appears to approach this limit of linear dispersion better than the translation-invariant spectrum of \eqref{EQ_H_ISING}, as the former does not gradually saturate.

\begin{figure}[t]
\centering\small

(a) \hspace{10pt} Higher bond dimension $\{3,7\}$ MTN at $n=3,N=87$ \\ 
\includegraphics[width=0.95\textwidth]{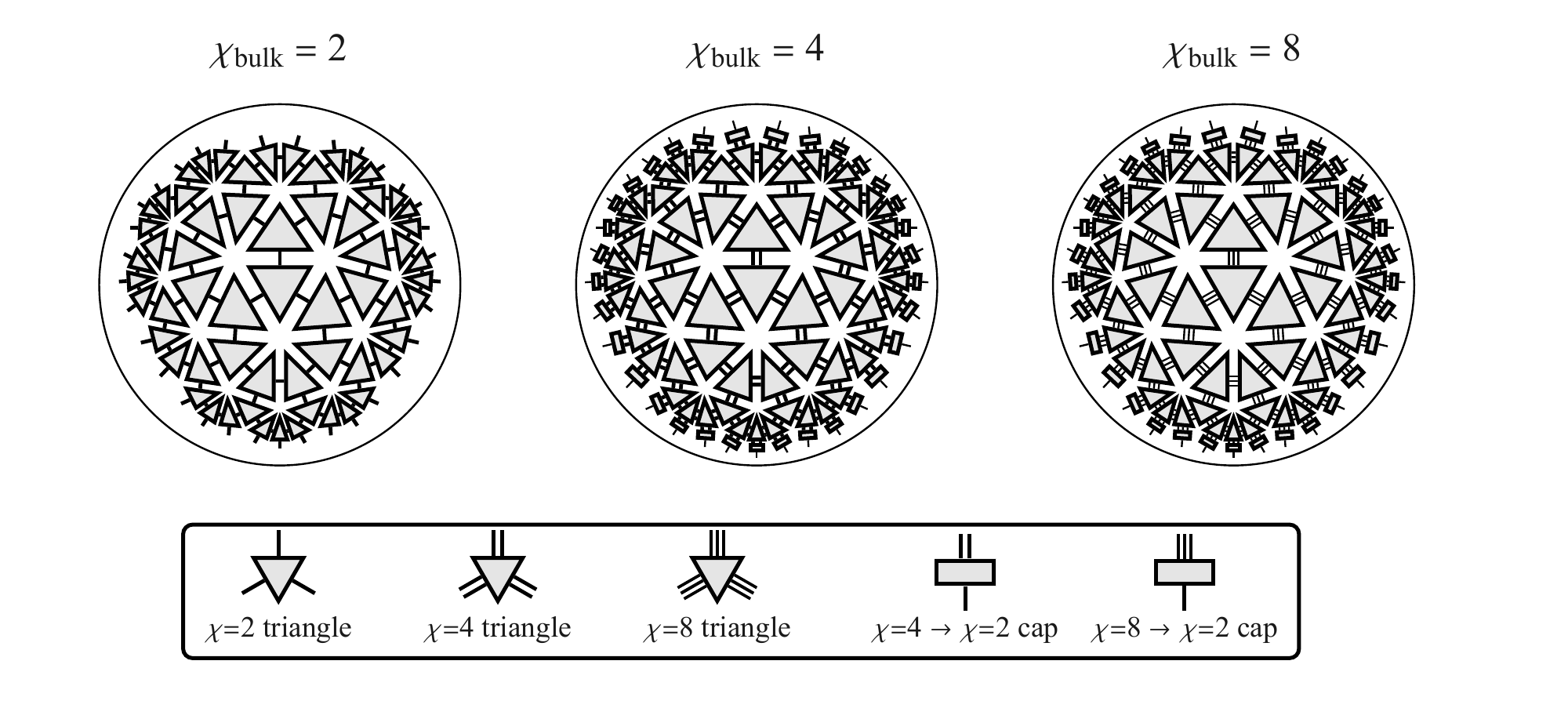}\\
(b) \hspace{10pt} Decay-adjusted covariance matrices \\ 
\includegraphics[width=0.95\textwidth]{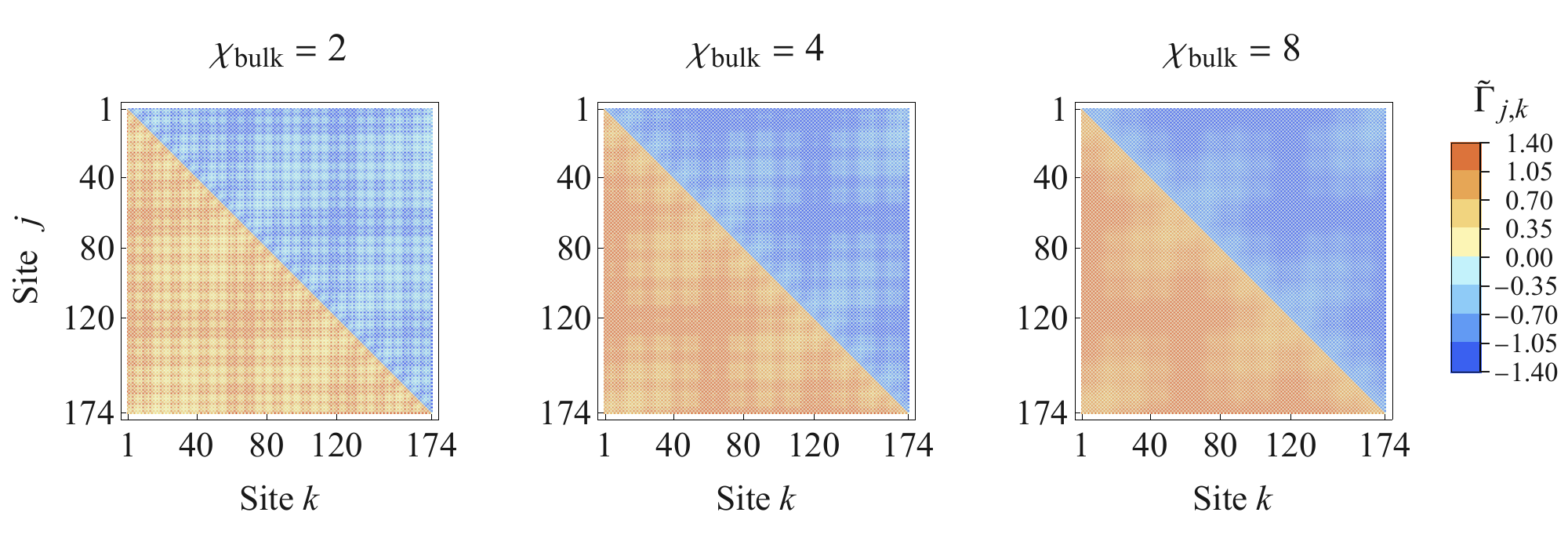} \\
(c) \hspace{10pt} Translation invariance for subsystem of length $\ell$, offset $\delta$ \\ 
\includegraphics[width=0.95\textwidth]{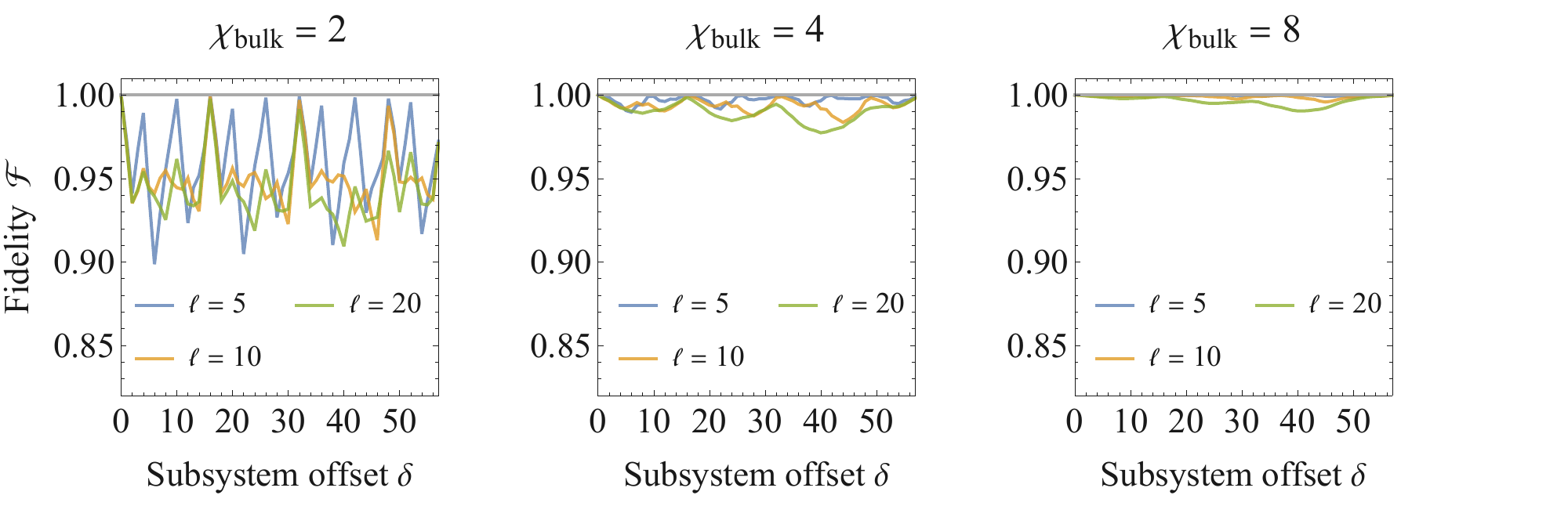}

\caption{Construction of matchgate tensor networks (MTN) with higher bond dimension $\chi_\text{bulk}$. 
(a) Tensor network construction on a $\{ 3,7 \}$ tiling with $\chi_\text{bulk}=2,4,8$, represented by $\log_2 \chi_\text{bulk}$ fermionic bonds in the matchgate formalism. For $\chi_\text{bulk} \geq 4$, additional cap tensors on the boundary legs reduce the boundary bond dimension to $\chi_\text{bdy}=2$.
(b) Decay-adjusted covariance matrix $\tilde{\Gamma}$ of the MTN boundary states at the critical Ising point (minimum energy with respect to $H^\text{I}$).
(c) Fidelity $\mathcal{F}$ of subsystems of the disorder vector $g$ extracted from each covariance matrix (via \eqref{EQ_DISORDER_37_EST}) under translations (compare Fig.\ \ref{FIG_QFT_TRL}). At large $\chi_\text{bulk}$ the disorder approaches the translation-invariant case $\mathcal{F}=1$.
}
\label{FIG_LARGE_CHI}
\end{figure}

We now consider the second limit necessary to approach continuum boundary states, that of large bond dimension $\chi$.
The previous MTN computations assume a tensor network with $\chi=2$, i.e., where each tiling edge is associated with a single spin or fermion degree of freedom.
We can extend the parameter space of this model by increasing the internal or bulk bond dimension $\chi_\text{bulk}=2^k$ to any $k \in \mathbb{N}$, effectively associating $k$ fermionic degrees of freedom with each $n$-gon edge.
At the same time, we wish to leave the size of the boundary local Hilbert space unchanged, as we wish to approach the continuum Ising model (rather than some higher-spin CFT).
We thus keep the boundary bond dimension fixed to $\chi_\text{bdy}=2$. This necessitates an additional local $k{+}1$-leg matchgate tensor on each of the $N$ boundary sites, called the \emph{cap tensor}, that reduces $\chi_\text{bulk}$ to $\chi_\text{bdy}$, introducing additional degrees of freedom.

Fig.\ \ref{FIG_LARGE_CHI} shows the resulting tensor network geometry as well as the (decay-adjusted) covariance matrix after an optimization with respect to the translation-invariant Ising Hamiltonian $H^\text{I}$.
As $\chi_\text{bulk}$ is increased, the quasiperiodic disorder is gradually washed out on all scales, approaching the translation-invariant result. 
This is also reflected in the disorder vector $g$ extracted from each covariance matrix via \eqref{EQ_DISORDER_37_EST}: Subsystems of $g$ become increasingly translation-invariant at larger $\chi_\text{bulk}$, with the fidelity $\mathcal{F}$ under subsystem translation approaching unity.
In addition, one can compute the \emph{state fidelity} $f$ between the exact Ising ground state on $N$ sites and the MTN result for varying $\chi_\text{bulk}$.
For Gaussian states with covariance matrices $\Gamma$ and $\Gamma^\prime$, it is defined as \cite{bravyi2016complexity}
\begin{equation}
\label{EQ_STATE_FID}
f(\Gamma,\Gamma^\prime) =  \left( \det \frac{\Gamma + \Gamma^\prime}{2} \right)^{1/2} \ . 
\end{equation}
This definition of the fidelity is exactly the overlap $\tr(\rho \rho^\prime)$ between the two pure Gaussian states with density matrices $\rho$ and $\rho^\prime$ (fulfilling $\tr \rho=\tr\rho^\prime = 1$).
For $N=87$ and $\chi_\text{bulk}=2$, the fidelity $f(\Gamma^\text{I},\Gamma^\text{MTN})$ between Ising and MTN  states is $0.641$; at $\chi_\text{bulk}=4$ and $8$ it increases to $f=0.970$ and $0.995$, respectively, showing a fast convergence of the MTN ansatz to the exact Ising ground state.
The numerical effort to compute these large $\chi_\text{bulk}$ models is still moderate: Using the tiling symmetries, the number of free parameters in the bulk triangle tensors and the boundary layer increases to $6$ and $12$ for $\chi_\text{bulk}=4$ and $8$, respectively, and numerical optimization over such a small set of parameters is still feasible within a few minutes of computation time.

To summarize, in the scaling limit $n\to\infty$ the $\{3,7\}$ MTN ansatz produces ground states of a nearest-neighbor parent Hamiltonian whose spectrum matches with the continuum limit of the Ising model. In the additional large bond dimension limit $\chi_\text{bulk} \to\infty$ these states appears to converge to the translation-invariant Ising ground state, with the disorder being gradually suppressed out on all scales.

\subsection{Disordered correlation functions}
\label{SS_DISORDERED_CORRELATION_FUNTIONS}

The boundary states of MTNs at the critical Ising point have been shown to reproduce the correct Ising correlation functions of CFT primary operators (or suitable lattice versions thereof) when performing site-averaging as defined in  Eq.~\eqref{EQ_AVG_CORR_FALLOFF}. \cite{Jahn:2017tls}.
We now show more precisely how the correlators of the boundary states with quasiperiodic disorder indeed, on average, reproduce the continuum CFT values.
Here we restrict ourselves to the case with minimal bulk bond dimension $\chi_\text{bulk}=2$, where the disorder is strongest.

\begin{figure}
\centering
\includegraphics[width=0.5\textwidth]{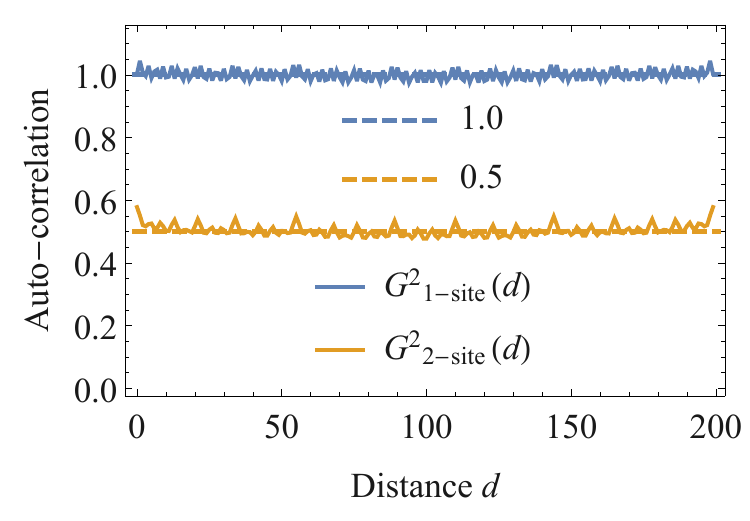}
\caption{Auto-correlation of the disorder vector $g$ for the $\{3,7\}$ MTN at the critical Ising point between $1$- and $2$-site blocks at distance $0 \leq d \leq N/3$, for a boundary system of $N=597$ sites. Both auto-correlations are independent of $d$ up to small fluctuations.
}
\label{FIG_G_AUTOCORR}
\end{figure}

We first consider, for illustration, the two-point functions of fermionic fields $\psi_k$ and $\bar{\psi}_k$, which can be defined in terms of complex combinations of the odd and even Majorana operators $\m_{2k-1}$ and $\m_{2k}$ in a consistent manner\footnote{Ref.\ \cite{Zou:2019dnc} recently suggested a more refined approach to identifying lattice operators in critical spin chains corresponding to CFT primaries, leading to additional corrections. These do not affect the averaging argument in this section.} 
so that \cite{Jahn:2017tls}
\begin{align}
\langle \psi_j \psi_k \rangle 
= \langle \bar{\psi}_j \bar{\psi}_k \rangle 
&= \frac{1}{4} \left( \Gamma_{2j,2k-1} + \Gamma_{2j-1,2k} \right) \\
&= \frac{1}{4} \left( g_{2j}\, g_{2k-1}\, \Gamma^\prime_{2j,2k-1} +  g_{2j-1}\, g_{2k}\, \Gamma^\prime_{2j-1,2k} \right) \ ,
\end{align}
where in the second line we have expressed the matchgate covariance matrix $\Gamma$ in terms of the approximately translation-invariant $\Gamma^\prime$ and the disorder vector $g$, following \eqref{EQ_ISING_COV_MOD}. Using this approximate translation invariance to rewrite $\Gamma^\prime_{j,k} \approx \Gamma^\prime(k-j)$, we find that boundary site averages over the two-point functions lead to
\begin{align}
\frac{1}{N}\sum_{k=1}^N \langle \psi_k \psi_{k+d} \rangle \approx \frac{1}{4N} \left( 
\Gamma^\prime(2d-1) \underbrace{\sum_{k=1}^N g_{2k}\, g_{2k+2d-1}}_{=: G^2_\text{1-site}(2d-1)} + 
\Gamma^\prime(2d+1) \underbrace{\sum_{k=1}^N g_{2k-1}\, g_{2k+2d}}_{=: G^2_\text{1-site}(2d+1)} \right) \ .
\label{EQ:G1SITE}
\end{align}
where we have defined the two-point auto-correlation $G^2_\text{1-site}(d)$ of the disorder vector $g$ between two single sites at distance $d$.
The quasiperiodicity of $g$ ensures that $G^2_\text{1-site}$ must be independent of $d$, as the converse would imply a periodicity inherent to $g$. Numerical evidence of this is shown in Fig.\ \ref{FIG_G_AUTOCORR}.
From this, we conclude that site-averages over the two-point correlator of the fields $\psi$ and $\bar{\psi}$ of the quasiperiodic Ising model yield the value of the correlator for the translation-invariant Ising model, up to numerical fluctuations at small scales where \eqref{EQ_ISING_COV_MOD} no longer holds. 

The same logic applies to higher-order correlation functions. Consider for example the two-point function of the $\epsilon$ field, which can be written as \cite{Jahn:2017tls}
\begin{align}
\langle \epsilon_j \epsilon_k \rangle - \langle \epsilon_j \rangle \langle \epsilon_k \rangle 
&= \frac{1}{4} \Gamma_{2j-1,2k} \Gamma_{2j,2k-1} \\
&= \frac{1}{4} g_{2j-1}\, g_{2j}\, g_{2k-1}\, g_{2k} \Gamma^\prime_{2j-1,2k} \Gamma^\prime_{2j,2k-1} \ ,
\end{align}
the average over which becomes
\begin{align}
\frac{1}{N}\sum_{k=1}^N \left( \langle \epsilon_k \epsilon_{k+d} \rangle - \langle \epsilon_k \rangle \langle \epsilon_{k+d} \rangle \right) \approx \frac{1}{4N} \Gamma^\prime(2d-1) \Gamma^\prime(2d+1) 
\underbrace{\sum_{k=1}^N g_{2k-1}\, g_{2k}\, g_{2k+2d-1}\, g_{2k+2d}}_{G^2_\text{2-site}(2d)} \ ,
\end{align}
which now depends on the auto-correlation $G^2_\text{2-site}$ between blocks of two sites.
Again, quasiperiodicity of $g$ ensures that $G^2_\text{2-site}$ remains independent of the distance $d$ (see Fig.\ \ref{FIG_G_AUTOCORR}) and the averaged correlation functions lead to the expected Ising behavior, as previously observed numerically. This logic can be extended to arbitrary expectation values, including non-local observables such as correlators of the $\sigma$ primary field, which then become dependent on higher-order auto-correlation functions of $g$ that also remain constant.
More generally, we therefore conclude that the MDI model preserves the correlation functions of the critical Ising model as long as the disorder is quasiperiodic. Note that this requirement is weaker than the specific multi-scale quasiperiodicity produced by our MTN ansatz, and should hold for a much more general class of models.

\subsection{Excitations and holography}
\label{SS_EXCITATIONS_AND_HOLOGRAPHY}
\begin{figure}[tb]
\centering
\includegraphics[width=1.0\textwidth]{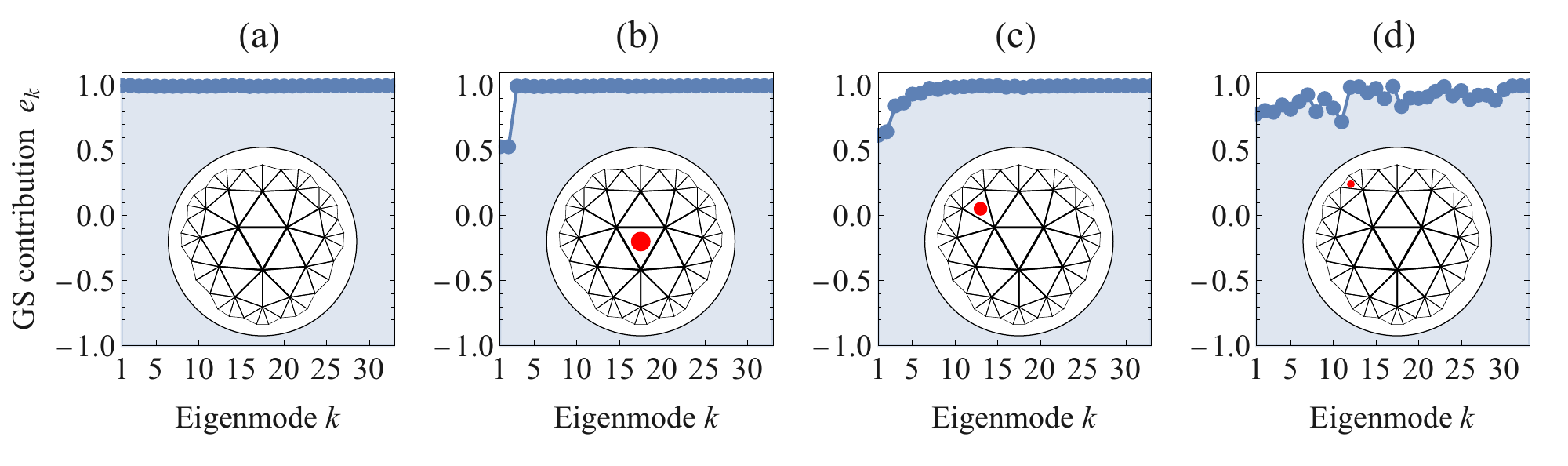}

\caption{Ground-state mode contributions $e_k$ for local bulk excitations in the $\{3,7\}$ matchgate tensor network at different bulk sites. (a) No excitation. (b)-(d) Excitation on single tile (red dot) belonging to different inflation layers.
Modes are sorted by their energy eigenvalue $\lambda_k$ relative to the mode-disordered Ising Hamiltonian $H^\text{MDI}$ corresponding to the tiling with $N=33$ boundary sites ($n=2$ inflation steps).
}
\label{FIG_QC_EXC}
\end{figure}

Our current discussion of the boundary states produced by hyperbolic MTNs focused on their relationship with \emph{ground states} of the critical Ising model.
On the bulk side, this led us to consider MTNs with the same tensor on each site of the hyperbolic $\{p,q\}$ lattice to preserve the maximum amount of lattice symmetry; in AdS/CFT language, this would correspond to the identification of ``pure AdS'' geometry without massive perturbations with the boundary CFT ground state.
To consider excited states in the MTN framework, let us therefore consider breaking some of these symmetries by modifying individual tensors in the hyperbolic lattice while leaving the rest with the same matchgate input that previously led to the MDI/MQI ground state.
We refer to the modified tensors as ``defects'' in the uniform tensor network. 
As we will show, there exists a simple correspondence between the energy of the excitations and the distance of the defect tensors from the boundary.
We illustrate this correspondence in both directions: First, we explore what boundary eigenstates are excited when we add simple defects in the $\{3,7\}$ MTN bulk; adding these defects creates superposition of low-energy eigenstates. Conversely, we explored what defects must be added in order to approximate specific excited eigenstates at the boundary.

To begin, we study which boundary eigenstates are generated when we modify only a single tensor in a simple way. As a first example, we apply sign flips $a \mapsto -a$ on the generating matrix component) for a single matchgate tensor as defined in \eqref{EQ_matchgate_TRI}. 
We now examine the covariance matrix $\Gamma^{\mbox{\tiny flip}}$ of the state resulting from such a defect MTN. To diagnose whether the defective state is an eigenstate, we transform $\Gamma^{\mbox{\tiny flip}}$ to the eigenmode basis $\Gamma^{\mbox{\tiny flip}} = O \tilde{\Gamma} O^\text{T}$ (see App.\ \ref{APP_DISORDERED_ISING} for details), where $O \in SO(2N)$ is the same transformation matrix that block-diagonalizes the mode-disordered Ising Hamiltonian $H^\text{MDI}$ for the ground state, i.e., the parent Hamiltonian of the boundary state without the defect. 
In other words, we determine whether including the defect preserves the eigenspace of $H^\text{MDI}$, in which case $\tilde{\Gamma}$ remains block-diagonal.

Numerically, we find that these flip defects largely preserve the block-diagonal structure of $\tilde{\Gamma}$, but generally act on many eigenmodes of $H^\text{MDI}$ at once, creating soft but complicated excitations.
We identify which eigenstates are excited by considering the $N$ eigenmode occupations numbers
\begin{equation}\label{eq:gscontribution}
e_k = \tilde{\Gamma}_{2k,2k-1}.
\end{equation}
The ground state is identified with $e_k=1$ for all $k$ (see \ref{FIG_QC_EXC}(a)), with all eigenmodes unoccupied. Similarly, the first excited state is characterized by $e_1=-1$ and $e_k=1$ for all $1<k\leq N$, occupying only the first mode.
Whenever a defect in the bulk does not produce an eigenstate, the eigenmode occupation numbers is somewhere between the values $+1$ and $-1$, similar to thermal states.
By varying the position of a single flip defect from the center of the tensor network towards the boundary, we find a simple relationship between excitation energy and the distance of the defect from the boundary, as illustrated in \ref{FIG_QC_EXC}(b)-(d):
A defect at the center of the tensor network only changed the two lowest-lying energy eigenvalues, but positioning the defect closer to the boundary affects eigenstates with increasingly higher energy.
Note, however, that the change in eigenstate contributions is always small relative to the ground state; that is, a single flip defect creates states that are only slightly different from the ground state.
We find more generally that as long as only the low-energy subspace is excited (as in \ref{FIG_QC_EXC}(b) and (c)), eigenmodes are always excited in pairs. Comparison with the MDI spectrum in Fig.\ \ref{FIG_QC_H_SPEC} shows that this occurs because these energy eigenvalues are twofold degenerate as modes come in left- and right-moving pairs, equally excited by our real-valued matchgate input. We therefore define the non-chiral energy eigenstate vectors $\ket{E_k}$ in the following manner: The ground state vector $\ket{E_0} = \ket{0}_1\ket{0}_2 \dots$ corresponds to all eigenmodes unoccupied ($e_k=1$ for all $k$). The first non-chiral excitation $\ket{E_1} = \ket{1}_1\ket{1}_2 \ket{0}_3 \ket{0}_4 \dots$ corresponds to occupying only the first two modes ($e_1=e_2=-1$, $e_k=1$ for $k>2$), while the next two excitations $\ket{E_2} = \ket{0}_1\ket{0}_2 \ket{1}_3 \ket{1}_4 \ket{0}_5 \dots$ and $\ket{E_3} = \ket{1}_1\ket{1}_2 \ket{1}_3 \ket{1}_4 \ket{0}_5 \dots$ continue the pairwise occupation patterns leading to increasingly higher energy.
\begin{figure}[tb]
\centering\small
(a) \hspace{10pt} Eigenstate contributions of tuned excitation \\
\includegraphics[width=0.9\textwidth]{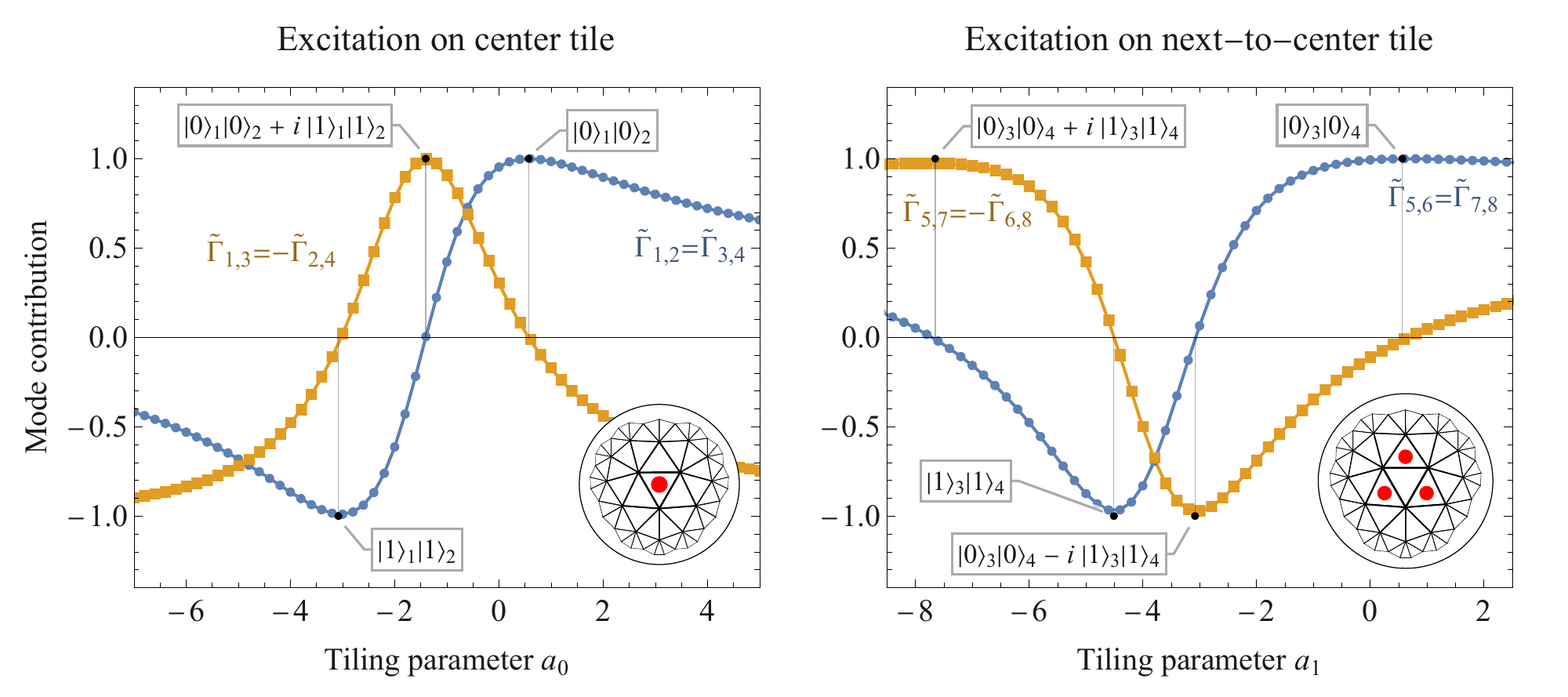} \\
(b) \hspace{10pt} Bulk parameters for boundary eigenstates \\
\includegraphics[width=0.9\textwidth]{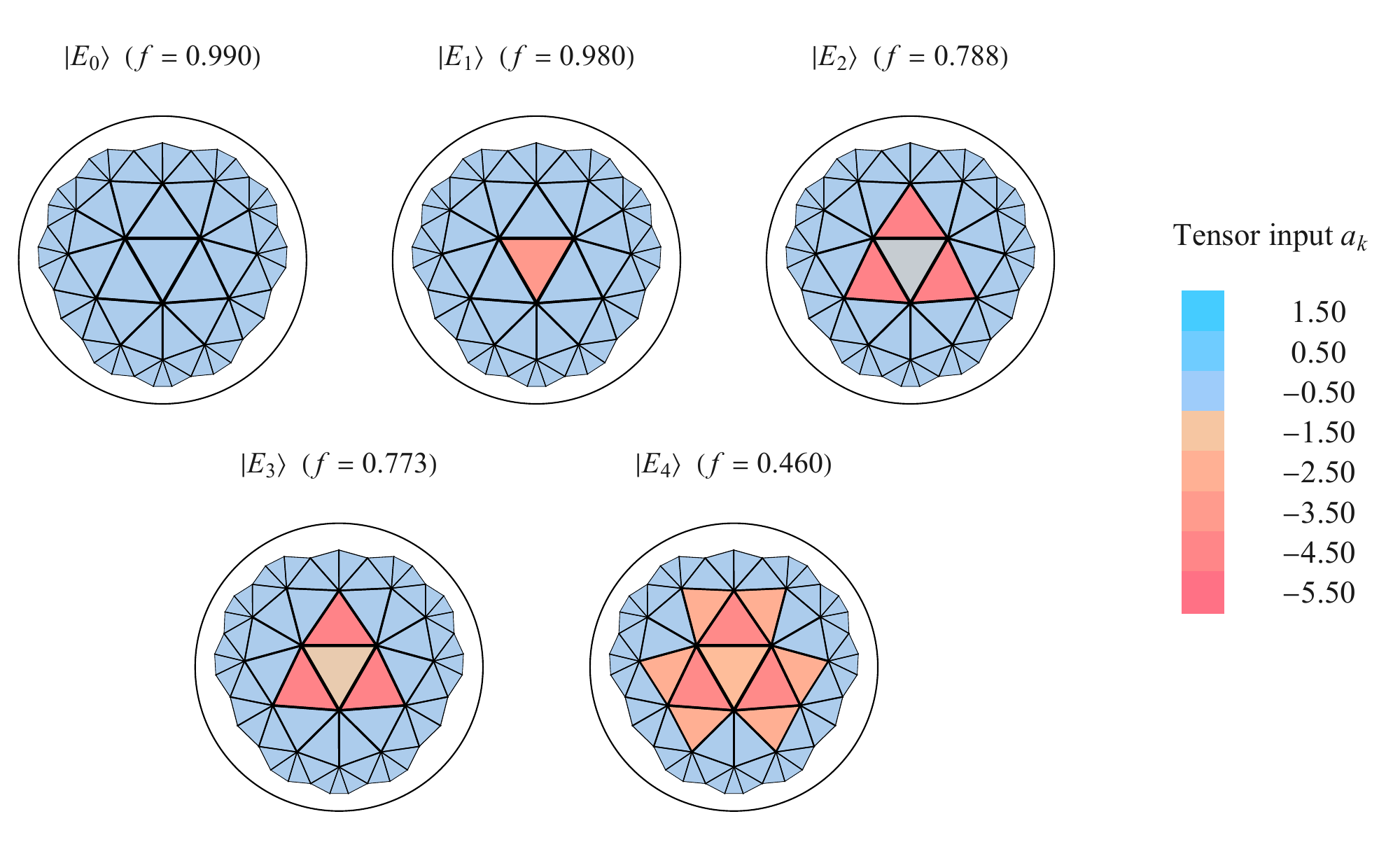}

\caption{(a)~Dependence of covariance matrix entries $\tilde{\Gamma}_{j,k}$ in the eigenbasis of $H_\text{QI}$ under variation of the tiling parameter on the central tile (left) or its neighboring tiles (right) in the $\{3,7\}$ tiling. The modified tiles are marked as red dots in the inset. The boundary consists of $N=33$ sites ($n=2$ inflation steps).
(b)~The bulk input corresponding to various excited states color-coded for each tile. The state fidelity $f$ with respect to the exact excited state of $H^\text{MDI}$ is also given.
}
\label{FIG_EXC_PARAM}
\end{figure}

We now consider defects with more a precise control over the specific bulk input to reproduce the $\ket{E_k}$ for small $k$.
Once again, we alter the parameter $a = a_0$ only on the central tensor of the $\{3,7\}$ MTN, but instead of just flipping its sign we consider a general real number as input. We find again that only the lowest two energy eigenstates are excited at the boundary, allowing us to manipulate the low-energy boundary subspace in a controlled way.
Denoting the resulting covariance matrix (in the eigenmode basis) $\tilde{\Gamma}$, we find that for any $a_0 \in \mathbb{R}$ 
\begin{align}
\tilde{\Gamma}_{1,2} &= \tilde{\Gamma}_{3,4} \ , &
\tilde{\Gamma}_{1,3} &= -\tilde{\Gamma}_{2,4} \ , &
\tilde{\Gamma}_{1,4} &= \tilde{\Gamma}_{2,3} \approx 0 \ .
\end{align} 
Tuning $a_0$ allows for a range of different boundary states to be produced, as shown in Fig.\ \ref{FIG_EXC_PARAM}(a). At $a_0 = a_\text{Ising} \approx 0.58$ we find the ground state $\ket{E_0}$, as excepted.
At $a_0 \approx -3.0$ the first two eigenmode contributions are flipped to the excited state vector $\ket{E_1}$, appearing as $\ket{1}_1\ket{1}_2$ along the first two eigenmodes (in a product state with the remaining ones).
Additionally, at $a_0 \approx -1.4$ we find a Bell pair superposition $\ket{E_0} + \i \ket{E_1}$ of these two outputs. 

Similar behavior is also seen when adding multiple identical defects around the central tensor. For example, by changing the tiling parameter $a=a_1$ on the three neighbors of the central tensor, also shown in Fig.\ \ref{FIG_EXC_PARAM}(a), we find that the third and fourth lowest-lying eigenmodes are simultaneously excited. 
By choosing an appropriate $a_1$, these two states can be superimposed or a specific eigenmode can be excited. However, tuning $a_1$ also affects the contributions to the first two eigenmodes.
If, for example, we wish to determine the value of $a_1$ for which the first four eigenmodes are fully occupied, i.e., to produce $\ket{E_3}$, we have to first tune $a_1 \approx -4.306$ and only then choose an optimal $a_0 \approx -0.069$. Note that this particular value for $a_0$ differs from the case where we only modified the central tensor.
We can also tune $a_0 \approx 0.071$ to occupy \emph{only} the second pair of eigenmodes,  thus creating the energy eigenstate vector $\ket{E_2}$.

Next, we invert this approach to determine the defects that produce specific excited eigenstates. For this we numerically optimize the matchgate tensors (their respective parameter $a$ on each tile) to maximize the state fildelity $f$ (as defined in \eqref{EQ_STATE_FID}) with respect to the exact excited states of $H^\text{MDI}$.
In our numerical optimization, we choose local parameters $a$ that preserve the $D_3$ symmetry group ($2\pi/3$ rotations and reflections along three axes) of the $\{3,7\}$ tiling centred around a central triangular tile.
The numerical results of this optimization are visualized in Fig.\ \ref{FIG_EXC_PARAM}(b) for $N=33$ boundary sites, where this optimization can be performed with small computational effort. We obtain a high fidelity for the lowest-lying excited states, but our numerical approach becomes less accurate for higher-energy states that involve optimizing more tensors closer to the boundary.

We conclude this section with some remarks pertaining to a holographic interpretation of these plots. In the interpretation as a holographic model, where the tensors encode the strength of local bulk correlations, one may interpret these bulk ``excited\ states'' as deformations from the pure AdS time-slice geometry. 
The excitations produced by local bulk operations on the matchgate tensor content only reliably change the low-energy eigenmode contributions while leaving the high-energy part of the ground state intact. We thus interpret these bulk operations as being equivalent to acting with low-energy operators on the boundary state. 
This follows from a simple counting argument: The number of bulk tiles (or bulk-symmetric sets thereof) is smaller than the number of boundary sites, so it is only possible to describe a boundary subspace using generic input in the bulk.
This a feature general for holographic dualities wherein the hyperbolic bulk is `smaller' than the boundary and hence boundary subspace restrictions follow, which may be interpreted as a code subspace \cite{Almheiri15}.
It should be noted that within our approach it is also possible to consider complex valued tensors $a\in \mathbb C$ and possibly an entire Bloch sphere of the two lowest lying eigenmodes $\{\alpha_{00}\ket 0_1\ket0_2 +\alpha_{11}\ket1_1\ket1_2\}$ can be realized via MTN.
This would have the meaning of encoding a single qubit in two lowest-lying eigenstates of the boundary CFT.
While various multi-mode superpositions are possible encoding more qubits can no longer be possible just within MTNs, as non-Gaussian states become necessary for generically entangled qubit states.
Nonetheless, one can consider the basis of a full quantum error correction code to arise from MTNs tuned to the vicinity of the Ising CFT and then a non-Gaussian tensor network could be used to explore the entire encoded space.
This is similar to the HaPPY code \cite{Pastawski2015} because MTNs suffice to study the product states of bulk qubits, as was shown in Ref.~\cite{Jahn:2019nmz}.
In other words, our MTN investigations point to the possibility that holographic quantum error correction-codes encoded in low-energy subspaces of critical quantum systems can be explored in a constructive way using tensor networks.

\section{Discussion}
\label{S_DISCUSSION}
In this work, we studied the physical theories arising from hyperbolic tensor network models based on matchgate tensors.
For suitable chosen parameters, these matchgate tensor networks (MTNs)  were shown to produce a boundary theory described by a critical mode-disordered Ising (MDI) model, which differs from the critical Ising model by a disorder applied to each individual Majorana mode in the fermionic representation of the model.
We found that the disorder vector for these states, which completely captures the disorder present in the system, exhibits quasiperiodicity on all length scales and can be captured by an analytical multi-scale quasicrystal Ising (MQI) model.
This model also includes the appearance of discretely broken symmetries of conformal ground states, concretely realizing the recent proposal of a quasiperiodic conformal field theory (qCFT) \cite{Jahn:2020ukq} which predicted specific symmetries of boundary states of tensor networks with regular hyperbolic tilings.

We calculated explicit MDI and MQI Hamiltonians whose ground states are given by the MTN boundary states using an analytical technique, providing a rare example of a tensor network ansatz whose parent Hamiltonian can be efficiently constructed: In the MDI approach, this Hamiltonian can be directly computed from the boundary state disorder, while the analytical MQI model relies only on a single parameter, assuming its consistency in the scaling limit. 
By an analytical argument, we have shown that the quasiperiodic nature of such states leads to site-averaged correlation functions that match the critical Ising model, implying that hyperbolic tensor networks are effective at studying the properties of translation-invariant critical models with a continuum CFT limit.
We also showed that the disorder disappears in the large bond dimension limit for suitably optimized MTNs, yielding good approximations of the translation-invariant ground states of the critical Ising model.
At large bond dimension, the effects of bulk discretization are thus diminished, leading to boundary states more alike those of a continuum CFT.

Finally, excitations in our tensor network ansatz can be related to tuning the matchgate parameters away from their critical value on individual tensors in the hyperbolic tiling, leading to a numerical dictionary between bulk tensor input and low-lying excited states on the boundary.
Here we found that the energy scale of the excitation can be related to the distance of such bulk deformations from the boundary.
In the accessible low-energy regime, the energy spectrum of the disordered models matches that of the Ising model, implying that the low-energy physics of these models is insensitive to the discrete nature of the bulk lattice.

Our MTN ansatz shares many of the features of the more widely studied MERA tensor networks, such as a hierarchical structure describing the boundary state on different length scales \cite{Vidal:2007hda,RigorousWavelet} and an effective map between low-energy boundary excitations and bulk input configurations \cite{Chua:2016cxo}. However, the regular hyperbolic tilings considered here naturally discretize AdS time-slices, i.e., the hyperbolic plane, and may thus be more useful for constructing new types of holographic models. Indeed the boundary symmetries resulting from such a bulk geometry largely determined the boundary state properties we studied in this work.

The results in this paper appear to hold generally for MTNs on regular hyperbolic $\{p,q\}$ tilings, though we focused on the $\{3,7\}$ case for explicit results. In App.\ \ref{APP_45TILING}, we show that the main results carry over to $\{4,5\}$ tilings analogously. Beyond our tensor network construction, it appears that disordered Ising models with properties like those discussed here can be constructed from any multi-scale quasiperiodic sequence; such theories may therefore be studied independently from the tensor network approach.

While our current numerical techniques assume Gaussianity, the symmetry setup is independent of the choice of tensors and we expect that tensor networks on these geometries can be used to study critical theories more complicated that the Ising model. The lack of constraints on the tensor content, as those required by the MERA, allows for much more general models to be constructed. We should expect such models to be more computationally challenging that the MERA in general, as the evaluation of local expectation values can no longer be simplified through the use of such constraints. Fortunately, our results show that averages over hyperbolic MTN boundary states at small bond dimension, requiring little computational effort, already capture crucial properties of the continuum model.
Another potential solution, in particular in the context of strongly-coupled models relevant for holography, may be to build experimental quantum architectures with qubits on regular hyperbolic geometries \cite{Koll_r_2019}.

As matchgate tensor networks can now be used to build holographic toy models with a well-defined local boundary Hamiltonian, an important question that we will explore in further work concerns time evolution, in particular whether boundary dynamics can be understood in terms of operations on the bulk tensors. Similar questions concerning bulk dynamics of excitations were previously explored using the MERA \cite{Chua:2016cxo}. This line of research is related to one of the key goals of the field of tensor network holography: Finding tensor network models reproducing aspects of gravity and gravitational dynamics. We hope that our work will contribute to such future endeavours.

\acknowledgments{We thank the DFG (EI 519/15-1, CRC 183, projects B01 and A03) 
 for support. A.~J.\ has been supported by the FQXi and the Simons Collaboration on It from Qubit: Quantum Fields, Gravity, and Information. This work has also received funding from the European Union's Horizon 2020 research and  innovation programme under grant agreement No.\ 817482 (PASQuanS). S.~S.\ acknowledges the support provided by the Alexander von Humboldt Foundation and the Federal Ministry for Education and Research through the Sofja Kovalevskaja Award while he was employed at the Max-Planck Institute for Gravitational Physics in Potsdam.}

\providecommand{\href}[2]{#2}\begingroup\raggedright\endgroup

\appendix

\section{Derivation of the mode-disordered Ising model}
\label{APP_DISORDERED_ISING}

We here show that ground states of the mode-disordered Ising (MDI) Hamiltonian \eqref{EQ_ISING_MOD} with disorder vector $g$ are approximately described by the covariance matrix \eqref{EQ_ISING_COV_MOD}.
To arrive at these findings, we start with a generic free-fermion (Gaussian) Hamiltonian
expressed as a quadratic polynomial in Majorana fermions
\begin{align}
\label{EQ_H_FREE_FERMIONS}
H = \i \sum_{j,k=1}^{2N} M_{j,k} \m_j \m_k \ ,
\end{align}
which can be block-diagonalized by transforming the $2N \times 2N$ coupling matrix $M$ into the form $M = O^\text{T} \tilde{M} O$, with
\begin{equation}
\tilde{M} = \bigoplus_{k=1}^N 
\begin{pmatrix}
0 & \lambda_k \\
-\lambda_k & 0
\end{pmatrix} \ 
\end{equation}
using a mode transformation $O\in SO(2N)$, with real $\{\lambda_k\}$. We can choose $O$ so that $\lambda_k>0$.
Expressed in terms of the new modes, the ground-state covariance matrix $\Gamma^0 = O \tilde{\Gamma}^0 O^\text{T}$ of $H$ takes the simple form 
\begin{equation}
\tilde{\Gamma}^0 = \bigoplus_{k=1}^N
\begin{pmatrix}
0 & -1 \\
1 & 0
\end{pmatrix} \ .
\label{EQ_COVMATRIX_BLOCKDIAG}
\end{equation}
As a specific choice for $M$, we first consider the translation-invariant Ising Hamiltonian \eqref{EQ_H_ISING}. Its coupling matrix is given by
\begin{equation}
M_{j,k} = \frac{1}{2} \left( \delta_{j+1,k} - \delta_{j,k+1} \right) \ .
\end{equation}
Here, we have defined the indices modulo $2N$ so that, e.g., $\delta_{2N+1,2N} \equiv \delta_{1,2N}$ (the lack of a sign flip corresponds to antisymmetric boundary conditions).
The coefficients $\lambda_k$ of the block-diagonal decomposition of $M$ correspond to the eigenvalues of the problem $\i\, M \vec{v}_k = \lambda_k \vec{v}_k$. Under the given boundary conditions, this problem is solved with an exponential ansatz
\begin{align}
\label{EQ_ISING_EVEC_EVAL}
\left(\vec{v}_k^\pm\right)_j &= e^{\mp \i\pi j (2k-1) / 2N} \ , &
\lambda_k^\pm &= \pm \sin\frac{(2k-1) \pi}{2N} \ .
\end{align}
Each eigenvalue is twofold degenerate as $\lambda_{N+1-k}^\pm =  \lambda_k^\pm$. 
One can explicitly construct the orthogonal transformation $O$ by stacking the real and imaginary parts of the eigenvectors $\vec{v}_k^\pm$, leading to
\begin{align}
\label{EQ_GAMMA_FROM_O_1}
O_{k,j} = 
\begin{cases}
\frac{1}{\sqrt{N}}\cos\left( \pi \frac{j\, k}{2N} \right) & \text{for odd } k \\
\frac{-1}{\sqrt{N}}\sin\left( \pi \frac{j (k-1)}{2N} \right) & \text{for even } k \\
\end{cases} \ .
\end{align}
Applied to the ground state covariance matrix $\tilde{\Gamma}^0$ in the mode basis, which contains equal contributions from all negative eigenmodes, this gives an analytical expression of the covariance matrix of the ground state of the free-fermionic Ising model,
\begin{align}
\label{EQ_GAMMA_FROM_O_2}
\Gamma^0_{i,j} = (O^\mathrm{T}\tilde{\Gamma}^0 O)_{i,j} 
&= \sum_{k=1}^{N} \left( O_{2k-1,i}O_{2k,j} - O_{2k,i}O_{2k-1,j} \right) \nonumber \\
&= \frac{-1}{N} \sum_{k=1}^{N} \sin\left( \pi \frac{(2k-1)(i-j)}{2N} \right) \nonumber \\
&= 
\begin{cases}
0 & \text{for even } i-j \\
\frac{-1}{N \sin\left(\frac{\pi}{2N}(i-j) \right)} & \text{for odd } i-j
\end{cases} \ .
\end{align}
Note that in the infinite system limit $N \to \infty$ we find $\Gamma^0_{i,i+2d-1} = 2/(\pi(2d-1))$, as expected from the two-point function decay of the fermion fields $\psi,\bar{\psi}$ with scaling dimension $\Delta=\frac{1}{2}$.
We also see that the covariance matrix is manifestly translation-invariant.

Now consider a mode-disordered Ising model with Hamiltonian \eqref{EQ_ISING_MOD} where each Majorana mode $\m_k$ is associated with a weight $g_k$. This leads to a family of coupling matrices 
$g\mapsto M(g)$ in \eqref{EQ_H_FREE_FERMIONS} of the form
\begin{equation}
M_{j,k}(g) = \frac{1}{2 g_j g_k} \left( \delta_{j+1,k} - \delta_{j,k+1} \right) .
\end{equation} 
An eigenvector $(\vec{v}_k)_j$ of $\i M$ has to solve the equation
\begin{equation}
-\i \lambda_k (\vec{v}_k)_j = \frac{(\vec{v}_k)_{j+1}}{2g_j g_{j+1}} - \frac{(\vec{v}_k)_{j-1}}{2g_{j-1}g_j} \ .
\end{equation} 
In the continuum limit where we replace indices $j$ by a continuous coordinate $x=j/(2N)$, leading to an eigenfunction $\vec{v}_k \to v_k(x)$ and a continuum disorder $g_j \to g(x)$ on $x \in [0,1]$, this equation becomes
\begin{equation}
-\i\, \lambda_k g(x) v_k(x) = \frac{\text{d}}{\text{d}x} \left( \frac{v_k(x)}{g(x)} \right) \ , 
\end{equation}
which can be solved by an exponential ansatz in $x\mapsto v_k(x)/g(x)$, yielding the solution 
\begin{align}
v_k(x) &= g(x) \exp\left( -\i\lambda_k h(x) \right) \ , & 
h(x) &= \int_0^x\mathrm{d}y\, g(y)^2 .
\end{align}
While the disorder function fulfills periodic boundary conditions $g(0)=g(1)$, the eigenfunctions need to fulfill anti-periodic ones, leading to
\begin{align}
\lambda_k = \frac{(2k-1)\,\pi}{h(1)-h(0)} \ ,
\end{align}
where we are keeping the number $N$ of eigenmodes finite. Now extending \eqref{EQ_GAMMA_FROM_O_1} and \eqref{EQ_GAMMA_FROM_O_2} to arbitrary eigenvectors, we find
\begin{align}
\Gamma^0[g](x,y) = \frac{1}{N} \sum_{k=1}^N \operatorname{Im}\left( v_k^\star(x) v_k(y) \right) 
&= \frac{-g(x) g(y)}{N} \sum_{k=1}^N \sin \left( (2k-1) \pi \frac{h(x) - h(y)}{h(1)-h(0)} \right) \nonumber\\
&= \frac{-g(x) g(y)}{N} \frac{\sin^2 \left(N \pi \frac{h(x) - h(y)}{h(1)-h(0)} \right)}{\sin\left( \pi \frac{h(x) - h(y)}{h(1)-h(0)} \right)}
\ .
\end{align}
If $x\mapsto g(x)$ varies mostly on small scales and stays approximately constant over large ones, 
we can write
\begin{align}
\label{EQ_QR_H_CONDITION}
\frac{h(x) - h(y)}{h(1)-h(0)} \approx x-y \ .
\end{align}
Reintroducing the lattice, this gives us the approximate covariance matrix
\begin{align}
\Gamma^0_{i,j}[g] \approx \frac{-g_i g_j}{N} \frac{\sin^2 \left( \frac{\pi}{2} (i-j) \right)}{\sin\left( \frac{\pi}{2N} (i-j) \right)}
&=
\begin{cases}
0 & \text{for even } i-j \\
\frac{-g_i g_j}{N \sin\left(\frac{\pi}{2N}(i-j) \right)} & \text{for odd } i-j
\end{cases}  \ .
\end{align}
We have thus arrived at \eqref{EQ_ISING_COV_MOD} and find that this approximation is valid at sites $i,j$ where $|i-j|$ is larger than the typical fluctuation scale of $g$. We can immediately check that this reproduces \eqref{EQ_GAMMA_FROM_O_2} at all scales for the case of no disorder.

Note that for more general deformations of an Ising model it may not be possible to find a faithful disorder function because the number of needed parameters scales linearly in the system size while a generic non-translation invariant covariance matrix depends on quadratically many independent parameters.
This compressibility is an early indication that tensor networks on hyperbolic tilings produce structured deviations from translation invariance.

It is also interesting to note that the disorder created by a replacement $\m_k \mapsto \m_k/g_k$ in the coupling matrix led to a transformation $\Gamma_{j,k} \mapsto g_j g_k \Gamma_{j,k}$ in the covariance matrix. If we introduced a new set of Majorana modes $\m^\prime_k = \m_k/g_k$ with anti-commutation relation 
\begin{equation}
\{\m^\prime_j, \m^\prime_k \} = \frac{2\delta_{j,k}}{g_j g_k}, 
\end{equation}
then it follows that $\Gamma_{j,k} = g_j g_k \Gamma^\prime$. The above approximation \eqref{EQ_QR_H_CONDITION} is then equivalent to approximating $\Gamma^\prime$ by the Ising ground state covariance matrix \eqref{EQ_GAMMA_FROM_O_2}, i.e., neglecting contributions from the non-canonical commutator. Ising models with such nonstandard fermionic algebras may provide an interesting direction for further study.

\section{Quasiperiodic parent Hamiltonians from convex optimization}
\label{APP_COUPLING_OPT}

\begin{figure}[t]
\centering

\includegraphics[width=1.0\textwidth]{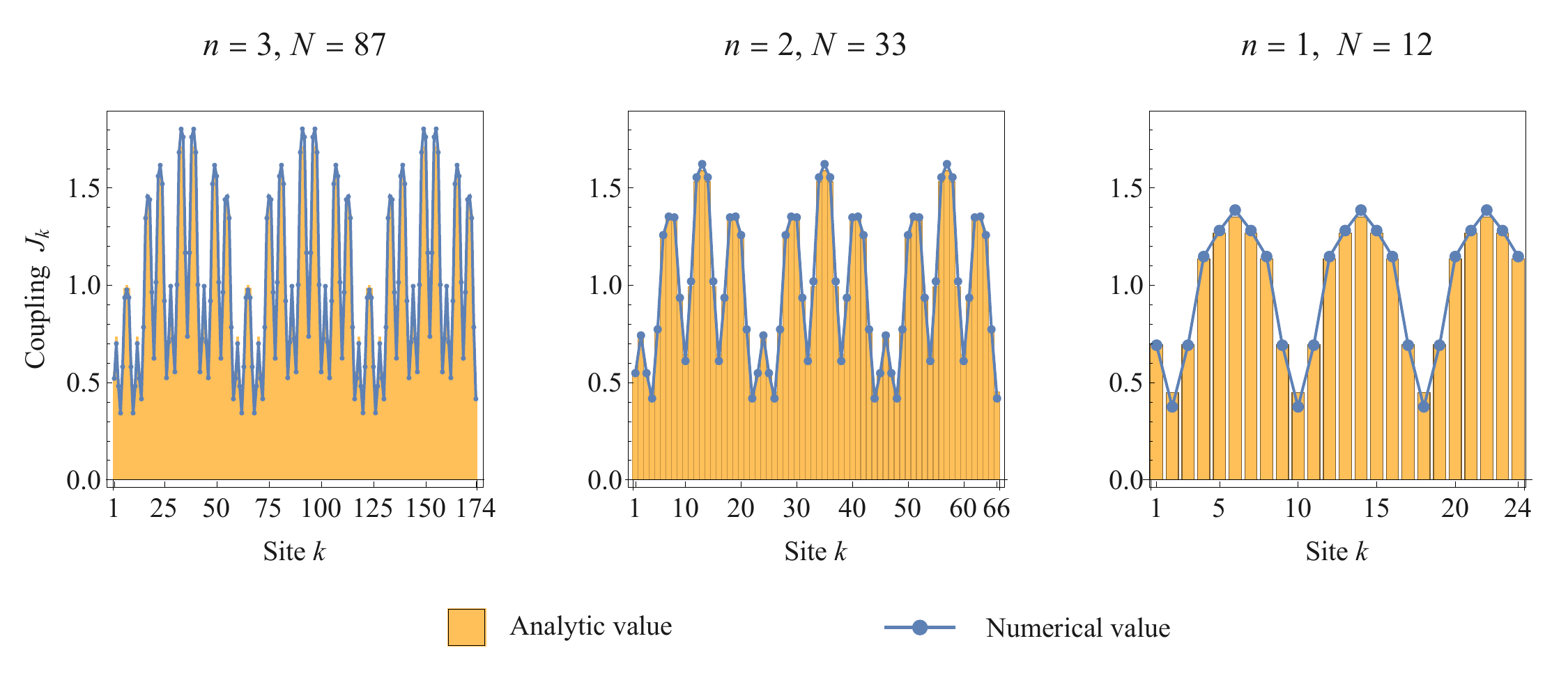}

\caption{Coupling terms $\{J_k\}$ of the parent Hamiltonian \eqref{EQ_H_QR} for the $\{3,7\}$ MTN ground state, obtained analytically as in \eqref{EQ_ISING_MOD} from the disorder vector $g$ (orange bars) and numerically through the optimization of a nearest-neighbor Hamiltonian (blue points). Both couplings closely match for systems of different size $N$ at various inflation steps. The couplings are normalized so that $\frac{1}{2N}\sum_k J_k = 1$.
	}
\label{FIG_PARENT_H}
\end{figure}

The boundary states of matchgate tensor networks on hyperbolic $\{3,7\}$ tilings at the critical Ising point were shown to be ground states of a Hamiltonian \eqref{EQ_ISING_MOD} with a disorder vector $g$ extracted from fluctuations in the state's covariance matrix. Such a \emph{parent Hamiltonian}, a Hamiltonian whose ground state matches the given quantum state (a concept that is ubiquitous in the theory of tensor network states), is not unique, as different Hamiltonians can give rise to the same ground state. Generally speaking, all possible parent Hamiltonians can be easily classified. Let $\Gamma^0$ be the covariance matrix of a pure fermionic Gaussian state, so that 
\begin{equation}
(\Gamma^0)^\text{T}\, \Gamma^0 = \mathbb{I}.
\end{equation}
Expressed in terms of these modes, the ground-state covariance matrix $\Gamma^0 = O \tilde{\Gamma}^0 O^\text{T}$ of $H$ takes the simple form
\begin{equation}
O^\text{T}\, \Gamma^0 O =
\tilde{\Gamma}^0 = \bigoplus_{k=1}^N
\begin{pmatrix}
0 & a_k \\
-a_k & 0
\end{pmatrix} 
\end{equation}
with $a_k\in\{-1,1\}$ for $k=1,\cdots, N$. The family of 
parent Hamiltonians of $\tilde{\Gamma}^0 $ can be easily identified
and in fact fully classified. Consider the family of Hamiltonians $\tilde H$
of the form
\begin{align}
\tilde H = \i \sum_{j,k=1}^{2N} M_{j,k} \m_j \m_k \ ,
\end{align}
with 
\begin{equation}
M= \bigoplus_{k=1}^N
\begin{pmatrix}
0 & \lambda_k \\
-\lambda_k & 0
\end{pmatrix}
\end{equation}
where
\begin{equation}\label{freedom}
	\lambda_k  \left\{
	\begin{array}{ll}
	>0 ,\, \text{ if } a_k=-1,\\
	<0 ,\, \text{ if } a_k=1.\\
	\end{array}
	\right.
\end{equation} 
The case of $\lambda_k=0$ for some $k$ leading to degenerate ground states is hence avoided. The parent Hamiltonians of the pure states governed by $\Gamma^0$
are hence all Hamiltonians $H$ of the form
\begin{equation}
	H =  O \tilde{\Gamma}^0 O^\text{T},
\end{equation}
for the above fixed given $O\in SO(2N)$ that back-rotates to the original basis.
That is to say, the freedom one has is precisely the 
one of choosing the $N$ real parameters $\{\lambda_k\}$, under the sign constraint as
 in Eq.\ (\ref{freedom}). This form allows to easily find parent Hamiltonians under additional
 constraints of locality. For example, one can ask for nearest-neighbor parent 
 Hamiltonians $H$ by solving the \emph{linear feasibility problem} that asks, given a 
 $O\in SO(2N)$ and a collection of $a_k\in \{-1,1\}$ for $k=1,\cdots, N$, 
  whether real 
 $\{\lambda_k\}$ can be found, subject to Eq.\ (\ref{freedom}), and the further linear constraints
 such that
 \begin{equation}
(O \tilde{\Gamma}^0 O^\text{T})_{j,k}=0,\, |j-k|>1.
\end{equation}
This linear program can be solved efficiently, and practically to any system size. What is more,
for a given vector $\vec{a}=(a_1,\cdots, a_N)$ and $O\in SO(2N)$,
one can minimize the vector $\vec{\lambda}=(\lambda_1,\cdots, \lambda_N)$
of Hamiltonian weights, to give rise to a \emph{semi-definite
problem} \cite{Boyd2004}, as 
\begin{eqnarray}
	\text{minimize }  &\|\vec{\lambda}\|_2,\\
	\text{subject to } & (O \tilde{\Gamma}^0 O^\text{T})_{j,k}=0,\,\forall j,k=1,\cdots, N 
	\text{ with }  |j-k|>1,\\
	& \lambda_k  \left\{
	\begin{array}{ll}
	>0 ,\, \text{ if } a_k=-1,\\
	<0 ,\, \text{ if } a_k=1.\\
	\end{array}
	\right.
	\forall k=1,\cdots, N.
\end{eqnarray}
This semi-definite program then produces the unique Gaussian parent Hamiltonian with local couplings being only non-zero between neighboring sites. 
Rather than this more general program, we now assume that a local Hamiltonian \eqref{EQ_H_QR} fulfilling these conditions already exists, and numerically determine its $2N$ nearest-neighbor coupling terms $\{J_k\}$ using a simple conjugate gradient code.
As our optimization function to be maximized, we choose the state fidelity $f$ (defined in \eqref{EQ_STATE_FID}) relative to the given boundary state of the matchgate tensor network. 
This algorithm produces a unique solution plotted in Fig.\ \ref{FIG_PARENT_H} for various system sizes, matching almost exactly with the values for the couplings $J_k=1/(g_k g_{k+1})$ with the disorder vector $g$ extracted from  of the corresponding boundary state.
This confirms our claim that there exists a unique nearest-neighbor model, a mode-disordered Ising model (or more specifically, a close approximation of a multi-scale quasiperiodic Ising model), which describes the boundary states of regular hyperbolic matchgate tensor networks.

\section{Non-Ising matchgate boundary states}

\begin{figure}
\centering
\includegraphics[height=0.22\textheight]{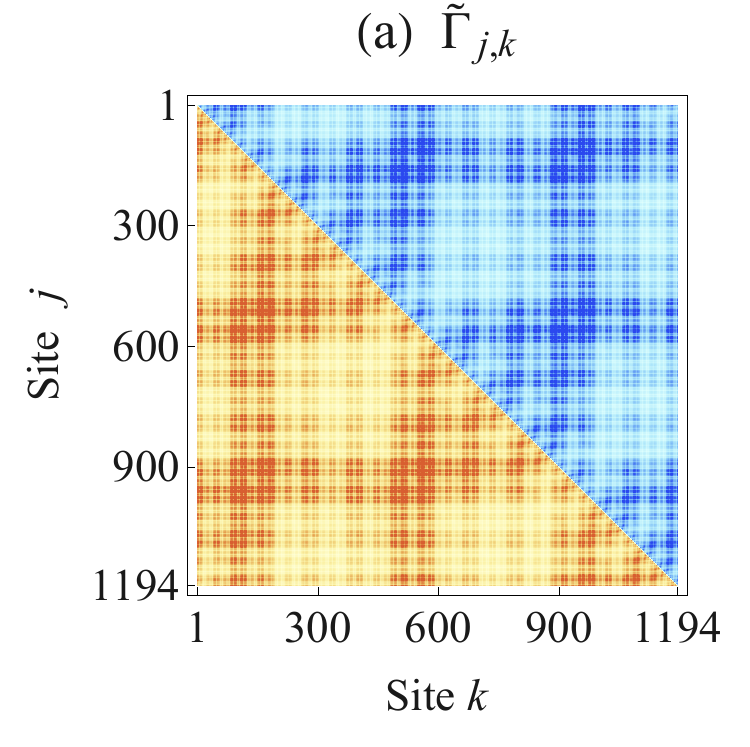}
\hspace{0.2cm}
\includegraphics[height=0.22\textheight]{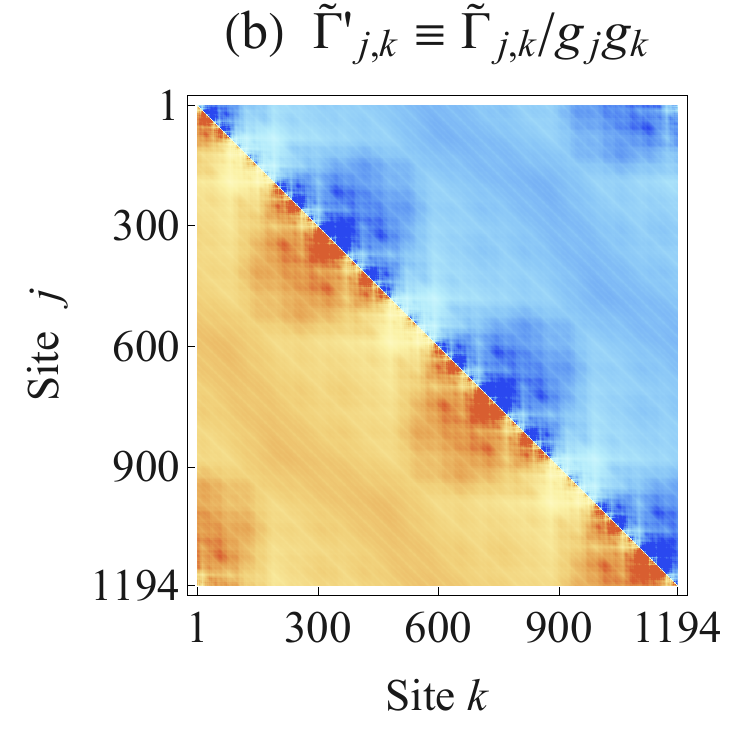}
\hspace{0.2cm}
\includegraphics[height=0.22\textheight]{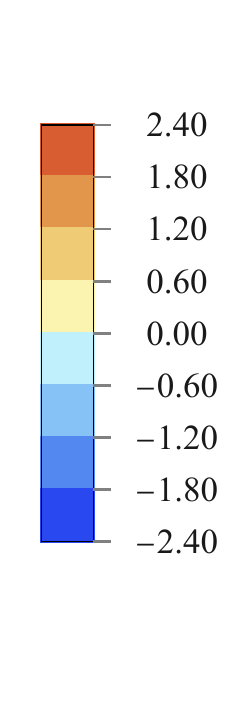}
\caption{Decay-adjusted covariance matrix $\tilde{\Gamma}$ of the $\{3,7\}$ MTN boundary state for a non-Ising bulk input $a=0.3$.
(a)~Original covariance matrix with disordered correlation pattern.
(b)~Nearly translation-invariant matrix $\tilde{\Gamma}^\prime$ after applying a disorder vector $g$ on every site. Deviations from translation invariance are stronger than in the critical Ising case of Fig.\ \ref{FIG_QR_SYMMETRIES}(b).
}
\label{FIG_QR_SYMMETRIES_NONISING}
\end{figure}

We have previously considered matchgate tensor networks at the critical Ising point, i.e., where average boundary correlations between fermionic modes decay inversely with the boundary distance.
However, the matchgate ansatz can produce much more general boundary states: For the case of hyperbolic regular tilings with isotropic bulk input, any polynomial decay of correlations with power $p \geq 1$ corresponds to some value of the matchgate parameter $a$ considered in the main text \cite{Jahn:2017tls}.
As an example, we consider a bulk input parameter $a=0.3$, which produces an average correlation falloff (as defined in \eqref{EQ_AVG_CORR_FALLOFF}) of approximately $c(d) \propto 1/d^5$.
Unsurprisingly, the resulting boundary states exhibit multi-scale quasiperiodic boundary symmetries very similar to the critical Ising case, as their geometrical construction still enforces the qCFT symmetries following from the multi-scale quasicrystal ansatz as shown in Fig.\ \ref{FIG_QCFT_SCALING}.
This allows us to extract a disorder vector $g$ in the same way as at the critical Ising point.
As shown in Fig.\ \ref{FIG_QR_SYMMETRIES_NONISING}, we can use $g$ to map the covariance matrix to one that is nearly translation-invariant at large scales, analogously to Fig.\ \ref{FIG_QR_SYMMETRIES}. In direct comparison of both plots, one finds that the resulting $g$ function varies more strongly away from the Ising point, and that the ``reordered'' decay-adjusted covariance matrix $\tilde{\Gamma}^\prime$ shows stronger deviations from translation invariance at small separations.

Unfortunately, the identification of such a model with a Hamiltonian is unclear. It appears plausible that various choices for $a$ can be used to approximate a translation-invariant chain of fermions at a critical point, for example of Hamiltonians with couplings more complicated than nearest-neighbor ones. However, we cannot a priori determine if the low-energy spectrum of such models matches with that of an equivalent model without disorder, as we have found for the critical Ising case. 
This leaves the non-Ising case to be explored further in future work.

\section{The $\{4,5\}$ matchgate tensor network}
\label{APP_45TILING}

\begin{figure}[t]
\centering
\includegraphics[width=1.0\textwidth]{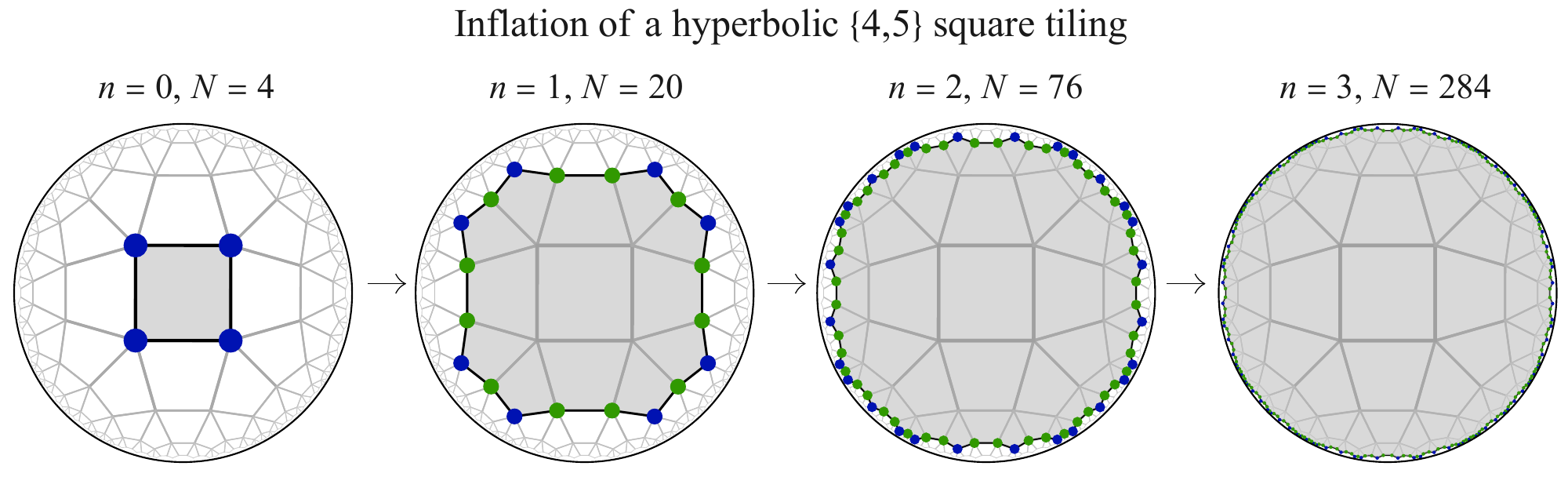}\\
\includegraphics[width=1.0\textwidth]{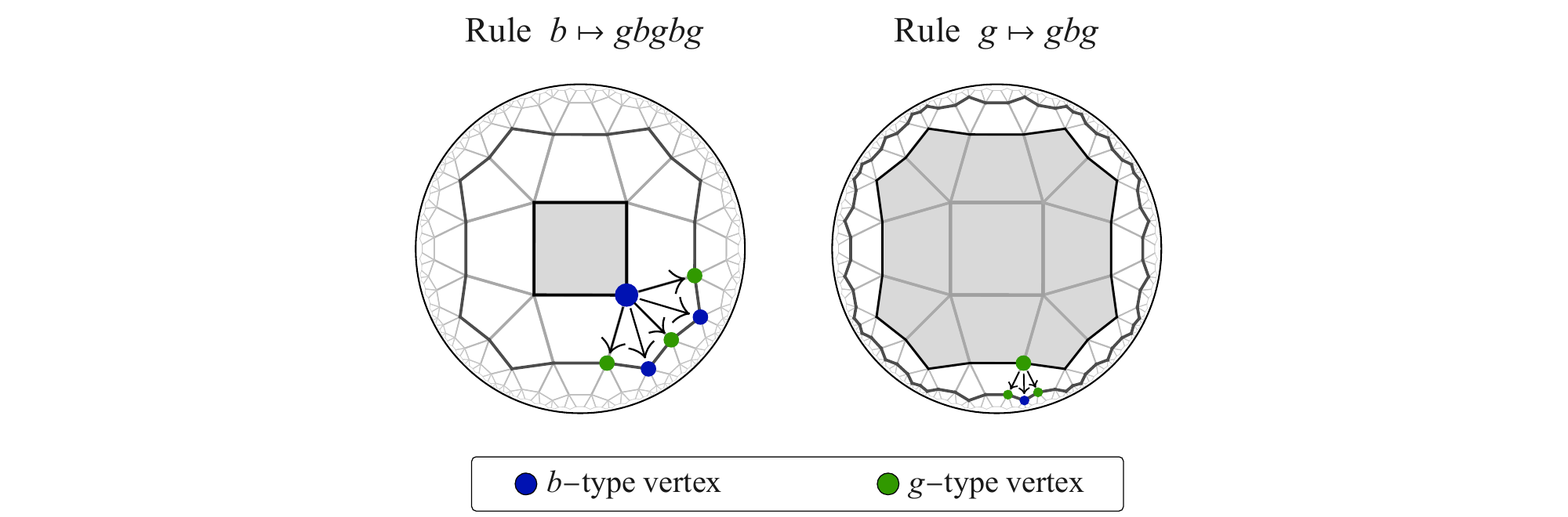}
\caption{Iterative construction of a regular hyperbolic tiling through inflation rules \eqref{EQ_V_INFLATION_37} of a $\{4,5\}$ square tiling.
As in the $\{3,7\}$ case in Fig.\ \ref{FIG_TILING_INFLATION}, the number of boundary sites $N$ increases exponentially with the number of iterations $n$.
The vertex types $b,g$ denote boundary vertices with zero and one adjacent edges connected to the previous inflation layer, respectively.
}
\label{FIG_TILING_INFLATION2}
\end{figure}

The MTN setup in the main text focused on a hyperbolic $\{3,7\}$ tiling with triangles corresponding to $3$-leg tensors.
In this appendix, we show that some key results of the $\{3,7\}$ case can also be reproduced with an MTN on a the hyperbolic $\{4,5\}$ tiling, i.e., with $4$-leg tensors.
We first review the vertex inflation rule for this tiling, shown in Fig.\ \ref{FIG_TILING_INFLATION2}. Unlike the $\{3,7\}$ case, the tiling boundary at each inflation step can be fully characterized in terms of only two letters $b$ and $g$. As in the $\{3,7\}$ case, we define a $b$ vertex as being only connected to vertices that are also boundary vertices, i.e., to zero ''interior vertices'' (considering only vertices in the tiling up to the given inflation step).
The $g$ vertex is again defined as a boundary vertex connected to one interior vertex. We find that no $r$ boundary vertices with more interior connectivity exist, as the boundary produced by vertex inflation is convex. 
One then finds the following vertex inflation rule \cite{Jahn:2019mbb}
\begin{align}
\label{EQ_V_INFLATION_45}
b &\mapsto gbgbg \ , & 
g &\mapsto gbg \ .
\end{align}
Starting with a single tile corresponds to the initial letter sequence $bbbb$ (Fig.\ \ref{FIG_TILING_INFLATION2}, top left). 
The asymptotic scaling factor of the number of boundary vertices in each inflation is step is found to be $\lambda_{\{4,5\}} = 2 {+} \sqrt{3}$.

\begin{figure}[ht]
\centering
\includegraphics[width=1.0\textwidth]{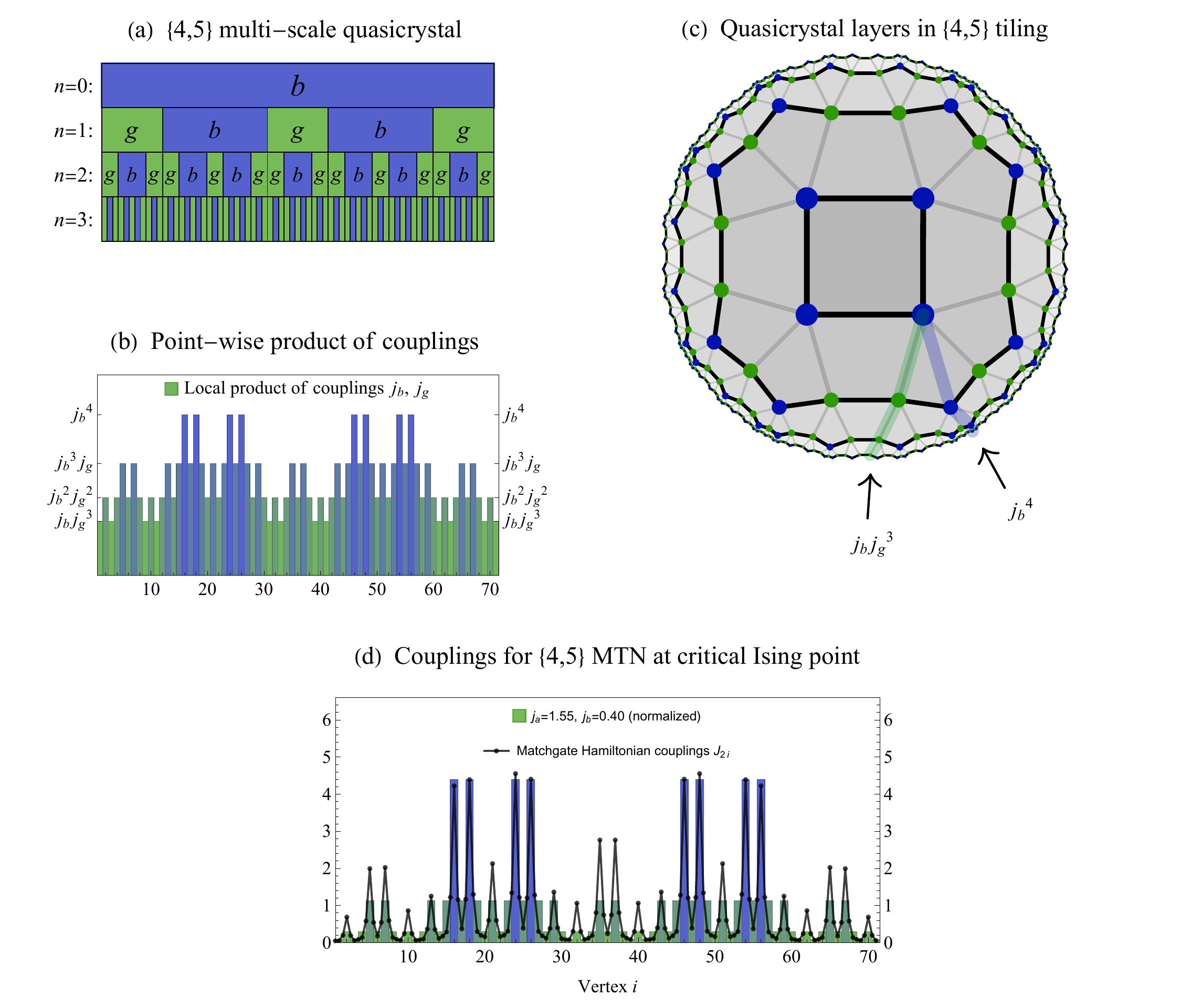}
\vspace{-0.5cm}
\caption{Construction of a multi-scale quasicrystal ansatz (MQA) for coupling sequences of the $\{4,5\}$ tiling.
(a) The MQA of Ref.\ \cite{Jahn:2020ukq} composed of inflation layers following the rule \eqref{EQ_V_INFLATION_45}.
(b) A point-wise (vertical) product of couplings $j_l$ corresponding to each letter $l \in (g,r)$ of the inflation sequence. Each bar corresponds to the product of blocks directly above it in (a).
(c) The inflation layers embedded into the $\{4,5\}$ tiling, with the path between layers highlighted for three point-wise products.
(d) For a suitable choice of the $j_l$ the couplings closely reproduce the effective Hamiltonian couplings for the $\{4,5\}$ MTN states at the critical Ising point.
The couplings are shown normalized so that $\langle v \rangle = 1$.
}
\label{FIG_MQI_45}
\end{figure}

The matchgate tensor corresponding to each square tile can be constructed similarly to the triangular case.
The tensor $T$ is defined by a $4 \times 4$ generating matrix $A$ as
\begin{align}
T_{i,j,k,l}(A) = \bra{i,j,k,l} c \exp\left( \frac{1}{2} \sum_{a,b=1}^3 A_{a,b} \fd_a \fd_b \right) \vacket_3 \ ,
\end{align}
with $c$ again being a normalization constant and the fermionic basis vector
\begin{align}
\ket{i,j,k,l} := (\fd_1)^i (\fd_2)^j (\fd_3)^k (\fd_3)^l \vacket_4 \ ,
\end{align}
each index being either zero or one.
Choosing a matrix $A$ that preserves rotational symmetry now leads to two free parameters $a$ and $b$ in the form
\begin{equation}
A = 
\begin{pmatrix}
\msp 0 & \msp a & \msp b & \msp a \\
-a & \msp 0 & \msp a & \msp b \\
-b &-a & \msp 0 & \msp a \\
-a &-b &-a & \msp 0
\end{pmatrix} \ .
\end{equation}
In the course of the optimization, we again choose $0 \leq a,b \leq 1$ to ensure a well-behaved decay of correlations.
Numerically minimizing the energy of the resulting MTN boundary states with respect to the Ising Hamiltonian $H^\text{I}$ yields a set of minimal solutions, one of which is given by $a=b\approx 0.46$. Boundary correlations of this state are again disordered with multi-scale quasiperiodicity, and it is possible to construct a mode-disordered Ising (MDI) Hamiltonian $H^\text{MDI}[g]$ whose ground state closely matches the $\{4,5\}$ MTN boundary state.
As an estimator for the disorder vector $g$, we generalize \eqref{EQ_DISORDER_37_EST} to
\begin{equation}
\label{EQ_DISORDER_45_EST}
g_j \approx \frac{\sum_{k=1}^{N} \tilde{\Gamma}_{j,\frac{N}{2}+j+k}}{N_g} \ , \quad
N_g = \frac{2}{N}\sum_{j=1}^{N/2} \sum_{k=1}^{N} \tilde{\Gamma}_{j,\frac{N}{2}+j+k} \ .
\end{equation}
As with the $\{3,7\}$ MTN, the $\{4,5\}$ MDI model represents a disordered Ising model whose average correlation functions closely approximate that of the original Ising model.
As we show in Fig.\ \ref{FIG_MQI_45}, the disorder vector $g$ corresponds to coupling terms $J_k = 1 / g_k g_{k+1}$ that can again be close approximated by a multi-scale quasicrystal ansatz (MQA) with two coupling parameters $j_b$ and $j_g$, associated with the vertex types $b$ and $g$, defining a multi-scale quasicrystal Ising (MQI) model.

Similarly to the analysis in Sec.\ \ref{SS_MQI}, the values of $j_b$ and $j_g$ can be shown to be  constrained by the requirement of finiteness in the scaling limit. 
For this purpose, we first define the (modified) substitution matrix
\begin{align}
M_{\{4,5\}} &= 
\begin{pmatrix}
2 & 3 \\
1 & 2 
\end{pmatrix} \ , &
M_{\{4,5\}}^\prime &= 
\begin{pmatrix}
2 j_b & 3 j_g \\
1 j_b & 2 j_g
\end{pmatrix} \ ,
\end{align}
which leads to an inflation step
\begin{equation}
\vec{v}_{\{4,5\}}^{(n)} \mapsto \vec{v}^{(n+1)}_{\{4,5\}} = \vec{v}^{(n)}_{\{4,5\}} M_{\{4,5\}}
\end{equation}
for a state vector $\vec{v}^{(n)}_{\{4,5\}}$ with two entries that counts the number of $b$ and $g$ vertices at each inflation step. For example, the starting sequence $b$ evolves through two inflation steps as
\begin{equation}
\vec{v}_{\{4,5\}}^{(0)} = \vectwo{1}{0} \mapsto \vectwo{2}{3} \mapsto \vectwo{7}{12} \ .
\end{equation}
We again define the analogous MQA inflation step as
\begin{equation}
\vec{v}^{\prime (n)}_{\{4,5\}} \mapsto \vec{v}^{\prime (n+1)}_{\{4,5\}} = \vec{v}^{\prime (n)}_{\{4,5\}} M^\prime_{\{4,5\}}
\end{equation}
for a modified state vector $\vec{v}^{\prime (n)}_{\{4,5\}}$ that stores the sum of all multi-scale products (see Fig.\ \ref{FIG_MQI_45}(a) and (b)) with a final $j_b$ and $j_g$ coupling, respectively.
By requiring that $\vec{v}^{\prime (n)}_{\{4,5\}}$ grows with the same asymptotic scale factor as $\vec{v}_{\{4,5\}}^{(n)}$, given by $\lambda_{\{4,5\}}$, we ensure finiteness of the average multi-scale coupling in the limit $n \to \infty$ of many inflation steps.
This leads to the constraint
\begin{align}
j_g = \frac{7 + 4\sqrt{3} - 2\left(\sqrt{3}+2\right) j_b}{4 + 2 \sqrt{3} - j_b} \ .
\end{align}
Again, a choice $v_b \approx 1.55, v_g \approx 0.40$ fulfilling this constraint leads to a final sequence of couplings closely resembling the matchgate results (Fig.\ \ref{FIG_MQI_45}(d)).

\end{document}